\documentclass[12pt, doublesided,onecolumn]{extbook}
\usepackage[a4paper, width=150mm, top=35mm, bottom=35mm, hmarginratio=3:2]{geometry} 
\usepackage{graphicx}
\usepackage{latexsym}
\usepackage{amsmath}
\usepackage{amssymb}
\usepackage{lipsum,float}
\usepackage{subfig, caption}
\usepackage{quotchap}

\usepackage{fancyhdr}
\usepackage{listings}
\usepackage{bm}
\usepackage{fancyheadings}
\usepackage{ulem}
\usepackage{xcolor}
\usepackage{tikz}
\usetikzlibrary{backgrounds}
\usetikzlibrary{calc}
\usepackage{slashed}
\usepackage{stackengine}
\newcommand\xrowht[2][0]{\addstackgap[0.5\dimexpr#2\relax]{\vphantom{#1}}}
\usepackage{array}
\usepackage{rotating}
\DeclareCaptionLabelFormat{adja-page}{\\#1 #2 \emph{(previous page)}}
\usepackage{upgreek}
\usepackage[square,numbers,sort&compress]{natbib}
\usepackage[nottoc]{tocbibind} 
\usepackage{enumitem}
\newlist{abbrv}{itemize}{1}
\setlist[abbrv,1]{label=,labelwidth=1.5in,align=parleft,itemsep=0.1\baselineskip,leftmargin=!}
\usepackage[printonlyused,withpage]{acronym}
\usepackage{xparse}    
\NewDocumentCommand{\DIV}{om}{%
    \IfValueT{#1}{\setcounter{#2}{\numexpr#1-1\relax}}%
    \csname #2\endcsname
}
\usepackage{widetext}
\usepackage{xspace}
\usepackage{accents}
\usepackage{wasysym}
\DeclareMathAlphabet{\bi}{OML}{cmm}{b}{it}
\fancyheadoffset{0.005\textwidth}

\definecolor{ThemeColor1}{cmyk}{0,1,0.65,0.34}
\definecolor{ThemeColor2}{cmyk}{0,0.08,0.27,0}

\usepackage{setspace} 
\usepackage{hyperref}
\hypersetup{
    colorlinks=true,
    linkcolor=blue,
    filecolor=magenta,      
    urlcolor=magenta,
}
\usepackage[utf8]{inputenc}
\usepackage[english]{babel} 

\begin{document}
\frontmatter
%
%
%
%
%
%

\begin{tikzpicture}
[remember picture, overlay] \draw[line width=3pt] ($(current page.north west)+(28mm,-21mm)$) rectangle ($(current page.south east)+(-19mm,21mm)$);
[remember picture, overlay] \draw[line width=1pt] ($(current page.north west)+(29mm,-22mm)$) rectangle ($(current page.south east)+(-20mm,22mm)$);
\end{tikzpicture}

\marginparsep 0pt
\textwidth 15.0 truecm
\setlength{\baselineskip}{0.780cm}
\setcounter{page}{1}
\pagestyle{empty}

\begin{center}
\vspace{-0.8cm}
\LARGE{\tikz[remember picture,overlay] \draw [fill,ThemeColor2!90] ($(current page.north west)+(29.23mm,-31mm)$) rectangle ($(current page.south east)+(-20.23mm,239.38mm)$);%
\tikz[remember picture,overlay] \draw [fill,purple!100] ($(current page.north west)+(29.23mm,-31mm)$) rectangle ($(current page.south east)+(-20.23mm,267.2mm)$);%
\tikz[remember picture,overlay] \draw [fill,purple!100] ($(current page.north west)+(29.23mm,-58.7mm)$) rectangle ($(current page.south east)+(-20.23mm,239.50mm)$);%
\bf 21 cm Line Astronomy and Constraining New Physics
}\\
\end{center}

\vspace{1cm}
\begin{center}
{\it \large A Thesis Submitted \\ in Partial Fulfilment of 
the Requirements \\ for the Degree of \\ \Large{\bf Doctor of Philosophy}} \\
\vspace{1cm}
\large{ by\\
	\vspace{0.1cm}
{\Large{\bf Pravin Kumar Natwariya}}\\}
\vspace{0.1cm}
{Roll No. \it 17330022}\\
\vspace{1.25cm}
Under the guidance of\\
\vspace{0.2cm}
{\large{\bf Prof. Jitesh R. Bhatt}}\\
Theoretical Physics Division\\
Physical Research Laboratory, Ahmedabad.\\
\vspace{1.35cm}
to the \\
\begin{figure}[h]
\begin{center}
\includegraphics[scale=.125]{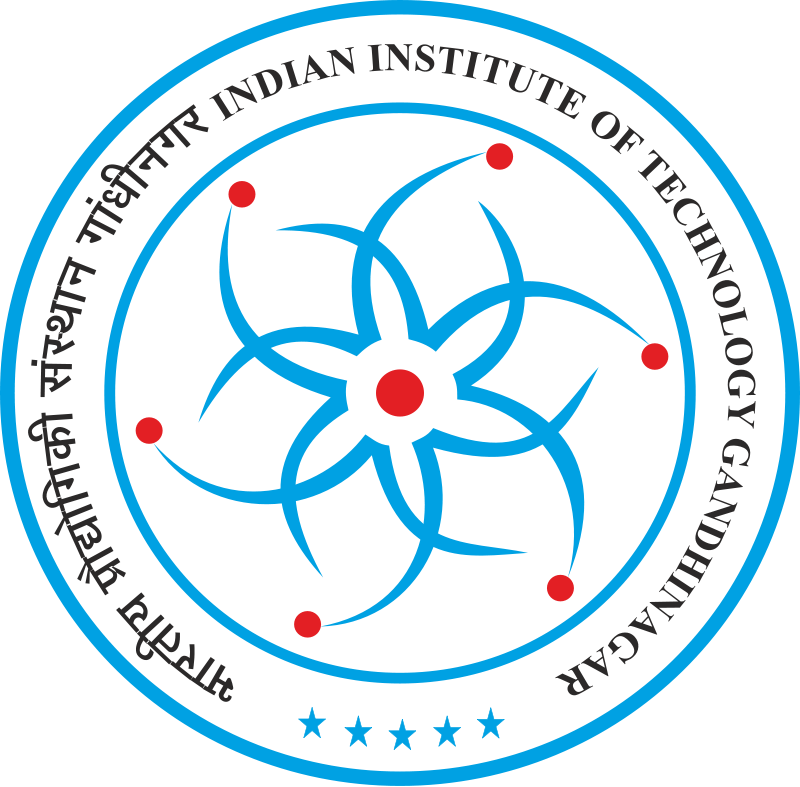}
\end{center}
\end{figure}  
\vspace{-0.5cm}
{\large { Discipline of Physics}\\}
{\large { Indian Institute of Technology Gandhinagar,\\ Gujarat 382355, India.}\\}
\vspace{0.65cm}
\vspace{-0.25cm}
\end{center}

\cleardoublepage
%
%
%
%
%
%
%
%
%
%
%
%
%
%
%
%
%
%
%
%
\captionsetup[table]{font={stretch=1.5}}
\captionsetup[figure]{font={stretch=1.5}}
\setstretch{1.5}
\clearpage
\pagestyle{empty}
\cleardoublepage
\phantomsection
\addcontentsline{toc}{chapter}{List of Acronyms} 
\centerline{\Large\bf\underline{List of Acronyms}}
\bigskip

\begin{abbrv}
	\item[$\Lambda$CDM] $\Lambda$ Cold Dark Matter
	\item[BBN] Big-Bang Nucleosynthesis
	\item[WMAP] Wilkinson Microwave Anisotropy Probe
	\item[COBE] COsmic Background Explorer
	\item[QCD] Quantum ChromoDynamic
	\item[EoR] Epoch of Reionization
	\item[IGM]		InterGalactic Medium
	\item[JWST] James Webb Space Telescope
	\item[EDGES] Experiment to Detect the Global Epoch of Reionization Signature
	\item[ISM] InterStellar Medium
	\item[CMB] Cosmic Microwave Background
	\item[SIDM]  Self-Interacting Dark Matter
	\item[WDM]  Warm Dark Matter
	\item[THESEUS] Transient High Energy Sky and Early Universe Surveyor 
	\item[NuSTAR] Nuclear Spectroscopic Telescope Array
	\item[CMBR] Cosmic Microwave Background Radiation
	\item[LIGO]		Laser Interferometer Gravitational-Wave Observatory
	\item[PBH]		Primordial Black Hole
	\item[AGN]       Active Galactic Nuclei
	\item[LUX]		Large Underground Xenon
	\item[PandaX]	Particle and astrophysical Xenon
	\item[CRESST] Cryogenic Rare Event Search with Superconducting Thermometers
	\item[PICO]	{\bf PI}CASSO and {\bf CO}UPP 
	\item[PICASSO]  Project in CAnada to Search for Super-symmetric Objects
	\item[FIRAS] Far Infrared Absolute Spectrophotometer
	\item[COUPP]	Chicagoland Observatory for Underground Particle Physics
	\item[AMEGO] All-sky Medium Energy Gamma-ray Observatory
	\item[PMFs] Primordial Magnetic Fields
	\item[MHD] MagnetoHydroDynamics 
	\item[ARCADE] Absolute Radiometer for Cosmology, Astrophysics and Diffuse Emission
	\item[LWA] Long Wavelength Array 
	\item[HESS] High Energy Stereoscopic System
\end{abbrv}
\cleardoublepage
\clearpage
\pagestyle{empty}
\cleardoublepage
\pagestyle{myheadings}
\markright{}
{\large {\tableofcontents}}
\cleardoublepage
\clearpage
%
%
%
%
%
%
%
%
%
%
%
%
\clearpage
\pagestyle{empty}
\cleardoublepage
\fancyhf{}
\pagestyle{fancyplain}
\renewcommand{\chaptermark}[1]{\markboth{#1}{}}
\renewcommand{\sectionmark}[1]{\markleft{\thesection\ #1}}
\renewcommand{\sectionmark}[1]{\markright{\thesection\ #1}}
\fancyhead{}
\fancyhead[RO,LE]{{\textbf{21 cm Line Astronomy and Constraining New Physics}}}
\renewcommand{\headrulewidth}{1pt}
\fancyfoot{}
\fancyfoot[RE,LO]{\textbf{\thepage}} 
\fancyfoot[CE,CO]{\textbf{\leftmark}}
\fancyfoot[LE,RO]{\textbf{Chapter \thechapter}}
\renewcommand{\footrulewidth}{1pt}
\fancypagestyle{plain}{%
	\fancyhead{}%
	\fancyfoot{}
	\renewcommand{\headrulewidth}{0pt}
	\renewcommand{\footrulewidth}{0pt}
}

\mainmatter
\clearpage
\pagestyle{empty}
\cleardoublepage
\pagestyle{fancy}

\begin{savequote}[75mm]
	``It might seem limited to impose our human perception to try to deduce the grandest cosmic code. But we are the product of the universe and I think it can be argued that the entire cosmic code is imprinted in us. Just as our genes carry the memory of our biological ancestors, our logic carries the memory of our cosmological ancestry. We are not just imposing human-centric notions on a cosmos independent of us. We are progeny of the cosmos and our ability to understand it is an inheritance.''
	\qauthor{Janna Levin, \textit{How the Universe Got its Spots (2002)}}
\end{savequote}

\chapter{Introduction}\label{chap1}
\vspace{-1.5cm}

In the 21st century, our knowledge of the Universe has proliferated--- thanks to the tremendous progress of observational instruments in the last three decades. Especially the precision cosmology has grown remarkably in the past three decades as a result of an ample amount of high-quality Cosmic Microwave Background (CMB) data, in addition to the data and comprehensive studies of supernovae, stars and nearby galaxies. It is the greatest triumph of precision cosmology that we now know the age of our Universe, and it is only the tip of the iceberg. As another example, we know that observable/baryonic matter in the Universe is only about 5 percent and the leftover energy component consists the dark matter ($\sim26$ percent) and dark energy ($\sim69$ percent) based on the $\Lambda$CDM model of cosmology--- the standard model of cosmology \cite{Planck:2018}. Here, $\Lambda$ represents the dark energy, and CDM represents the cold dark matter.
\begin{figure}[]
    \begin{center}
        {\includegraphics[width=5.1in,height=2.4in]{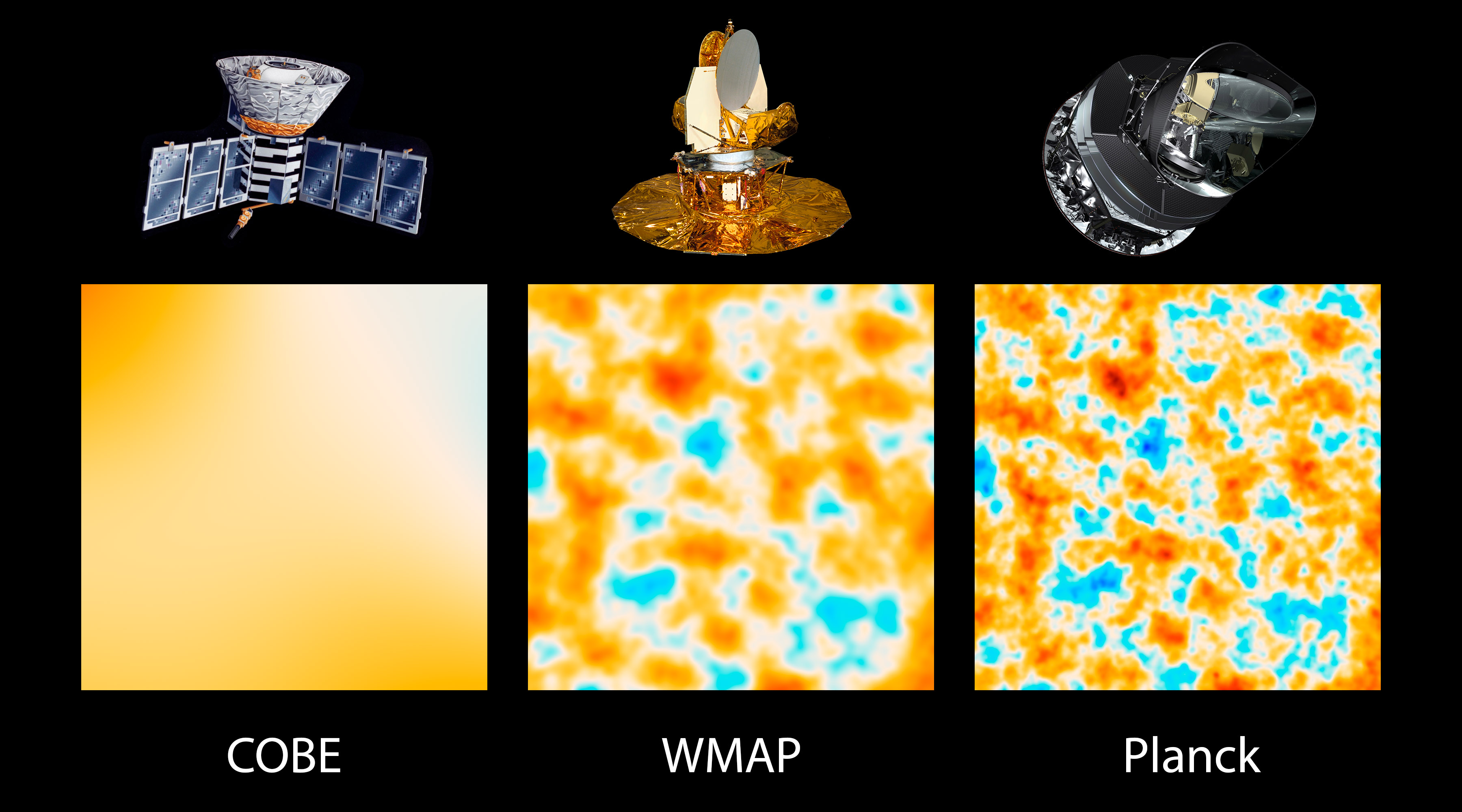}} 
    \end{center}
    \caption{The graphic represents the evolution of precision cosmology over three decades--- comparison between CMB temperature maps reported by each satellite. Left to right: Cosmic Background Explorer (COBE)\protect\footnotemark--- launched in 1989, WMAP--- launched in 2001 and Planck--- launched in 2009. {\it Image credits: NASA/JPL-Caltech/ESA,  \href{https://www.nasa.gov/mission\_pages/planck}{https://www.nasa.gov/mission\_pages/planck}}.}\label{CoWmPl}
\end{figure}
\footnotetext{\href{http://lambda.gsfc.nasa.gov/product/cobe/}{http://lambda.gsfc.nasa.gov/product/cobe/}}
The $\Lambda$CDM model, together with the cosmological inflation, can provide a complete picture of the evolution of our Universe from the beginning. The CMB data from Wilkinson Microwave Anisotropy Probe (WMAP)\footnote{\href{https://map.gsfc.nasa.gov/}{https://map.gsfc.nasa.gov/}} played a crucial role in establishing the $\Lambda$CDM model. It is also supported by the Planck\footnote{\href{https://www.esa.int/Science_Exploration/Space_Science/Planck}{https://www.esa.int/Science\_Exploration/Space\_Science/Planck}} observations \cite{Planck::2016, Planck:2018}. The $\Lambda$CDM model is widely accepted now, and there are various good reasons to believe this model: N-body simulations of structure formation based on the $\Lambda$CDM framework can explain the observed large scale structure of the Universe \cite{Springel:2005},  it can also explain the CMB anisotropies \& polarization \cite{Spergel:2003, Spergel:2007, Jarosik:2011, Planck::2016, Planck:2018} and accelerating expansion of the Universe caused by cosmological constant $\Lambda$ \cite{Riess:1998, Perlmutter:1999, Planck:2018}\footnote{Saul Perlmutter with Brian P. Schmidt and Adam G. Riess received the Nobel Prize in Physics for 2011 ``for the discovery of the accelerating expansion of the Universe through observations of distant supernovae."}. In addition to this, the predictions for the helium and deuterium fractions by the standard Big-Bang Nucleosynthesis (BBN) for $\Lambda$CDM cosmology agree very well with observations.

\section{Evolution of our Universe}\label{EoU}
\begin{figure}[]
    \begin{center}
        {\includegraphics[width=5.85in,height=2.8in]{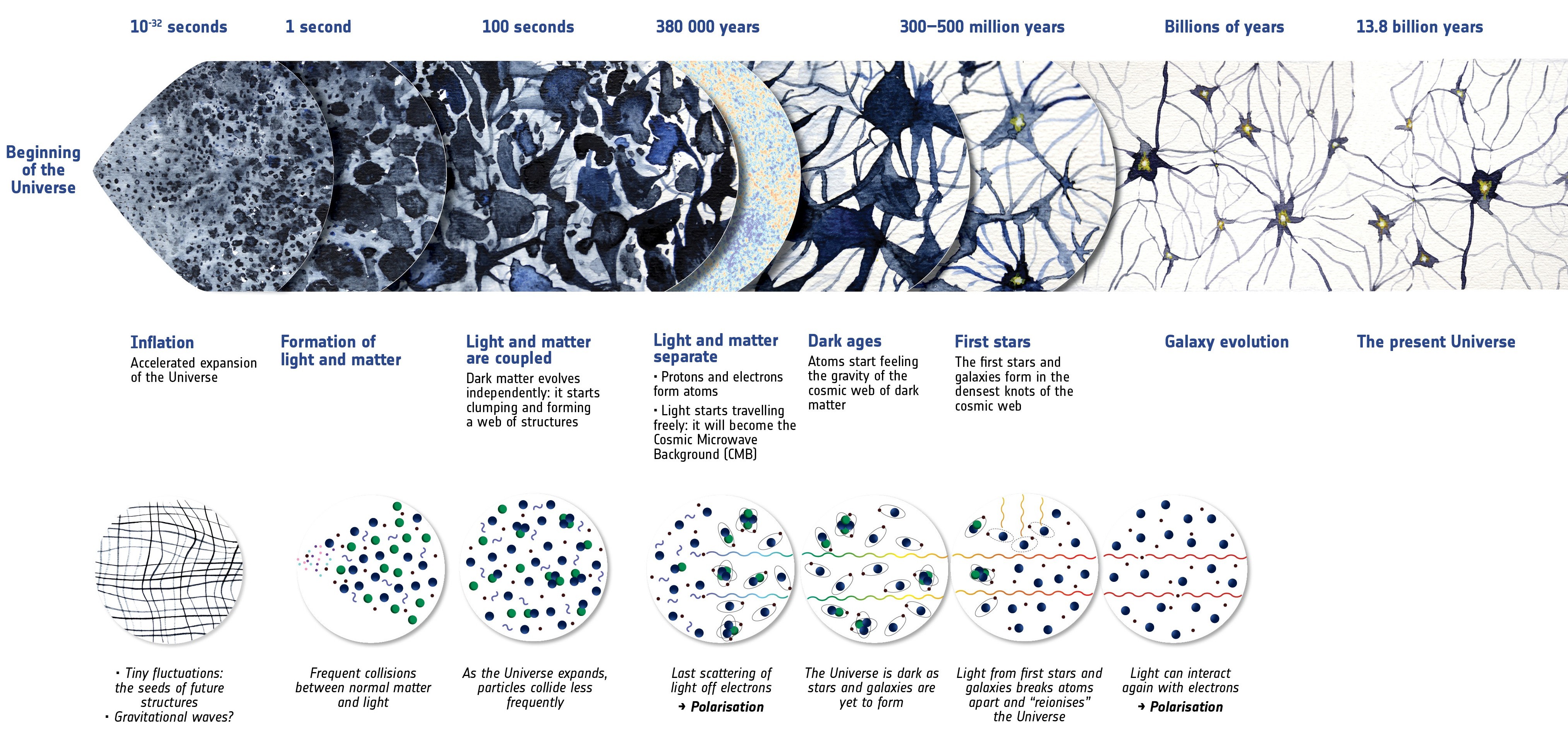}} 
    \end{center}
    \caption{The evolution of the Universe from it's beginning. {\it Image credits: European Space Agency} ({\it ESA})\protect\footnotemark.}\label{History}
\end{figure}
\footnotetext{\href{https://www.esa.int/}{https://www.esa.int/}} 

Before going into 21 cm cosmology,  we briefly review cosmic history from the beginning of our Universe to the present day. The figure \eqref{History}, shows a schematic picture of the evolution of our Universe from the beginning. The best current widely agreed model of the origin and evolution of our Universe is the Big-Bang model. According to this model, our Universe came into existence with a Big-Bang about 13.8 billion years ago \cite{Planck:2018}. Observations of CMB also support this theory \cite{COBE:1993, Planck:2018}\footnote{Results from the COBE were honoured with the Nobel Prize in Physics 2006.}. According to our best present-day understanding, the early Universe had an exponential expansion after the Big-Bang--- it is known as the inflationary epoch\footnote{Alan H. Guth, Andrei D. Linde and Alexei A. Starobinsky received 2014 KAVLI prize in Astrophysics ``for pioneering the theory of cosmic inflation."} \cite{Guth:1981, Linde:1982}. 
There are several reasons to believe the inflation model: It can solve the three technical problems of the Big-Bang model--- the horizon problem, flatness problem and the magnetic monopole problem \cite{Guth:1981, Linde:1982}:
\begin{itemize}
    \item The Big-Bang model fails to explain why causally disconnected regions appear homogeneous. The observation shows that the CMB temperature is uniform up to a scale of $\Delta T/T\approx10^{-5}$ even when observed in opposite directions. Here, $\Delta T$ is the temperature difference between the two regions of the sky, and $T$ is the average temperature over the whole sky. Assuming the standard Big-Bang model, opposite directions were so far separated that they always have been acausal. Then, why does CMB appear so uniform? It is known as the horizon problem. 
    \item The second one is the flatness problem: The present-day total energy density of the Universe is equal to the critical energy density of the Universe. Any departure from the critical density will result in the curvature of the Universe. The observation shows dimensionless curvature energy density of the Universe $\Omega_k=0.001\pm 0.002$ \cite{Planck:2018}. It implies a flat Universe. A slight deviation of total energy density from critical energy density would have resulted in extreme effects on the flatness of the Universe over the cosmic time. Therefore, a flat universe like ours requires extreme fine-tuning conditions in the beginning. It is known as the flatness problem. 
    \item The Grand Unified Theories (GUT) predict the existence of magnetic monopoles as at a very high temperature as the electromagnetic, weak and strong forces are not fundamental forces. Therefore, there can exist many stable magnetic monopoles in the Universe. No monopoles have been observed yet. It is known as the monopole problem. 
\end{itemize}
These problems of the Big-Bang model can be circumvented by introducing the cosmic inflation model \cite{Guth:1981, Linde:1982}. Additionally, the inflation can give an idea of the origin of the observed structures in the Universe. The quantum fluctuations, prior to inflation, embedded in the initial energy density might have grown to astronomical scales over the cosmic time. Later, the dense regions might have condensed into structures like stars, galaxies and clusters of galaxies. The inflation epoch ends when inflation potential steepens, and the inflation field acquires kinetic energy. Then inflation sector energy creates the standard model particles. This process is known as reheating. As the Universe expands continuously, it cools down. Then, Baryogenesis (excess of baryons over antibaryons)\footnote{The exact time and mechanism for Baryogenesis are not exactly known yet.}, electroweak phase transition (100 GeV) and QCD phase transition (150 MeV) takes place. The table \eqref{tab:1}, represents the time scale, redshift and temperature for various events in the Universe. 
\begin{table}
\begin{center}
    \begin{tabular}{|| m{5.5cm} | m{2.4cm} | m{2.3cm} | m{2.8cm} ||} 
        \hline\xrowht[()]{15pt}
        Event & time & redshift & Temperature \\ [2ex] 
        \hline\hline\xrowht[()]{15pt}
        Inflation & $10^{-36}$~sec & - & - \\ [1ex]
        \hline\xrowht[()]{15pt}
        Baryogenesis & ? & ? & ? \\[1ex]
        \hline\xrowht[()]{15pt}
        Electroweak phase transition & 20 ps & $10^{15}$ & 100~GeV \\[1.5ex]
        \hline\xrowht[()]{15pt}
        QCD phase transition & 20 $\mu$s & $10^{12}$ & 150~MeV \\[1ex]
        \hline\xrowht[()]{15pt}
        Dark matter freeze-out & ?& ? & ? \\[1ex]
        \hline\xrowht[()]{15pt}
        Neutrino Decoupling & 1 sec & $6\times10^9 $ & 1~MeV \\[1ex]
        \hline\xrowht[()]{15pt}
         Electron-positron annihilation & 6 sec & $2\times10^9$ & 500 KeV \\[1ex]
        \hline\xrowht[()]{15pt}
         Big-Bang nucleosynthesis & 3 minute & $4\times10^8$ & 100 KeV \\[1ex]
        \hline\xrowht[()]{15pt}
        Matter-radiation equality & 60 Kyr & 3400 & 0.75 eV \\[1ex]
        \hline\xrowht[()]{15pt}
        Recombination & $260-380$ Kyr & $1400-1100$ & $0.33- 0.26$ eV \\[1ex]
        \hline\xrowht[()]{15pt}
        Photon decoupling & $\sim380$ Kyr & $\sim1100$ & $\sim0.27$ eV \\[1ex]
        \hline\xrowht[()]{15pt}
        First stars formation & $\sim100$ Myr& $\sim30$ & $\sim7$~meV \\[1ex]
        \hline\xrowht[()]{15pt}
        Reionization & $\sim400$ Myr& $\sim11$ & $\sim2.6$~meV \\[1ex]
        \hline\xrowht[()]{15pt}
        Dark energy-matter equality & 9 Gyr & 0.4 & 0.33 meV \\[1ex]
        \hline\xrowht[()]{15pt}
        Present & 13.8 Gyr & 0 & 0.24 meV \\[1ex]
        \hline
    \end{tabular}
\end{center}
\caption{Approximate time scale, redshift and temperature for various events in the Universe. {\it Table credit: Daniel Baumann, ``Lecture notes on cosmology: Part III Mathematical Tripos." }}\label{tab:1}
\end{table}
The decoupling and freeze-out of various species can be understood by comparing the rate of interaction ($\Gamma$) and Hubble expansion ($H$). If $t_\Gamma\ll t_H$, then particle interactions dominates over expansion. Here, $t_*\equiv1/*$ is the time scale for corresponding rate ($*\equiv \Gamma$ or $H$). Therefore, local thermal equilibrium can be reached. As Universe cools down, the value of $t_\Gamma$ increases faster than $t_H$. At $t_\Gamma \sim t_H$ particles starts to decouple from thermal equilibrium. Different species decouple at different times as $t_\Gamma$ varies from species to species. If the mass ($m$) of particles becomes larger than their temperature ($T$), the distribution function is exponentially suppressed, $\propto e^{-m/T}$ and particles freeze out. For example, the cross-section for weak interaction is $\sigma\sim G_F^2\,T^2$; $G_F= 1.17\times10^{-5}~{\rm GeV^{-2}}$ is Fermi constant. It implies $\Gamma/H\sim(T/{\rm MeV})^3$. For example, the neutrinos interact through weak interaction only and they decouple around $T\sim {\rm 1 ~MeV}$ from primordial plasma.

\subsection{Big-Bang nucleosynthesis}

When plasma cools down below $\sim100$~KeV, around three minutes after the beginning of the Universe, Big-Bang nucleosynthesis takes place. In this phase, light elements were formed. The neutrons and protons start to form deuterium via the process,
\begin{alignat}{2}
n+p \leftrightarrow D+\gamma\,.\label{eq_np}
\end{alignat}
Now, these formed nuclei can form the heavier nuclei via the process,
\begin{alignat}{2}
D+p \leftrightarrow {\rm He^3}+\gamma\qquad{\rm and}\qquad 
D+{\rm He^3} \leftrightarrow {\rm He^4}+p\,.
\end{alignat}
The number density ratio of these elements can be found easily. For example: in equation \eqref{eq_np}, $\mu_n+\mu_p=\mu_D$ as $\mu_\gamma=0$. Here, $\mu$ is the chemical potential for the corresponding species. It implies the number densities ratio to be,
\begin{alignat}{2}
\left[\frac{n_D}{n_n\,n_p}\right]_{\rm eq}=\frac{3}{4}\, \left[\frac{m_D}{m_n\,m_p}\,\frac{2\,\pi}{T}\right]^{3/2}\ e^{-(m_D-m_n-m_p)/T}\,,
\end{alignat}
here, $T$ is the plasma temperature. $m_D$, $m_n$ and $m_p$ are masses of deuterium, neutron and proton, respectively.

\subsection{Recombination and photon decoupling}

Within the $\Lambda$CDM cosmology, free electrons and protons cool sufficiently after $\sim3\times 10^{5}$ years of Big-Bang to form neutral hydrogen atoms. Recombination occurs around redshift 1100. During this epoch, electrons and protons combine to form hydrogen atoms via the process,
\begin{equation}
e^-+p\leftrightarrow {\rm H}+\gamma\,.\label{p1}
\end{equation}
When the plasma temperature was above 1 eV, there were still free electrons and protons in the plasma. Photons remain tightly coupled to electrons due to Compton scattering, and electrons were coupled to protons due to Coulomb scattering. In turn, there was only a small density of neutral hydrogen atoms. When the plasma temperature decreased sufficiently, electrons and protons combined and formed hydrogen atoms. Subsequently, the free electron density fell rapidly. 
As the number density of free electrons decreased adequately, the mean free path of photons increased sharply; and photons decoupled from plasma. As discussed in the end of section \eqref{EoU}, one can estimate the photon decoupling redshift by relation, $t_{\Gamma_{\gamma}}(z_{\rm dec})\sim H(z_{\rm dec})\,$. Here, $\Gamma_{\gamma}(z_{\rm dec})=n_e(z_{\rm dec})\, \sigma_T$ is the photon interaction rate or photon mean free path at the time of decoupling, $z_{\rm dec}$ is the redshift of photon decoupling from plasma and $\sigma_T$ is Thomson cross-section. The electron number density ($n_e$) can be found by using the Saha equation for the process in equation \eqref{p1}. After solving the relation, we can find $z_{\rm dec}\sim 1100\,$ or corresponding time to 380,000 yr after Big-Bang. We can also estimate that the free electron fraction in the plasma remains only about one percent--- the plasma becomes mostly transparent for photons. This time is known as the surface of last-scattering. After decoupling, these photons stream freely and are known as CMB radiation (CMBR).

\subsection{Through the dark ages to the present day}

After photon decoupling from baryonic matter, there were no luminous objects--- this epoch is known as the dark ages. The Universe was predominantly neutral during this era. This period of darkness ensued until the first luminous object was not formed in the Universe for about a hundred million years after the Big-Bang. During this era, overdensity was growing in the dark matter perturbations already. Later, these overdensities reached a critical value and collapsed to form dark matter halos--- a gravitationally bound structure \cite{Kolb:1990}. The first generation of luminous objects sprung up around redshift 30 inside dark matter halos--- this period is known as the Cosmic Dawn. As of now, it is not clear that these objects were either quasars or stars. As the first stars formed in very different circumstances, they probably were very different from our nearby stars. After the formation of the first luminous objects, their radiation start to ionize the gas in the Universe. This era is known as the epoch of reionization (EoR). Three-year WMAP observations of CMB  suggest that reionization starts around redshift 11 and ends by $\sim7$ \cite{Spergel:2007}. Planck observations suggest instantaneous reionization with mid-point redshift of reionization $ 7.68 \pm0.79\,$ \cite{Planck:2018}. Supernovae observations suggest that the Universe enters into an accelerated expansion phase around redshift $\sim 0.5\,$ \cite{Perlmutter:1999}. This accelerating expansion can not be explained only by matter in the Universe. To explain, one requires the existence of dark energy \cite{Perlmutter:1999, Peebles:2003}. Then, we reach the present-day after 13.8 billion years from the Big-Bang. 

The first complexity in the physics, after the dark ages, emerged with the event of the formation of the first luminous objects. As of now, this era is not observed due to the lack of our instrumental capability. The recently launched James Webb Space Telescope (JWST)\footnote{\href{https://jwst.nasa.gov/}{https://jwst.nasa.gov/}} will be able to probe the Universe back to redshift $\sim20$. One of the best pre-eminent and promising methods to probe the cosmic dawn era is the observation of the redshifted radiation from the hyperfine transition in the ground state of the neutral hydrogen atom. The low-frequency radio telescopes, sensitive to a frequency of as low as 40 MHz, can help to explore this era.

\section{21 cm line as a probe during end of darkness}\label{sec212}

The 21 cm signal appears to be a  treasure trove to provide an insight into the period when the first luminous objects were formed; hereafter we will refer these objects as first stars. The 21 cm line has been actively used to trace the neutral hydrogen in Milky Way for more than seven decades since its first observation in 1951 \cite{Ewen:1951}. It was first suggested by H. C. van de Hulst in 1945 that a 21 cm line might be observable in the galactic radiation spectrum \cite{Hulst:1945}. However, probing the neutral hydrogen during and pre cosmic dawn via the 21 cm signal is different. These periods are observed in the form of absorption/emission by the neutral hydrogen medium relative to the CMBR or background radiation at a reference wavelength of 21 cm. It is referred as the 21 cm differential brightness temperature--- we will discuss it later. 

\begin{figure}[hb!]
    \begin{center}
        {\includegraphics[width=5.2in,height=2.1in]{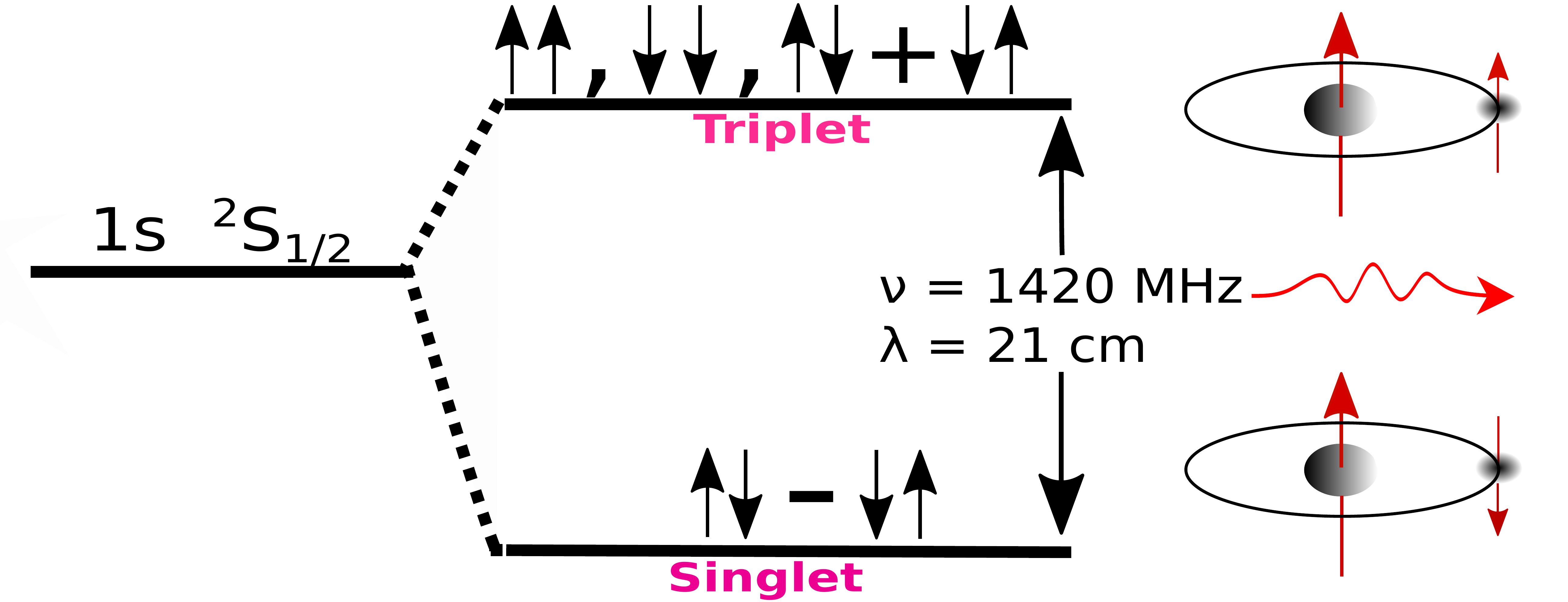}} 
    \end{center}
    \caption{A schematic diagram for hyperfine transition in ground state of neutral hydrogen atom.}\label{Hiperfine}
\end{figure}

The 21 cm line corresponds to the wavelength for hyperfine transition between 1S singlet and triplet states of the neutral hydrogen atom. The corresponding frequency for the 21 cm line is 1420.4 MHz.  For a transition at redshift $z$, the frequency can be mapped for a present-day observed frequency as $1420.4/(1+z)$. Hydrogen is the dominating fraction in the Inter-Galactic-Medium (IGM) during cosmic dawn. Therefore, it is convenient and advantageous to study IGM using the 21 cm signal. The transition probability for the hyperfine state is once in $\sim10^7$ years in the absence of any external sources. The presence of any exotic source of energy can significantly affect the hyperfine transition, thus spin temperature of the hydrogen gas. The spin temperature ($T_S$) is characterized by the number density ratio in 1S singlet and triplet states of the neutral hydrogen atom, 
\begin{alignat}{2}
\frac{n_{\rm T}}{n_{\rm S}}=\frac{g_{\rm T}}{g_{\rm S}}\times\exp\left[-\frac{2\pi\nu_{\rm \scriptscriptstyle TS}}{T_S}\right],\ \nu_{ \rm \scriptscriptstyle TS}=1420.4~{\rm MHz}\simeq1/(21~{\rm cm})\,,\label{r_pop}
\end{alignat}
here, $n_{\rm T}$ and $n_{\rm S}$ are the population of triplet and singlet states, respectively. Hyperfine splitting suppresses the singlet and lifts the triplet state. $g_{\rm T}=3$ and $g_{\rm S}=1$ are the statistical or spin degeneracies of triplet and singlet states, respectively. In the cosmological scenarios, there are three processes that can affect the spin temperature: background radio radiation, Ly$\alpha$ radiation from the first stars and collisions of a hydrogen atom with another hydrogen atoms, residual electrons or protons. In the presence of all these three effect, we can write the rate of change in the population density of singlet state,
\begin{alignat}{2}
\frac{dn_{\rm S}}{dt}=-n_{\rm S}(P_{\rm \scriptscriptstyle ST}^{\rm R}+P_{\rm \scriptscriptstyle ST}^\alpha+P_{\rm \scriptscriptstyle ST}^{\rm C})+n_{\rm T}(P_{\rm \scriptscriptstyle TS}^{\rm R} +P_{\rm \scriptscriptstyle TS}^\alpha+P_{\rm \scriptscriptstyle TS}^{\rm C})\,,\label{pdn}
\end{alignat}
here, $P_{\rm \scriptscriptstyle ST}$ and $P_{\rm \scriptscriptstyle TS}$ are  excitation and de-excitation coefficients, respectively. $\rm R$, $\alpha$ and $\rm C$ superscripts represent the excitation/de-excitation due to background radio radiation\footnote{$P_{\rm \scriptscriptstyle TS}^{\rm R}$ includes both the induced emission due to background radio radiation and spontaneous emission--- equation \eqref{a4}.}, Ly$\alpha$ radiation from first stars and collisions, respectively. In the detailed balance between the population of $1$S singlet and triplet states, by solving the equation \eqref{pdn}--- see appendix \ref{appendA1}, one can find the spin temperature as \cite{Field, Pritchard_2012},
\begin{alignat}{2}
T_{\rm S}^{-1}=\frac{T_{\rm R}^{-1}+x_\alpha\,T_{\alpha}^{-1}+x_c\,T_{\rm gas}^{-1}}{1+x_\alpha+x_c}\,,\label{Ts-1}
\end{alignat}
here, $T_\alpha$ and $T_{\rm R}$ is the colour temperature of Ly$\alpha$ radiation from first stars and background radio radiation temperature, respectively. $T_{\rm gas}$ is the gas temperature. It refers to the temperature of either neutral species, ions, electrons or protons--- all remain in thermal equilibrium. Before the first luminous objects formation, there was no Ly$\alpha$ radiation implying $x_\alpha\ \& \ T_\alpha=0$. After the first luminous objects formation, their Ly$\alpha$ photons started repeatedly scatter with the gas, and brought the Ly$\alpha$ radiation into a local thermal equilibrium with the gas. Therefore, during the cosmic dawn era the colour temperature can be taken as gas temperature, $T_\alpha\simeq T_{\rm gas}$ \cite{Field, 1959ApJ...129..536F, Pritchard_2012}. $x_\alpha={P_{\rm \scriptscriptstyle TS}^{\rm \alpha}}/{P_{\rm \scriptscriptstyle TS}^{\rm R}}$ is the Ly$\alpha$ coupling coefficient due to Wouthuysen-Field effect \cite{1952AJ.....57R..31W, Field}. Here, $P_{\rm \scriptscriptstyle TS}^{\rm R}=\left(1+{T_R}/{T_{\rm \scriptscriptstyle TS }}\right)A_{10}$, $T_{\rm \scriptscriptstyle TS}=2\,\pi\,\nu_{\rm \scriptscriptstyle TS}=0.068$~K and $A_{10}=2.85\times 10^{-15}$~sec$^{-1}$ is the Einstein coefficient for spontaneous emission from triplet to singlet state. For the all presented scenarios in the thesis: $T_{\rm R}\gtrsim49~{\rm K}\gg T_{\rm \scriptscriptstyle TS}$ at required redshift $z\sim17$. Thus, one can approximate $P_{\rm \scriptscriptstyle TS}^{\rm R}\simeq A_{10}\times\left({T_R}/{T_{\rm \scriptscriptstyle TS }}\right)$. ${P_{\rm \scriptscriptstyle TS}^{\rm \alpha}}=4\,P_\alpha/27$ and $P_\alpha$ is the rate of scattering of Ly$\alpha$ photons \cite{Pritchard_2012}. $x_c={P_{\rm \scriptscriptstyle TS}^{\rm C}}/{P_{\rm \scriptscriptstyle TS}^{\rm R}}$ is the collisional coupling coefficient due to scattering  between hydrogen atoms or scattering of hydrogen atoms with other species such as electrons and protons. Hence, the  Ly$\alpha$ and collisional coupling coefficients \cite{Pritchard_2012},
\begin{alignat}{2}
x_\alpha&=\frac{P_{\rm \scriptscriptstyle TS}^{\rm \alpha}}{P_{\rm \scriptscriptstyle TS}^{\rm R}}=\frac{4\,P_\alpha}{27\,A_{10}}\times  \frac{T_{\rm \scriptscriptstyle TS }}{T_R}\,,\label{xalph}\\
&~~\nonumber\\
x_c&=\frac{P_{\rm \scriptscriptstyle TS}^{\rm C}}{P_{\rm \scriptscriptstyle TS}^{\rm R}}=\frac{P_{\rm \scriptscriptstyle TS}^{\rm C}}{A_{10}}\times  \frac{T_{\rm \scriptscriptstyle TS }}{T_R}\,.\label{xc}
\end{alignat}
Here, the de-excitation coefficient due to collisions in gas: $P_{\rm \scriptscriptstyle TS}^{\rm C}=n_{\rm HI}\,k_{10}^{\rm HH}+n_e\,k_{10}^{{\rm H}e}+n_p\,k_{10}^{{\rm H}p}$. $n_{\rm HI}$, $n_e$ and $n_p$ are the number density of neutral hydrogen, electrons and protons in the medium, respectively. $k_{10}^{\rm HH}$ is the rate of scattering between hydrogen atoms. $k_{10}^{{\rm H}e}$ is the rate of scattering between hydrogen atoms and electrons. $k_{10}^{{\rm H}p}$ is the rate of scattering between hydrogen atoms and protons. For a more detailed review, see the review article by Pritchard and Loeb \cite{Pritchard_2012}.

\subsection{21 cm differential brightness temperature}
\begin{figure}[hb!]
    \begin{center}
        {\includegraphics[width=4.5in,height=2.3in]{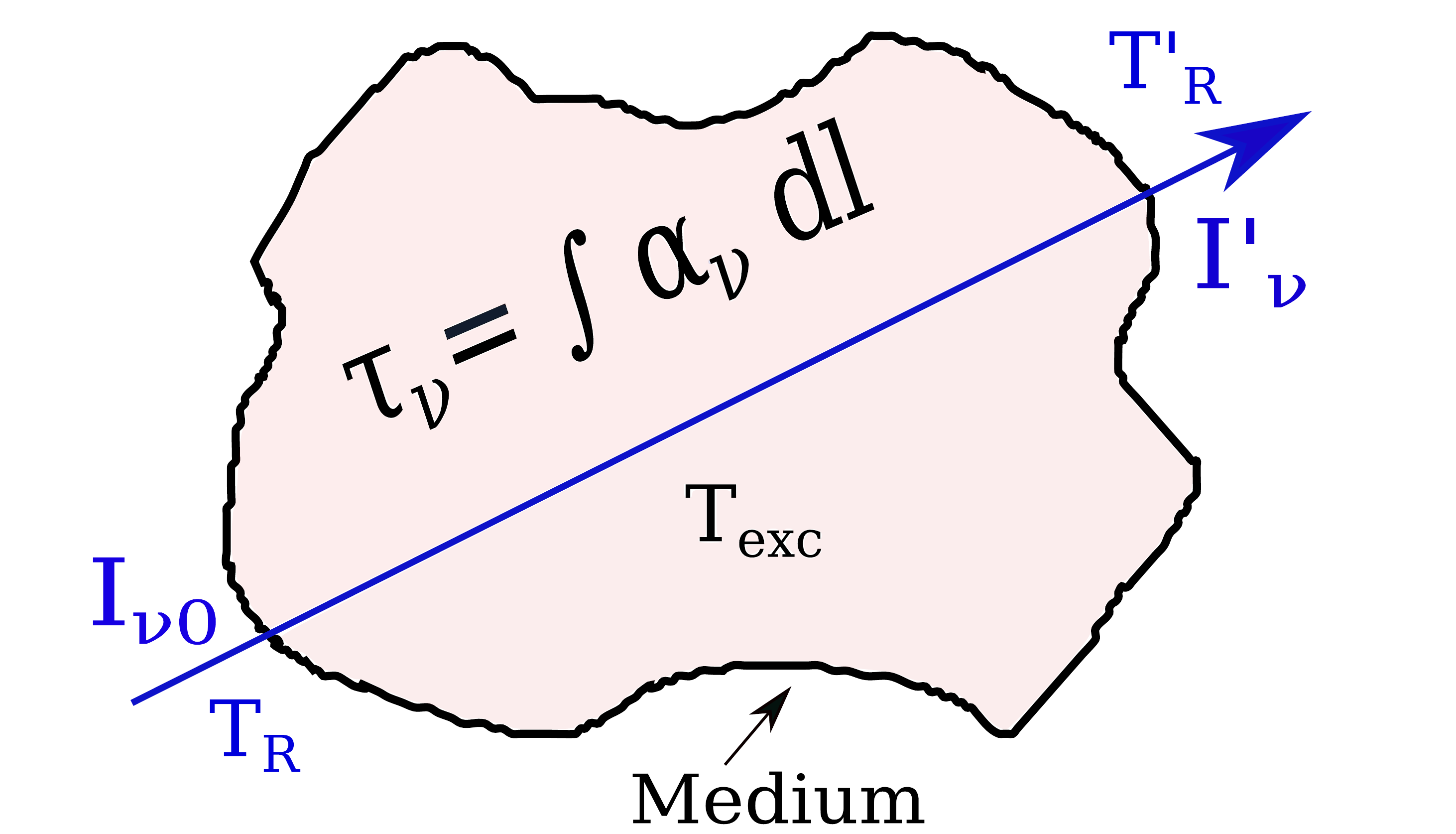}} 
    \end{center}
    \caption{A schematic diagram for the change in brightness temperature of a light when it passes through a medium.}\label{Differential}
\end{figure}
As discussed above, the 21 cm signal is observed in the form of differential brightness temperature during the cosmic dawn era. If a light with initial intensity ($I_{\nu0}$) \& brightness temperature ($T_{\rm R}$) passes through a medium having optical depth ($\tau_{\nu}$) \& excitation temperature ($T_{\rm exc}$), there can be an absorption or emission by the medium resulting in a different final/emergent intensity ($I_{\nu}'$) and brightness temperature ($T_{\rm R}'$). The divergence of the emergent brightness temperature ($T_{\rm R}'$) from the initial brightness temperature ($T_{\rm R}$) is known as the differential brightness temperature (observed temperature by antennas),
\begin{alignat}{2}
\delta T_B=T'_{\rm R}-T_{\rm R}\,.\label{brt}
\end{alignat}
In observation, we measure the specific intensity of radiation at some frequency. As discussed above, the initial frequency $\nu$ of light at redshift $z$ changes with time due to the expansion of the Universe. For present-day, it will modify to $\nu/(1+z)$. Accordingly, the frequency of $1420.4$~MHz of a light originated in the redshift range $z=15-10$ will suppress to $\mathcal{O}{(10^5~{\rm Hz})}$. While the CMB peak occurs around a frequency of $\mathcal{O}{(10^{8}~{\rm Hz})}$--- this is much higher than the 21 cm line. Therefore, we can approximate the blackbody spectrum as the Rayleigh-Jeans limit. In this limit the observed specific intensity of radiation at a frequency $\nu$,
\begin{alignat}{2}
I_\nu=\frac{4\,\pi\,\nu^3}{\exp(2\,\pi\,\nu/T)-1}\quad\xrightarrow[]{2\pi\nu/T\,\ll\, 1}\quad I_\nu \equiv 2\,\nu^2\,T\,,\label{Inu}
\end{alignat} 
$T$ is the brightness temperature of the blackbody. The emergent brightness temperature, $T_{\rm R}'$ in equation \eqref{brt}, is a combination of $T_{\rm R}$ and $T_{\rm exc}$. We can find $T_{\rm R}'$ by solving the equation of radiative transfer. If a light passes through a medium--- figure \eqref{Differential}, the change in its intensity ($dI_\nu$) due to the absorption or emission with travelled distance ($dl$), 
\begin{alignat}{2}
\frac{dI_\nu}{dl}=j_\nu-\alpha_\nu I_\nu\,,\label{RTe1}
\end{alignat}
where, $j_\nu$ is emission of light by spontaneous, stimulated emission, etc. $\alpha_\nu$ is the absorption coefficient of medium at frequency $\nu$. Here, we follow the review articles by Pritchard et al. \cite{Pritchard_2012} and Furlanetto et al. \cite{Furlanetto:2006}. Writing equation \eqref{RTe1} as,
\begin{alignat}{2}
\frac{dI_\nu}{d\tau_\nu}=S_\nu-I_\nu\,,\label{RTe2}
\end{alignat}
here, $d\tau_\nu=\alpha_\nu\, dl$ and $S_\nu=j_\nu/\alpha_\nu$. Therefore,
\begin{alignat}{2}
\tau_\nu=\int \alpha_\nu\ dl\,,\label{RTe3}
\end{alignat}
is the optical depth. Optical depth is a function of the absorption of light by the medium with travelled distance in the medium. 
By solving the equation \eqref{RTe1} and using equation \eqref{Inu}, we can find the $T_{\rm R}'$--- see the appendix \ref{appendA2},
\begin{alignat}{2}
T_{\rm R}'=T_{\rm exc}\,(1-e^{-\tau_\nu})+T_{\rm R}\,e^{-\tau_\nu}\,.\label{Tr''}
\end{alignat}
The differential brightness temperature, by equation \eqref{brt},
$\delta T_B=(T_{\rm exc}-T_{\rm R})\times(1-e^{-\tau_\nu}).$ 
For the expending Universe, the temperature of radiation is $\propto(1+z)$. Thus, the redshifted differential brightness temperature for present-day,
\begin{alignat}{2}
\delta T_B=\frac{T_{\rm exc}-T_{\rm R}}{1+z}\times(1-e^{-\tau_\nu})\,.\label{deltbz}
\end{alignat}
In our case, the medium is hydrogen gas and the $T_{\rm exc}$ for the 21 cm line is $T_{\rm S}$--- defined in equation \eqref{Ts-1}. The $\tau_\nu$ is $\ll1$ for neutral hydrogen gas--- optically thin. Hereafter, we will write $\delta T_B$ as $T_{21}$ for the 21 cm line. Therefore, the 21 cm differential brightness temperature \cite{Pritchard_2012},
\begin{alignat}{2}
T_{21}\simeq\frac{T_{\rm S}-T_{\rm R}}{1+z}\times\tau_\nu\,.\label{deltbz1}
\end{alignat}
The optical depth can be found by solving the equation \eqref{RTe3} for a hydrogen medium and a line profile--- see appendix \ref{appendA3},
\begin{alignat}{2}
\tau_\nu\simeq 27\,x_{\rm HI}\,(1+z)\,  \left(\frac{\rm mK}{T_{\rm S}}\right)\,\left(\frac{0.15}{\Omega_{\rm m }\,h^2}\,\frac{1+z}{10}\right)^{1/2}\left(\frac{\Omega_{\rm b}\,h^2}{0.023}\right)\,,\label{tau}
\end{alignat} 
here, $x_{\rm HI}=n_{\rm HI}/n_{\rm H}$ is the fraction of neutral hydrogen in the Universe, and $n_{\rm H}$ is the total number density of hydrogen. $\Omega_{\rm m }=\rho_{\rm M}/\rho_{\rm cr}$ and $\Omega_{\rm b}=\rho_{\rm b }/\rho_{\rm cr}$ are the dimensionless energy density parameters for total matter and baryons in the Universe, respectively. $\rho_{\rm M}$ and $\rho_{\rm b }$ are the energy density for total matter and baryons, respectively. $\rho_{\rm cr}=3\,H^2/(8\,\pi\,G_{\rm N})$ is the critical energy density and $G_{\rm N}$ is the gravitational constant. $h=H_0/(100~{\rm Km\ sec^{-1}\,Mpc^{-1}})$ and $H_0$ is the present-day value of Hubble parameter. Substituting the value of $\tau_\nu$ from equation \eqref{tau} into equation \eqref{deltbz1}, we get the final expression for the global 21 cm differential brightness temperature \cite{Zaldarriaga:2004, Mesinger:2007S, Mesinger:2011FS, Pritchard_2012, Mittal:2020},
\begin{alignat}{2}
T_{21}\simeq27\,x_{\rm HI}\,  \left(1-\frac{T_{\rm R}}{T_{\rm S}}\right)\,\left(\frac{0.15}{\Omega_{\rm m }\,h^2}\,\frac{1+z}{10}\right)^{1/2}\left(\frac{\Omega_{\rm b}\,h^2}{0.023}\right)~{\rm mK}\,.\label{t21f}
\end{alignat}
Depending on the ratio ${T_{\rm R}}/{T_{\rm S}}$, there can be three scenarios for 21 cm signal:  If $T_S=T_R$ then $T_{21}=0$ and there will not be any signal; for the case when $T_S> T_R$, emission spectra can be observed, and when  $T_S< T_R$, it leaves an imprint of absorption spectra. 

\subsection{Evolution of the global 21 cm signal}\label{21cme}

Usually, in the $\Lambda$CDM model of cosmology, the contribution in the background radiation is assumed to be solely by the CMB radiation, $T_{\rm R}\equiv T_{\rm CMB}\,$; $T_{\rm CMB}$ is the CMBR temperature. Therefore, in this subsection, we discuss the evolution of the global 21 cm signal when only CMBR is present as background radiation.
\begin{figure}[]
    \begin{center}
        {\includegraphics[width=5.6in,height=2.7in]{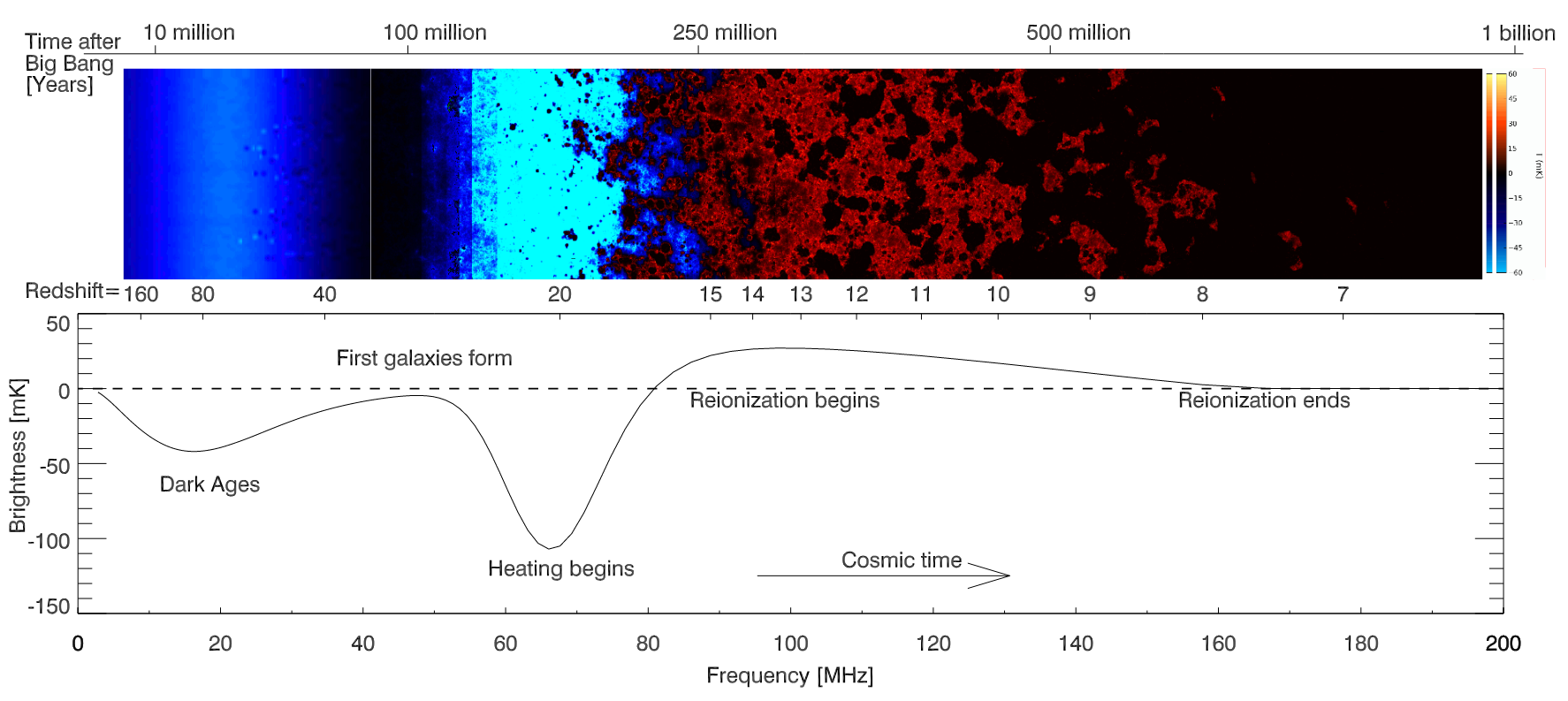}} 
    \end{center}
    \caption{The figures represents the evolution of fluctuation in the 21 cm signal (above) and global 21 cm signal (below) when the background radiation is CMBR\protect\footnotemark. {\it Image credits: Pritchard \& Loeb, Rep. Prog. Phys., 75, 086901, (2012)} \cite{Pritchard_2012, Pritchard_2010}.}\label{Evo}
\end{figure}
\footnotetext{The position and amplitude of the second dip from the left (between redshift $30-15$) may modify depending on models of first-stars formation or x-ray heating of the gas.}

At the end of recombination, the baryon number density of the Universe is dominated mainly by the neutral hydrogen, a small fraction of helium, residual free electrons and protons. After recombination ($z\sim1100$) down to $z\sim 200$, the residual free electrons undergo Compton scattering and maintain thermal equilibrium between electrons and CMBR. The free electrons remain in thermal equilibrium with other gas components implying $T_{\rm gas}\sim T_{\rm CMB}\,$ 
\cite{Peebles:1993}. Using equations \eqref{Ts-1} and \eqref{t21f}, we can find that $T_{21}=0\,$, and the 21 cm signal is not present during this era. From $z\sim 200$ until 40, the number density of free electrons decreases significantly and this  makes the Compton scattering insufficient. As a result, the gas decouples from CMBR, and its temperature falls adiabatically: $T_{\rm gas}\propto (1+z)^2$. The gas temperature falls below CMBR implying an early 21 cm absorption signal--- known as the collisional absorption signal \cite{Pritchard_2012}. During this period, collisions among the gas components dominate, i.e.  $x_c\gg1$, which implies $T_{\rm S}\sim T_{\rm gas}\,$--- equation \eqref{Ts-1}. Nevertheless, this signal is not observed yet due to the poor sensitivity of present-day available radio antennas as the  sensitivity of antennas falls dramatically below $\sim50$~MHz. After $z\sim40$ to the formation of the first star\footnote{The redshift of first stars formation is not well known and it could be around 35 to 25.}, number density and temperature of the gas are very small, hence, $x_c\rightarrow 0$. Therefore, $T_{21}\sim0$ and no signal is present there \cite{Barkana:2018nd, Pritchard_2012}. After the first star formation, gas temperature couples again to the spin temperature due to Ly$\alpha$ radiation emitted from the first stars by Wouthuysen-Field (WF) effect \cite{1952AJ.....57R..31W, 1959ApJ...129..536F, Field}. Therefore, $x_\alpha\gg 1,x_c$ and absorption spectra can be seen--- equations (\ref{Ts-1} and \ref{t21f}). After $z\sim 15$,  the gas temperature starts to rise due to x-ray radiation emitted from the first stars\footnote{It is also not very clear when x-ray heating begins to dominate the temperature of the gas. We use the fiducial models for x-ray heating considered in references \cite{Mirocha:2015G, Harker:2015M, Zygelman:2005, Kovetz2018}.}. Consequently, the temperature of gas rises above CMB temperature and the emission spectra can be seen. As the reionization ends, neutral hydrogen fraction becomes very small and no signal is observed. The small fraction of neutral hydrogen were left only in dense regions of collapsed structures. These regions can be analysed by 21 cm forest--- an analogy to Ly$\alpha$ forest.

\section{21 cm line as a probe of new physics}\label{21cmp}

\begin{figure}[hb!]
    \begin{center}
        {\includegraphics[width=4.05in,height=4.0in]{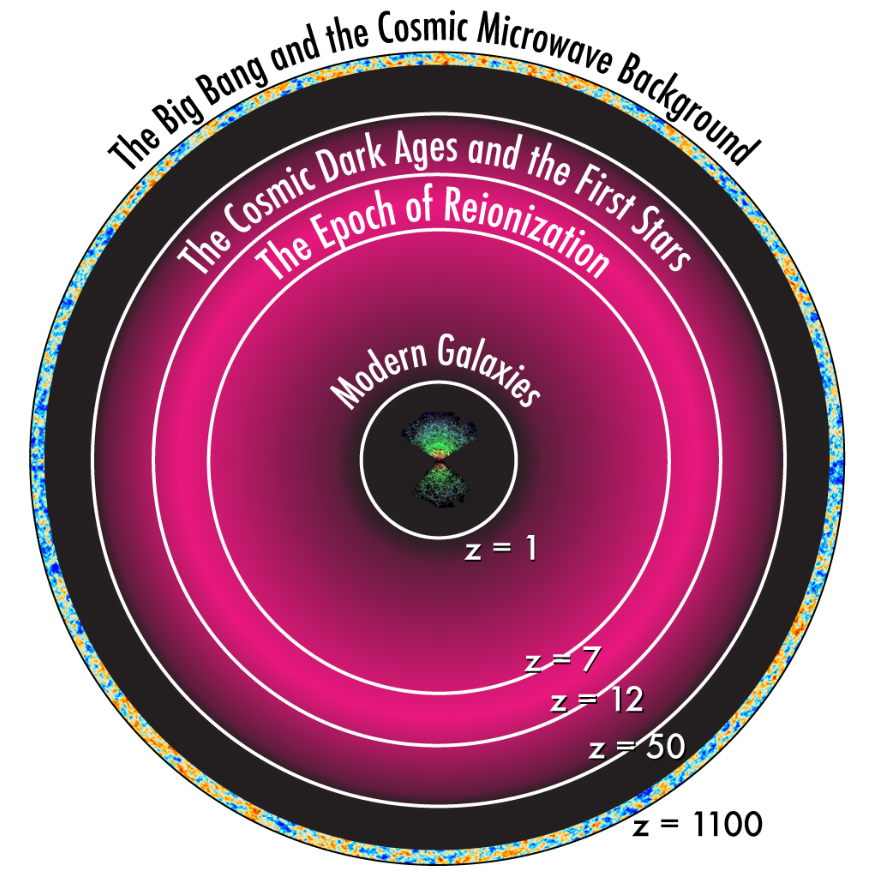}} 
    \end{center}
    \caption{The CMB observations can only probe the thin outer shell ($z\sim 1100$), and the observation of large scale structures can probe a small fraction of volume near the centre. We expect that the future advanced technology for the 21 cm signal observation will be able to probe the entire pink region. In the thesis, we focus on the 21 cm signal from the redshift 30 to 15.  {\it Image credits: With the permission of Josh Dillon} \cite{Dillon:2015}; {\it originally reproduced from Tegmark \& Zaldarriaga (2009)} \cite{Tegmark:2009}.} \label{age21}
\end{figure}

As shown in the figure \eqref{age21}, the 21 cm signal can probe a large volume of the history of our Universe--- pink region. Currently, we are not able to probe the high redshift Universe  ($z\gtrsim30$) as the sensitivity of presently available radio antennas becomes very low below $\sim50$ MHz. We expect that the future advanced technology for the 21 cm signal observation will be able to probe the Universe above the redshift 25. In the thesis, we focus on the 21 cm signal between the redshift range of 30 to 15.

After $z\sim200$ gas temperature falls adiabatically and reaches to $\simeq7$~K at $z=17.2$, while the  CMB temperature reaches to $\simeq 49.6$~K. From the equation \eqref{t21f}, this implies a value of absorptional amplitude of $T_{21}$ to $\sim-220$~mK in absence of any heating effects on the IGM gas due to first stars. Here, to calculate $T_{21}$, we have taken $x_{\rm HI}$ to unity. The $x_{\rm HI}$ can be written as $1-x_e\,$. In our case, at $z\sim17$ the ionization fraction, $x_e\lesssim\mathcal O (10^{-3})$ implying $x_{\rm HI}\simeq1$. Here, $x_e=n_e/n_{\rm H}$ is the ionization fraction and $n_e$ is the number density of residual free electrons.
The presence of any exotic source of energy can inject energy into IGM and heat the gas. This in turn can modify the absorption amplitude in the global 21 cm signal. This feature can provide a robust bound on the properties of such sources of energy injection into IGM. In the thesis, the following four works has been considered: sterile neutrinos and primordial black holes as dark matter candidates and constrain their properties in the light of the global 21 cm signal. Another two works discussed in the thesis are related to the constraining strength of primordial magnetic fields that might have been generated in the early Universe.


In 2018, the Experiment to Detect the Global Epoch of Reionization Signature (EDGES)\footnote{\href{https://www.haystack.mit.edu/astronomy/astronomy-projects/edges-experiment-to-detect-the-global-eor-signature/}{https://www.haystack.mit.edu/astronomy/astronomy-projects/edges-experiment-to-detect-the-global-eor-signature/}} collaboration reported an absorption profile for the 21 cm signal in the redshift range $15-20$ \cite{Bowman:2018yin}. The EDGES collaboration reported $T_{21}$ to be $-500^{+200}_{-500}$~mK in the redshift range $15-20$ centred at $78\pm 1$~MHz and in symmetric ``U" shaped form. This absorption amplitude is nearly two times smaller than predicted by theoretical models based on $\Lambda$CDM framework ($\sim-220$~mK). It is argued that to explain the EDGES observation, for the best fitting amplitude at the centre of the ``U" profile,  either the cosmic background radiation temperature $T_{\rm CMB}\gtrsim104$~K for the standard  $T_{\rm gas}$ evolution or $T_{\rm gas}\lesssim3.2$~K in the absence of any non-standard evolution of the $T_{\rm CMB}$ \cite{Bowman:2018yin}. Recently, many articles have questioned the EDGES measurement \cite{Saurabh:2021, Tauscher:2020, Hills:2018, Saurabh:2019, Bradley:2019}. For example, in Ref. \cite{Hills:2018}, the authors have questioned the fitting parameters for the foreground emission and data. There is a possibility that the absorption feature in the EDGES observation can be a ground screen artifact \cite{Bradley:2019}. The absorption amplitude may modify depending on the modelling of  the foreground \cite{Saurabh:2019, Tauscher:2020}. In a recent article \cite{Saurabh:2021}, authors claimed that the EDGES observation might not be of an astrophysical origin. We revisit the EDGES observation and controversies over it in the chapter \eqref{chap6} also. In the light of these controversies, in the recent two articles (\ref{sndm2} \& \ref{pbhdm}), we do not consider the absorption amplitude reported by the EDGES collaboration. In these articles, we take 21~cm differential brightness temperature such that it does not change, from its standard theoretical value ($\sim-220$~mK), by a factor of more than 1/4 (i.e. $-150$~mK) or 1/2 (i.e. $-100$~mK) at redshift 17.2\,. While in the older two articles \eqref{pmfs45}, we have considered the absorption amplitude reported by the EDGES collaboration.

\subsection{Sterile neutrino dark matter --- Chapter \ref{chap2}}\label{sndm2}
In the warm dark matter models, one of the theoretically well-motivated candidates is KeV mass sterile neutrinos. Sterile neutrinos are radiatively unstable and can inject photon energy into the IGM. The injection of energy into the IGM can modify the temperature and ionization history of the IGM gas thus absorption amplitude of 21 cm signal during cosmic dawn era. Therefore one can constraint the lifetime of sterile neutrinos and the mixing angle of sterile neutrinos with active neutrinos. 

The article has been published as: Pravin Kumar Natwariya  and Alekha C. Nayak, ``Bounds on sterile neutrino lifetime and mixing angle with active neutrinos by global 21 cm signal", \href{https://doi.org/10.1016/j.physletb.2022.136955}{\textit{Physics Letters B} 827 (2022) 136955}.

\subsection{Primordial black hole dark matter --- Chapter \ref{chap3}}\label{pbhdm}

Primordial black holes (PBHs) have attracted much interest in recent years and have been a part of intense studies for more than five decades. As PBHs are massive, interact only gravitationally and are formed in the very early Universe, they can be considered as a potential candidate for non-particle dark matter. Hawking evaporation of PBHs can inject energy into the IGM and therefore be constrained by the absorption feature in the global 21 cm signal. The mass and spin are fundamental properties of a black hole, and they can substantially affect the evaporation rate of the black hole. In this work, we derive an upper bound on the dark matter fraction in the form of the primordial black holes with a non-zero spin. 

The article has been published as: Pravin Kumar Natwariya, Alekha C. Nayak and Tripurari Srivastava, ``Constraining spinning primordial black holes with global 21-cm signal", \href{{https://doi.org/10.1093/mnras/stab3754}}{\textit{Mon Not R Astron Soc} 510, 4236--4241 (2022)}.

\subsection{Primordial magnetic fields --- Chapter \ref{chap4} {\it \&} \ref{chap5}}\label{pmfs45}

Observations suggest that the magnetic fields (MFs) are ubiquitous in the Universe--from the length scale of planets and stars to the cluster of galaxies. The origin and evolution of PMFs are one of the outstanding problems of cosmology. Decaying PMFs can inject magnetic energy into thermal energy of the IGM and heat the gas. As briefly mentioned earlier, one requires to cool the IGM gas during cosmic dawn below the standard evolution or increase the radio background at required redshift to explain the EDGES observation. Here, we explore the upper bounds on the present-day strength of the PMFs in both the scenarios by considering different models. The articles have been published as:  

\begin{itemize}
\item   Pravin Kumar Natwariya, ``Constraint on Primordial Magnetic Fields In the Light of ARCADE 2 and EDGES Observations",     \href{https://doi.org/10.1140/epjc/s10052-021-09155-z}{\textit{Eur. Phys. J. C} 81 (2021) 5, 394}.
    
\item  Jitesh R. Bhatt, Pravin Kumar Natwariya, Alekha C. Nayak and Arun Kumar Pandey, ``Baryon-Dark matter interaction in presence of magnetic fields in light of EDGES signal", \href{https://doi.org/10.1140/epjc/s10052-020-7886-x}{\textit{Eur. Phys. J. C} 80 (2020) 4, 334}.
\end{itemize}




\vspace{0.5cm}
{\bf {\Large C}hapter \ref{chap6}} summarises the main results of the thesis. We also discuss possibilities of further extensions and future scopes of the results obtained in the thesis.


%
\clearpage
\pagestyle{empty}
\cleardoublepage
\pagestyle{fancy}

\begin{savequote}[75mm]
``Would you tell me, please, which way I ought to go from here?'
`That depends a good deal on where you want to get to,' said the Cat''
	\qauthor{Lewis Carroll, \textit{Alice in Wonderland}}

``It is the nature of all greatness not to be exact''
\qauthor{Edmund Burke, \textit{speech ``On American Taxation"}}
\end{savequote}

\chapter[Sterile Neutrino Dark Matter]{Sterile Neutrino Dark Matter}\label{chap2}
\vspace{-1.5cm}

Despite the searching for decades, the nature of dark matter is still unknown. It is one of the biggest mysteries in particle physics and cosmology. Although $\Lambda$CDM model of cosmology is highly successful in explaining Big-Bang nucleosynthesis, CMB anisotropies and large scale structures of the Universe, it faces challenges at a smaller length scale, $\lesssim1$~Mpc (for a detailed review see \cite{Bullock:2017B} and references therein). These  problems include the missing satellite or dwarf galaxy problem \cite{Klypin:1999uc, Moore:1999nt},  the too-big-to-fail problem \cite{Boylan:2011sm, Boylan:2012bs} and the core-cusp problem \cite{block:2010wj}. In the simulations, the cold dark matter scenario clusters hierarchically and predicts a large number of satellite galaxies. However, the observations show less number of satellite galaxies \cite{Klypin:1999uc, Moore:1999nt}. For example, the Milky Way size halo simulations show around 500 satellites, while observations show a far less number of satellite galaxies \cite{Moore:1999nt, Drlica:2020}. Subsequently, the missing satellite creates a new problem also: The simulation of Galactic size haloes predicts a larger number of big satellites that are so massive that there is no way not to host visible stars. Therefore, these massive satellites should be visible. In contrast, the observations show no such satellites consistent with the simulations \cite{Boylan:2011sm, Boylan:2012bs, Papastergis:2016}. N-body simulations of cold dark matter also show the cuspy profile for dark matter density at the halo centre, while the observation of rotation curves suggest the flat profile \cite{block:2010wj}. In the light of these problems, alternatives to the cold dark matter model have been proposed, for e.g.  self-interacting dark matter \cite{Spergel:2000DSP, Tulin:2017Y, Kaplinghat:2016TY, Natwariya:2020V}, fuzzy cold dark matter \cite{Hu:2000, Schive:2014}, warm dark matter (WDM) \cite{Blumenthal:1982, Dodelson:1994, Colombi:1996, Sitwell:2014, Brdar:2018}, etc. The difference between cold, warm and hot dark matter can be characterized in the form of their thermal velocities, $v=\sqrt{(3\,T/m)}\,$. Here, $v$, $T$ and $m$ are the speed, temperature and mass of the particle, respectively. One can see that a larger speed implies a higher temperature for a fixed mass of particles. Roughly, if their speed is less than ten percent of the light speed ($v\lesssim0.1$), they can be considered cold dark matter candidates. If $v$ is $\gtrsim0.1$, they can be considered hot dark matter candidates \cite{Seigar:2015}. The WDM lies in between the hot and warm dark matter. The WDM behaves similar to CDM on large length scales. This scale can be characterized in the form of ``free-streaming length"--- the other important concept to differentiate between hot, cold or warm dark matter. Typically, the free-streaming length can be estimated by how far a particle has travelled from beginning to matter-radiation equality \cite{Schneider:2012},
\begin{alignat}{2}
\lambda_{\rm fs}=\int_{0}^{t_{\rm eq}}\frac{v}{a}\,dt\,,
\end{alignat}
here, $t_{\rm eq}$ is the matter-radiation equality time. For a length scale larger than $\lambda_{\rm fs}$, WDM behaves as CDM--- i.e. it makes structures hierarchically above $\lambda_{\rm fs}$. While below the length scale $\lambda_{\rm fs}$, there is a possibility that WDM may create structures ``top-down"--- i.e. small structures may emerge via the fragmentations of large structures \cite{Colombi:1996, Schneider:2012, Seigar:2015}. The free-streaming length can be found as \cite{Schneider:2012},
\begin{alignat}{2}
\lambda_{\rm fs}\sim0.4 \left(\frac{m_{\rm WDM}}{\rm KeV}\right)^{-4/3} \ \left(\frac{\Omega_{\rm WDM}\,h^2}{0.135}\right)^{1/3}\, {\rm Mpc}/h\ ,
\end{alignat}
here, $m_{\rm WDM} $ is the mass of WDM particle and $\Omega_{\rm WDM}$ is the dimensionless energy density parameter for WDM. The free-streaming scale is inversely proportional to mass of particle. It implies that the size of formed-first-structures will increase for a smaller particle mass--- the numbers of small-length-scale structures will suppress. For example, if one considers the mass of the WDM particle to be 10 KeV, then the free-streaming scale will be $\sim2\times10^1$~Kpc. Therefore, one can overcome the missing satellite problem by considering an adequate mass of WDM. In the hot dark matter scenario, the free-streaming length typically is so large that density fluctuations below cluster scale would get washed up, and formed-first-structures would have been the size of superclusters. Later, their fragments might have formed the clusters, then galaxies. While the observation shows that galaxies formed first, then emerged as clusters and then superclusters due to their mutual gravitational attraction \cite{Seigar:2015}. As discussed above, the nature of dark matter has significant effects on structure formation. The WDM can also solve the angular momentum problem--- galaxies have smaller specific angular momenta in CDM simulation compared to observations \cite{Martin:2003, Bode:2001}. Additionally, by including the baryonic feedback with WDM can address the too-big-to-fail and core-cusp problems also \cite{Lovell:2017, Andrea:2012, Lovell:2012}.  The two popular candidates for WDM are sterile neutrinos and gravitinos. The presence of sterile neutrino warm dark matter having KeV mass can also explain the recently observed unexpected and unidentified emission line around 3.5~KeV in x-ray spectra of nearby galaxies and clusters  \cite{Bulbul:2014, Boyarsky:2014, Boyarsky:2015, Brdar:2018, Silich:2021}. In this chapter, we consider sterile neutrino and study its lifetime and mixing angle with active neutrinos \cite{Natwariya:2022}. 
%
%
%
%
\section{Sterile neutrinos as dark matter}

Sterile neutrino with KeV mass is one of the exciting and well-motivated candidates for WDM (Ref. \cite{Adhikari:2017, Abazajian:2017, Boyarsky:2019} and Refs. therein). The standard model of particle physics considers the neutrinos as massless. However, experiments and theoretical models questioned the standard model of particle physics over the past years. One well-studied example is neutrino oscillation \cite{Pontecorvo:1958, Maki:1962, Pontecorvo:1968, Bahcall:1976}. To explain the observations of neutrino oscillations, one has to extend the standard model to introduce the massive neutrinos (for more details, see the reviews \cite{Smirnov:2006, Lesgourgues:2006})\footnote{Takaaki Kajita with Arthur B. McDonald received the Nobel Prize in Physics for 2015 ``for the discovery of neutrino oscillations, which shows that neutrinos have mass" \cite{Super-K:2004, Super-K:2005, SNO:2005, SNO::2005}.}. There are three flavours of active neutrinos--- electron, muon and tau neutrino, but absolute value of their masses are not very well known. Nevertheless, the square mass difference between different flavours has been constrained by various oscillation experiments, such as solar, atmospheric, reactor and accelerator (see the Ref. \cite{Capozzi:2016} and reviews \cite{Group:2020, Dasgupta:2021SN}).  In the standard model of particle physics, all particles get their mass via Higgs Mechanism, but neutrinos remain massless. One of the mechanisms via which neutrinos can get their mass is the Seesaw mechanism. As of now, active neutrinos have been observed with only left-handed chirality \cite{DREWES:2013}. To give mass to neutrinos, we also require the right-handed counterpart of active neutrinos. The right-handed neutrinos can have mass from a few eV to GUT scale \cite{DREWES:2013}. In the Seesaw mechanism, sterile neutrino naturally appears as an eigenstate of the neutrino mass matrix. Introducing a new Yukawa interaction with new Weyl fermions $N^{\beta}$ \cite{Dasgupta:2021SN}, 
\begin{alignat}{2}
\mathcal{L}_{\rm Y}\supset -y ^{\alpha\beta}(i\,\sigma^2\,H^*)\,L^{\alpha}N^{\beta}+h.c.\,,
\end{alignat}
here, $\alpha$ and $\beta$ are summed over $e,\,\mu,\,\tau$ and $1,\,2,\,...,\,n$\,, respectively; $n$ is the number of fields of $N^{\beta}$. $i\,\sigma^2\,H^*$ and $L^\alpha=(\nu^\alpha,\,e^\alpha)^T$ are $SU(2)_L$ doublet and carry opposite hypercharges: +1/2 and -1/2, respectively. Therefore, their combination is total singlet, implying $N^\beta$ to be total singlet also \cite{Dasgupta:2021SN}. When Higgs field ($H$) acquires vacuum expectation value ($v$), the neutrino mass term can be written as,
\begin{alignat}{2}
\mathcal{L}_{\rm mass}\supset - M_D^{\alpha\beta}\,\nu^\alpha\,N^{\beta}+h.c.\,,\label{neq2}
\end{alignat}  
the Dirac mass term $M_D^{\alpha\beta}\equiv y ^{\alpha\beta}\,v/\sqrt{2}$\,. Since $N^\beta$ does not have any strong, electromagnetic or weak coupling, it is called the sterile; and  it can be considered a dark matter candidate. $\nu^\alpha$ has weak coupling with standard model particles, and it is called active neutrino. As sterile neutrinos are singlet, in principle we can write a Majorana mass term for $N^{\beta}$: $\mathcal{L}_{\rm mass}=-({1}/{2})\,M_M^{\alpha\beta}\,N^\alpha\,N^\beta+h.c.\,$. From equation \eqref{neq2}, 
{\small \begin{alignat}{2}
	\mathcal{L}_{\rm mass}\supset -\frac{1}{2}\,n^T\,M\,n\equiv -\frac{1}{2}\,n^T\,\left[\begin{matrix}
	0 &M_D\\
	M_D^T&M_M
	\end{matrix}\right]\,n+h.c.\,,
	\end{alignat}}
here, $n=(\nu^e...\nu^\tau,\,N^1...N^n)^T$, $M_D=M_D^{\alpha\beta}$ and $M_M=M_M^{\alpha\beta}$. Assuming $||M_M||\gg||M_D||$, as $M_M$ is not protected by any symmetry and $M_D$ can not be larger than electroweak scale because it will require Yukawa coupling $\gg1$, the eigenvalues of mass matrix: $m_1^\nu={\mathcal{O}}\left({M_D^2}/{M_M}\right)$ and $ m^N=\mathcal{O}(M_M)\,$. We get the light neutrino mass to $m^\nu\sim0.1$~eV by taking $||M_D||\sim100$~GeV and $||M_M||\sim10^{14}$~GeV. The sterile neutrinos are stable--- have a larger lifetime compared to the age of the Universe. Therefore, they can make an excellent candidate for the warm dark matter if they also have mass in the KeV range \cite{Group:2020}.

One of the minimal extensions of the standard model of particle physics, where neutrino mass and KeV sterile neutrinos in the context of dark matter are widely explored via the Seesaw mechanism, is the Neutrino Minimal Standard Model ($\nu$MSM)  \cite{Shaposhnikov:2005, DREWES:2013, Group:2020, Dasgupta:2021SN}. In this model, we can lower one of the eigenvalues of $M_M$ to get KeV scale sterile neutrino while keeping others super-heavy. In the basis ($\nu_a,\,\nu_s,\,N$); $N$ represents the heavier sterile states, the mass matrix \cite{Dasgupta:2021SN},
{\small	\begin{alignat}{2}
	M=\left(\begin{matrix}
	0&0&0&M_s^1&M_D^{11}&M_D^{12}\\
	0&0&0&M_s^2&M_D^{21}&M_D^{22}\\
	0&0&0&M_s^3&M_D^{31}&M_D^{32}\\
	M_s^1&M_s^2&M_s^3&\mu_s&0&0\\
	M_D^{11}&M_D^{21}&M_D^{31}&0&M_M^1&0\\
	M_D^{12}&M_D^{22}&M_D^{32}&0&0&M_M^2\\
	\end{matrix}\right)
	\end{alignat}} 
here, $\nu_s$ and $\nu_a$ are sterile and active neutrinos, respectively. Applying the Seesaw mechanism, we can find the mass of neutrinos: $M_\nu\simeq-M_D\,M_M^{-1}\,M_D^T-M_s\,\mu_s^{-1}\,M_s^T\,;$ and the mass of light sterile neutrino: $m_s\simeq\mu_s$.

The $\nu$MSM model is a minimal extension of the standard model of particle physics--- with only three additional sterile neutrinos up to the Planck scale. One having KeV scale mass--- can account for dark matter. The other two heavier sterile neutrinos can account for the observed light neutrino masses by the Seesaw mechanism. They can also explain the baryon asymmetry in the Universe through oscillation-induced leptogenesis if they are nearly degenerate in the mass range 150 MeV$-$100 GeV \cite{Shaposhnikov:2005, Adhikari:2017}. More details about KeV sterile neutrino models can be found in the review article by A. Merle \cite{Merle:2013}.


\section{Existing bounds on sterile neutrinos}

The possibility of KeV mass range sterile neutrinos as a WDM candidate can be explored and constrained by the observation of large scale structures in the Universe \cite{Dodelson:1994}. There are several model-dependent mechanisms that can produce sterile neutrinos in the early Universe  \cite{Boyarsky:2019, Shi:1999, Dolgov:2002}. In recent years, various techniques have been proposed to probe the unexplored sterile neutrino dark matter parameter space, for example, by mapping of x-ray intensity at different redshift \cite{Andrea:2020}, by observing KeV energy photons using instruments onboard Transient High Energy Sky and Early Universe Surveyor (THESEUS) mission 
(for the details of instruments sensitivity of THESEUS, see the Ref. \cite{Morgan:2020}), by exploring the imprints of sterile neutrino on solar neutrino fluxes \cite{Lopes:2020}, by testing the hypothesis of decaying-sterile-neutrino \cite{Gouva:2020, Seto:2020}, etc. The lower bound on the mass of sterile neutrinos can be obtained by the Pauli exclusion principle \cite{Dasgupta:2021SN, Boyarsky:2019}. These bounds depends on momentum distribution and the dwarf galaxy used for astronomical data \cite{Dasgupta:2021SN, Boyarsky:2019}. The authors of the ref. \cite{Boyarsky:2009}, finds the lower bound on the mass of non-resonantly produced sterile neutrino to be $>1.7$~KeV when all the dark matter is composed of sterile neutrinos. Additionally, the parameter space of sterile neutrino dark matter has been constrained by various observations and theoretical studies. The observations from Nuclear Spectroscopic Telescope Array (NuSTAR)\footnote{\href{https://heasarc.gsfc.nasa.gov/docs/nustar/index.html}{https://heasarc.gsfc.nasa.gov/docs/nustar/index.html}} did not find any sign of anomalous x-ray lines for sterile neutrino mass range $10-40$~KeV. The future updated version of NuSTAR will be able to probe for sterile neutrino mass range $6-10$~KeV \cite{Brandon:2020}. In the context of EDGES signal, authors of the reference \cite{vipp:2021},  put a constraint on the Dodelson-Widrow sterile neutrinos mass to $63_{-35}^{+19}$~KeV. The WMAP,  Ly$\alpha$ forest and x-ray observations constrain the sterile neutrino mass in the range from $\sim2$~KeV to $\sim50$~KeV \cite{Viel:2005, Abazajian:2006, Seljak:2006, Boyarsky:2009J, Viel:2013}. The authors of the Refs. \cite{Andrea:2010, Polisensky:2011} compare the observed satellite galaxy with conferred from WDM simulations of Galaxy-sized halo and constrain the mass of sterile neutrino $\gtrsim2$~KeV. Further, individual bounds on the sterile neutrino parameter space can be found in the Refs. \cite{Boyarsky:2019I, foster:2021, salvio:2021, Abazajian:2012, Das:2018, Shakeri:2020, Honorez:2017, Vegetti:2018, Rudakovskyi:2018, Bezrukov:2017}. 


\section{Radiative decay of sterile neutrinos}

Sterile neutrinos with KeV mass can decay to active neutrinos via two channels: $\nu_{s}\rightarrow \nu_a\,\nu_a\,\bar\nu_a$  and  $\nu_s\rightarrow \nu_a\, \gamma$. In this work, we study the effect of radiative decay of sterile neutrinos on the thermal and ionization history of the Universe, and constrain the sterile neutrino decay time and mixing angle with active neutrinos. The decay of sterile neutrino to active neutrino via the radiative process can inject the photon energy into IGM and modify the absorption amplitude of the 21~cm signal during cosmic dawn. Hence, we can constrain the sterile neutrino decay time and mixing angle with the active neutrino using the 21 cm absorption signal. In this process, half of the total energy of a sterile neutrino ($m_{\nu_{s}}/2$) is carried away by a photon and remaining by an active neutrino. The decay width of sterile neutrino  for radiative process can be written as  (\cite{Boyarsky:2007, Boyarsky:2019} and reference cited therein),
\begin{alignat}{2}
\Gamma_{\nu_s}=\Gamma_{\nu_s\rightarrow \nu_a \gamma}=\frac{9\ \alpha\ G_F^2}{1024\,\pi^4}\, \sin^2(2\,\theta)\,m_{\nu_s}^5\,,\label{GS}
\end{alignat}
here, $\theta\equiv\sum_{i=e,\,\mu,\,\tau} |\theta_i|^2$ is the total mixing angle between sterile and active neutrinos. 
In equation \eqref{GS}, $G_F$ and $\alpha$ are the Fermi and fine structure constant, respectively. $m_{\nu_s}$ stands for the mass of the sterile neutrino. The mixing angle $\theta \lll 1$, therefore $\sin^2(2\,\theta)\simeq 4\,\sin^2(\theta)$. We can write the decay width as \cite{Boyarsky:2007, Boyarsky:2019},
\begin{alignat}{2}
\Gamma_{\nu_s}=\tau_{\nu_s}^{-1}\simeq 5.52\times 10^{-22}\,\sin^2(\theta)\ \left[\frac{m_{\nu_s}}{\rm KeV}\right]^5\ \left[\frac{1}{\rm sec}\right]\,,\label{s1}
\end{alignat}
here, $\tau_{\nu_s}$ is the lifetime or decay time of sterile neutrinos. For sterile neutrinos to be dark matter candidate, their lifetime must be larger than age of the Universe, $4.4\times10^{17}$~sec. Using this fact and equation \eqref{s1}, one can estimate the upper bound on the total mixing angle.


\section{Impact on the thermal and ionization history}

Evolution of the ionization fraction with redshift in the presence of energy injection by decaying sterile neutrinos \cite{Seager1999, Seager,Liu:2018uzy, Mitridate:2018, Amico:2018, AliHaimoud:2010dx, Galli:2009},
\begin{alignat}{2}
\frac{dx_e}{dz} = \frac{\mathcal{P}}{H\,(1+z)}&\times\,\Big[\,n_{\rm H} x_e^2\,\alpha_{B}(T_{\rm gas})-(1-x_e)\,\beta_{B}(T_{\rm gas})\,e^{-E_{\alpha}/T_{\rm gas}} \Big]\nonumber\\
&- \frac{1}{H\,(1+z)}\, \Bigg(\,\frac{1}{E_0}-\frac{1-\mathcal{P}}{E_\alpha}\,\Bigg)\,\frac{(1-x_e)\ \mathcal{E}}{3\,n_{\rm H}}\,,
\label{s4}
\end{alignat}
where $x_e=n_e/n_{\rm H}$ is the ionization fraction, $n_e$ is the free electron number density and $n_{\rm H}$ is the total hydrogen number density in the Universe. $\alpha_{B}$ and $\beta_{B}$ are the case-B recombination coefficient and photo-ionization rate, respectively \cite{Seager1999, Seager, Mitridate:2018}. $E_0=13.6$~eV and $E_\alpha=(3/4)\,E_0$ are ground state binding energy and Ly$\alpha$ transition energy for the hydrogen atom, respectively. $\mathcal{P}$ is the Peebles coefficient \cite{Peebles:1968ja, Mitridate:2018, Amico:2018},
\begin{alignat}{2}
\mathcal{P}=\frac{1+K_{\rm H}\,\Lambda_{\rm H}\,n_{\rm H}\,(1-x_e)}{1+K_{\rm H}\,(\Lambda_{\rm H}+\beta_{\rm H})\,n_{\rm H}\,(1-x_e)\,}\,,
\label{s5}
\end{alignat}
here, $K_{\rm H}=\pi^2/(E_\alpha^3\, H)$ and $\Lambda_{\rm H}=8.22/{\rm sec}$  account for the redshifting of Ly$\alpha$ photon due to expansion of the Universe and the 2S-1S level two photon decay rate of the hydrogen atom, respectively \cite{Tung:1984}. The last term in equation \eqref{s4}, describes the additional effect of sterile neutrinos decay on the ionization fraction. $\mathcal{E}\equiv\mathcal{E}(z,m_{\nu_s})$ is the energy deposition rate per unit volume into IGM gas due to decaying sterile neutrinos. It can be written as \cite{Mitridate:2018, Amico:2018, Ripamonti:2006},
\begin{alignat}{2}
\mathcal{E}(z,m_{\nu_s})= {\mathcal{F}}_{S}\ f_{\rm abs}(z,m_{\nu_s})\ \times\ \frac{\rho_{\nu_s,\rm o}}{\tau_{\nu_s}}\ (1+z)^3\label{s6}
\end{alignat}
here, $\tau_{\nu_s}$ is the lifetime of sterile neutrino to decay in a active neutrino and a photon. ${\mathcal{F}}_{S}$ is the fraction of the sterile neutrinos that are decaying. We consider that all sterile neutrinos are decaying, i.e. ${\mathcal{F}}_{S}=1\,$.  $\rho_{\nu_s,0}= m_{\nu_s}\,n_{\nu_s,0}$ is the present day energy density of sterile neutrino. $n_{\nu_s,0}$ is the present day number density of sterile neutrinos. For the present work, we consider that all the dark-matter is composed of sterile neutrinos, $\rho_{\nu_s,0}\equiv\rho_{\rm DM,0}\ $, and $\rho_{\rm DM,0}$ is the present day dark-matter energy density \cite{Ripamonti:2006, Mapelli:2005, Dolgov:2002, Boyarsky:2019}. $f_{\rm abs}(z,m_{\nu_s})$ is the energy deposition efficiency into IGM by decaying sterile neutrinos. The energy deposition happens due to only radiative decay of sterile neutrino as active neutrinos interact very weakly with matter. Therefore, we consider only radiative decay of sterile neutrinos. $f_{\rm abs}(z,m_{\nu_s})$ depends on the redshift and mass of sterile neutrino \cite{Ripamonti:2006}. The mass of decaying particles enters only through $f_{\rm abs}(z,m_{\nu_s})$. In the presence of energy deposition  into IGM, the gas temperature evolution with redshift \cite{Seager1999, Seager, Liu:2018uzy, Mitridate:2018, Amico:2018, Galli:2009}, 
\begin{alignat}{2}
\frac{dT_{\rm gas}}{dz}  =  2\frac{T_{\rm gas}}{(1+z)} +& \frac{\Gamma_{C}}{(1+z)\,H}\, (T_{\rm gas}-T_{\rm CMB}) \nonumber\\
&\ \ -\frac{2}{3\,H\,(1+z)}\times \, \frac{(1+2\,x_e)\ \mathcal{E}}{3\,n_{\rm tot}}
\,,
\label{s7} 
\end{alignat}
here, $n_{\rm tot}=n_{\rm H}\,(1+f_{\rm He}+x_e)$ is the total number density of gas, $f_{\rm He}=n_{\rm He}/n_{\rm H}$ is the helium fraction, $n_{\rm He}$ is the helium number density. The first term in this equation comes due to the expansion of the Universe. The matter temperature falls with redshift adiabatically: $\propto(1+z)^2$ when Compton scattering (second term) becomes insufficient ($z\lesssim200$) and $\tau_{\nu_s}\rightarrow\infty$. The  Compton scattering rate is defined as,
\begin{equation}
\Gamma_{C}= \frac{8\, \sigma_T\, a_r T_{\rm CMB}^4\, x_e}{3\,(1+f_{\rm He}+x_e)\,m_e}\,,\label{s8}
\end{equation}
where, $\sigma_T$, $a_r$ and $m_e$ are the Thomson scattering cross-section, Stefan-Boltzmann radiation constant and mass of electron, respectively. Above the redshift $z\sim200$, the gas remains in thermal equilibrium with photons due to Compton scattering as $\Gamma_{C}\gg H$. At $z=200$, one can find that $\Gamma_{C}\approx1.4\times10^{-14}$~${\rm sec^{-1}}$ when $\mathcal{E}=0$, while, $H=3.6\times10^{-15}$~${\rm sec^{-1}}$. As $\Gamma_{C}\propto(1+z)^4$ and $H\propto(1+z)^{3/2}$ for matter dominated era, the Compton scattering rate will dominate over $H$ as one increase $z$ above 200. Therefore, the gas and CMB share same temperature above $z\sim200\,$--- as second term dominates over the first term. Below $z\sim200$, the Compton scattering rate becomes smaller compared to $H$ resulting in an adiabatic evolution of the gas when there is no last term present in equation \eqref{s7}. The last term corresponds to the energy deposition into IGM due to radiative decay of sterile neutrinos.  Following the Refs. \cite{Amico:2018, Mitridate:2018,Chen:2004,Shull:1985}, we consider the `SSCK' approximation--- in which $(1-x_e)/3$ fraction of deposited energy goes into ionization, nearly same amount goes into excitation, and remaining $(1+2x_e)/3$ fraction goes into IGM heating. We also discuss the projected bounds on sterile neutrinos after the inclusion of the process of gas heating in the cosmic dawn era by CMBR using Ref. \cite{Venumadhav:2018}, in subsequent discussion we call this process VDKZ18. Here, the energy transfer between gas and CMBR is mediated by Ly$\alpha$ photons from the first stars. The authors claim that it can increase the gas temperature by the order of ($\sim 10\%$) at $z\sim17$. Here, it is to be noted that we do not include the x-ray heating of the gas due to the uncertainty of known physics of the first stars. For a fix value of $T_{21}$ at a redshift, if we include the x-ray heating of the gas, the projected bounds becomes stronger. Including the heating due to VDKZ18 effect, equation \eqref{s7} will modify as,
\begin{alignat}{2}
\frac{dT_{\rm gas}}{dz}=\frac{dT_{\rm gas}}{dz}\Bigg|_{[{\rm eq. \eqref{s7}}]}-\frac{\Gamma_{R}}{(1+z)\,(1+f_{\rm He}+X_e)}\,,\label{s9}
\end{alignat}
where, ${dT_{\rm gas}}/{dz}\big|_{[{\rm eq. \eqref{s7}}]}$ represents the  temperature evolution in equation \eqref{s7}, and heating rate due to energy transfer from CMB photons to the thermal energy of gas by Ly$\alpha$ photons,
\begin{alignat}{2}
\Gamma_{R}=x_{\rm HI}\,\frac{A_{10}}{2\, H}\,x_{R} \left[\frac{T_R}{T_S}-1\right]\,T_{10}\,,\label{seq13}
\end{alignat}
here, $A_{10}=2.86\times 10^{-15}$~sec$^{-1}$ is the Einstein coefficient for spontaneous-emission from triplet state to singlet state. $x_R=1/\tau_{21}\times[1-\exp(-\tau_{21})]$ and $\tau_{21}=8.1\times10^{-2}\,x_{\rm HI}\,[(1+z)/20]^{1.5}\,(10~{\rm K}/T_S)$ is the 21~cm optical depth. $T_{10}=2\pi\nu_{10}=0.0682$~K and $x_{\rm HI}\simeq 1 - x_e\,$ is the neutral hydrogen fraction in the Universe.


\section{Bounds on the sterile neutrinos}\label{BonSN}

As described in the \eqref{21cmp}, we get an absorption profile in the 21 cm signal around redshift $z\sim17$ with an amplitude of $T_{21}\sim-220$~mK in the theoretical models  based on $\Lambda$CDM framework of cosmology. 
We take 21~cm differential brightness temperature such that it does not change, from its standard value ($\sim-220$~mK), by more than about a factor of 1/4 (i.e. $-150$ mK) or 1/2 (i.e. $-100$ mK) at redshift 17.2\,. We solve the coupled equations \eqref{s4} and \eqref{s7} for different mass and lifetime of sterile neutrino to get $x_{\rm HI}$ and $T_{\rm gas}$ at redshift $z=17.2\,$. To get any absorption signal in redshift range $15-20$, the gas temperature should be less than CMB temperature in shaded region. By requiring $T_{21}\simeq-150$~mK or $-100$~mK at z=17.2, we can put the projected constraints on the lifetime of sterile neutrinos. Subsequently, using equation \eqref{s1}, we can also put projected constraints on the mixing angle of sterile neutrinos with active neutrinos.

\begin{figure}[]
	\begin{center}
		\subfloat[] {\includegraphics[width=2.9in,height=2.1in]{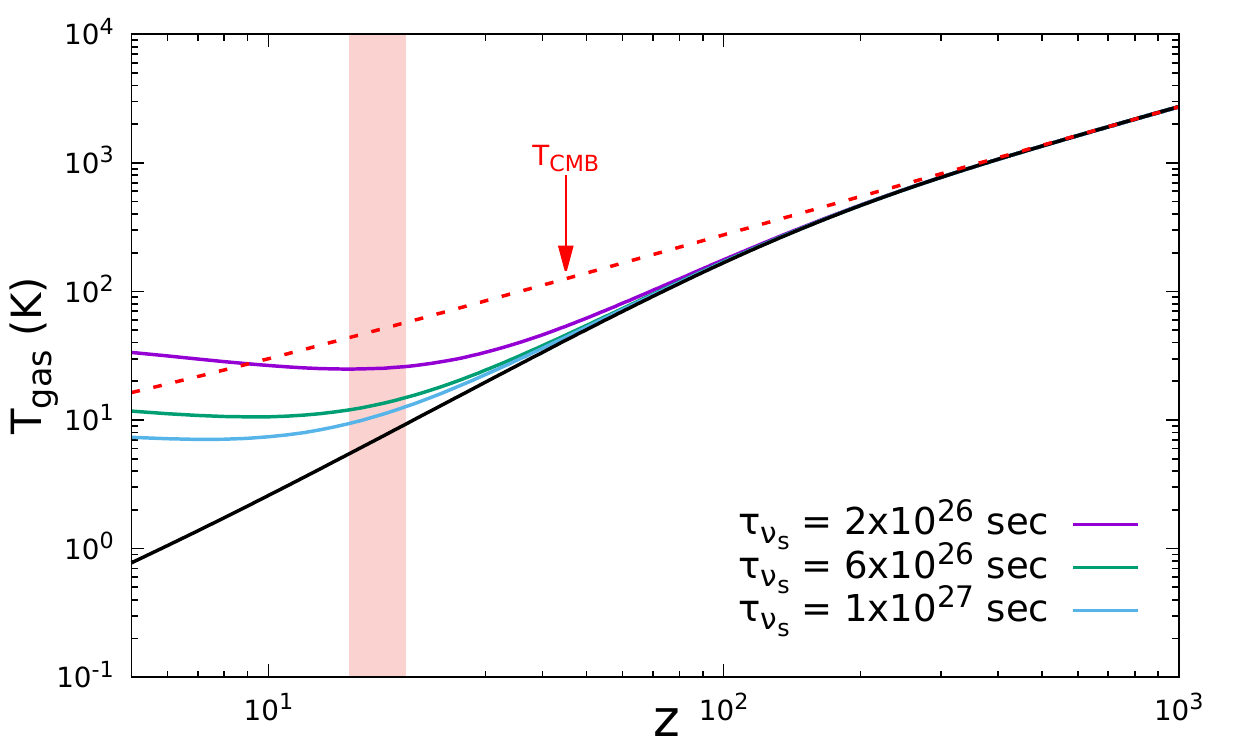}\label{plot:s1a}}
		\subfloat[]
		{\includegraphics[width=2.9in,height=2.1in]{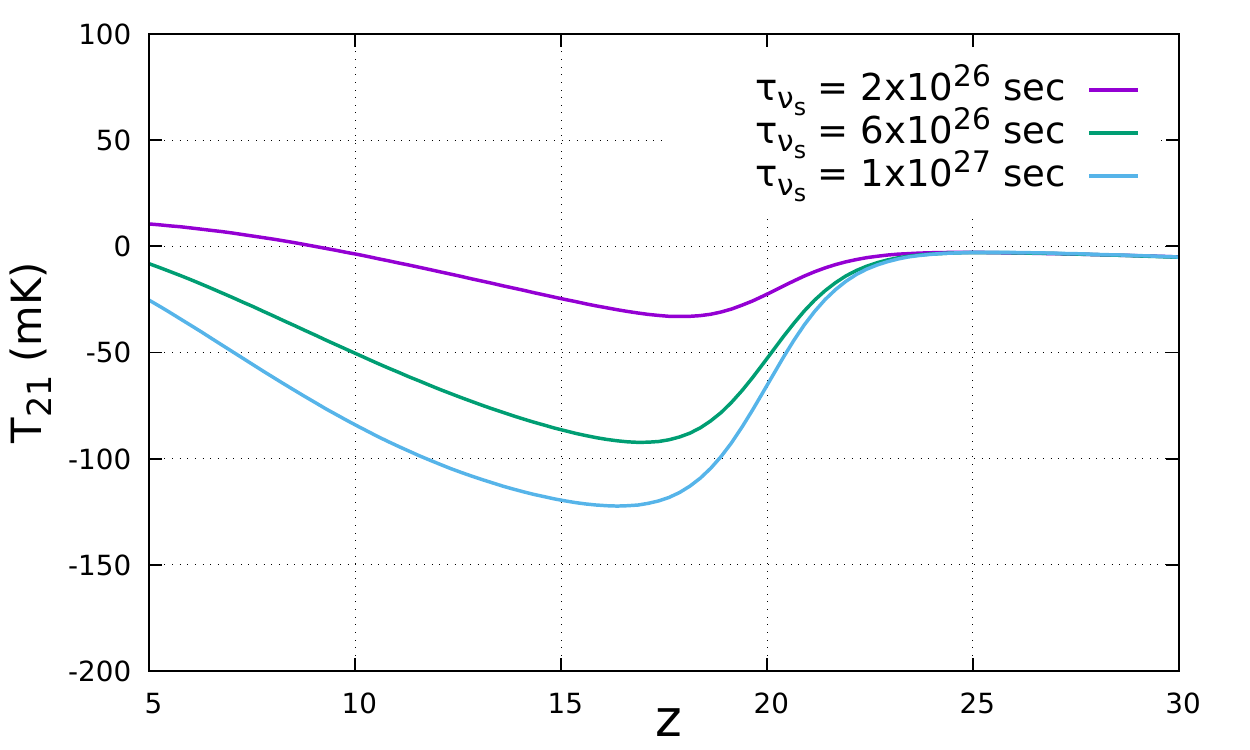}\label{plot:s2_1a}} 
	\end{center}
	\caption{ The gas temperature evolution with redshift in the presence of decaying sterile neutrinos. The red dashed line represents the CMB temperature evolution. The black solid line depicts the $T_{\rm gas}$ when there is no sterile neutrino decay. The shaded region corresponds to EDGES absorption signal, i.e. $15\leq z \leq 20$. In these figures, we keep mass of sterile neutrino fix to 10~KeV and vary lifetime. In figure \eqref{plot:s2_1a}, we plot evolution of 21 cm differential brightness temperature as a function of redshift for the cases represented in figure \eqref{plot:s1a}.}\label{plot:s1}
\end{figure}
\begin{figure}[]
	\begin{center}
		\subfloat[]  {\includegraphics[width=2.9in,height=2.1in]{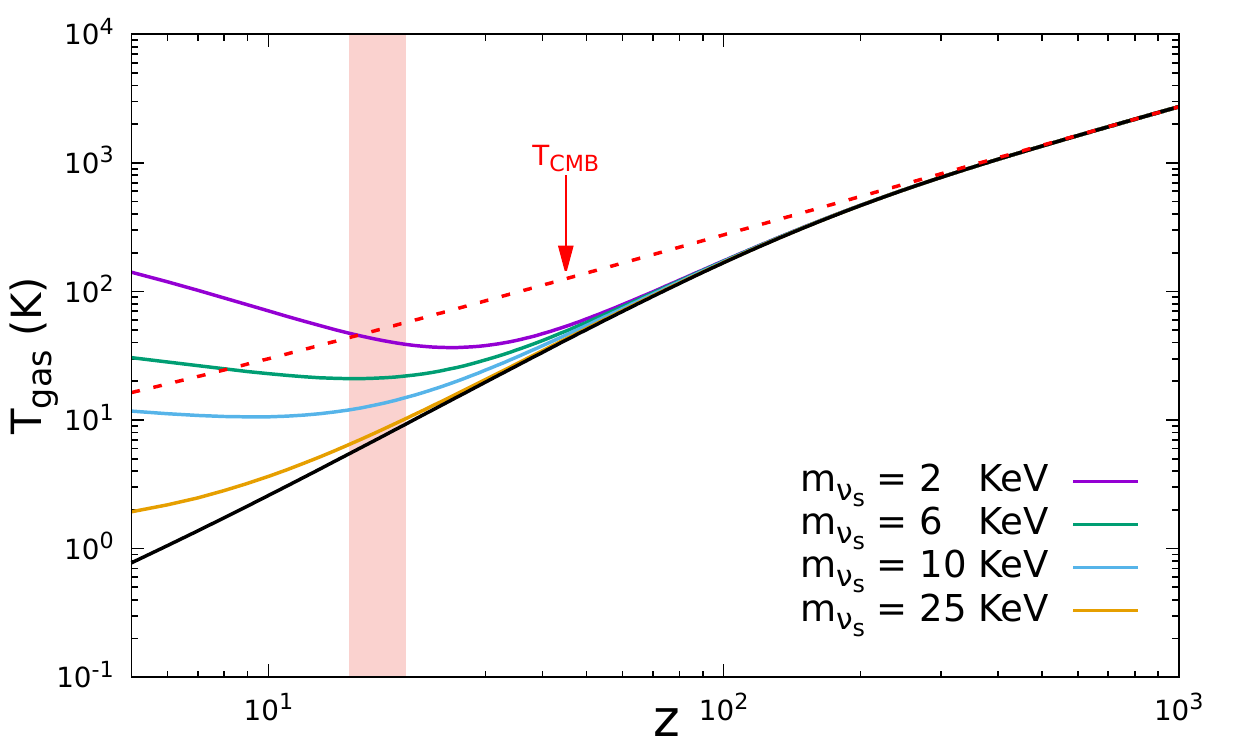}\label{plot:s1b}} 
		\subfloat[] 
		{\includegraphics[width=2.9in,height=2.1in]{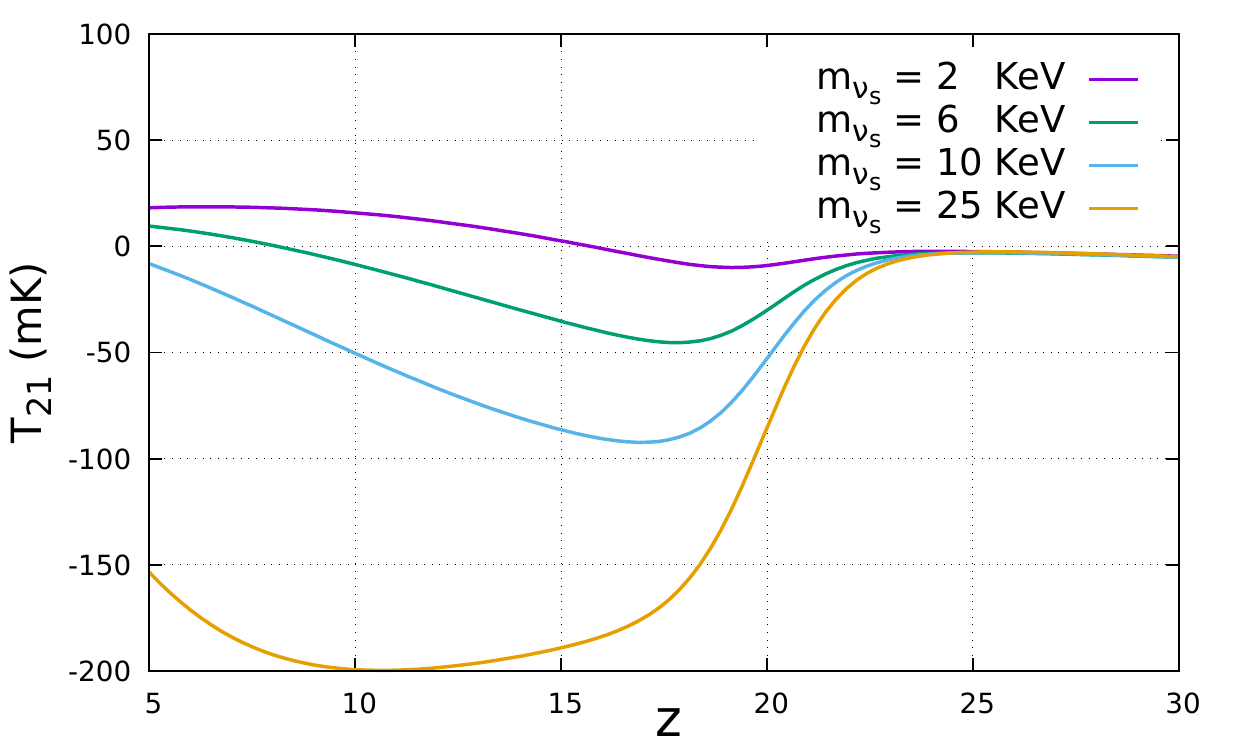}\label{plot:s2_1b}} 
	\end{center}
	\caption{The figure caption is same as in figure \eqref{plot:s1}, except here, we consider $\tau_{\nu_s}$ constant to $6\times10^{26}$~sec and vary mass of sterile neutrino. }\label{plot:s2}
\end{figure}

\begin{figure}[]
	\begin{center}
		\subfloat[] {\includegraphics[width=2.9in,height=2.1in]{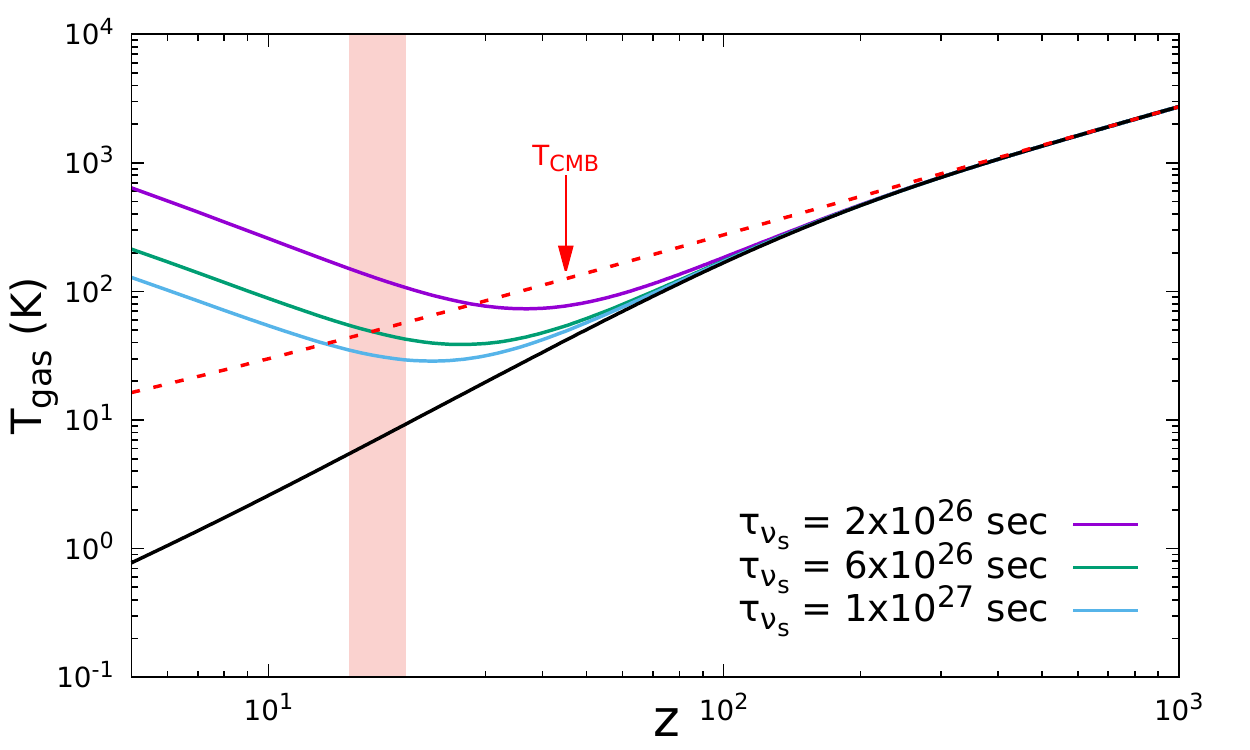}\label{plot:s1c}} 
		\subfloat[] {\includegraphics[width=2.9in,height=2.1in]{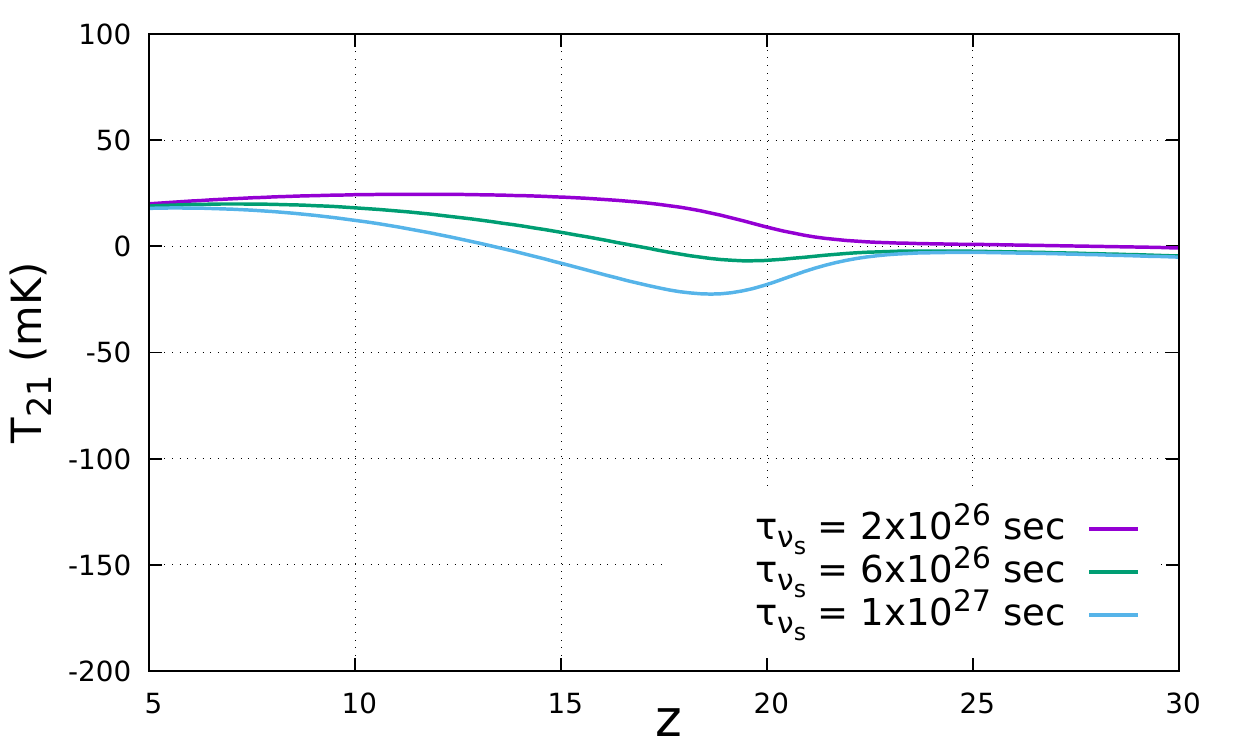}\label{plot:s2_1c}} 
	\end{center}
	\caption{ The figure caption is same as in figure \eqref{plot:s1}, except here, we keep $f_{\rm abs}(z,m_{\nu_s})=1/2$ and vary lifetime of sterile neutrino.}\label{plot:s3}
\end{figure}

In the figures \eqref{plot:s1a}, \eqref{plot:s1b} and \eqref{plot:s1c}, we plot the gas temperature evolution as a function of redshift for different mass and lifetime of sterile neutrino. The red dashed line in all plots represents the CMB temperature evolution with redshift. The black solid line represents the gas temperature evolution when there is no effect of decaying sterile neutrino on the IGM gas. The shaded pink region corresponds to redshift range $15\leq z \leq 20\,$. We obtain these results by considering $f_{\rm abs}(z,m_{\nu_s})$ from Ref. \cite{Ripamonti:2006}. In figures \eqref{plot:s2_1a}, \eqref{plot:s2_1b} and \eqref{plot:s2_1c}, we plot the evolution of the 21 cm differential brightness temperature as a function of redshift for the scenarios discussed in figures \eqref{plot:s1a}, \eqref{plot:s1b} and \eqref{plot:s1c}, respectively. We consider the $tanh$ parametrization model for the  Wouthuysen-Field coupling coefficient ($x_\alpha$) to get $T_{21}$ profiles \cite{Kovetz2018, Mirocha:2015G, Harker:2015M}. In the shaded region of the figures, the spin temperature can be approximated as gas temperature. Therefore, when the gas temperature is lower than CMB temperature, we get the absorption profile, i.e. $T_{21}<0$. When the gas temperature rises above the CMB temperature, $T_{21}$ becomes positive, and we see an emission profile. In all figures \eqref{plot:s2_1a}, \eqref{plot:s2_1b} and \eqref{plot:s2_1c}, above the redshift $z\gtrsim25$, $x_\alpha,\,x_c<1$, therefore, the spin temperature is dominated by CMB temperature, i.e. $T_{21}\approx0$. To get the absorption profile at $z\sim17$, one has to keep $T_{\rm gas}<T_{\rm CMB}$.

In figure \eqref{plot:s1a}, we keep the mass of sterile neutrino ($m_{\nu_s}$) fix to 10~KeV. The violet solid line depicts the gas temperature evolution when lifetime of sterile neutrino is $2\times10^{26}$~sec. As we increase the lifetime of sterile neutrino from $2\times10^{26}$~sec to $1\times10^{27}$~sec, the gas temperature decreases--- shown by green and cyan curves. It happens because by increasing the $\tau_{\nu_s}$, the radiative decay of sterile neutrinos decreases and the number of photons injected into IGM also decreases. Therefore, we get less heating of IGM by increasing the $\tau_{\nu_s}$, and it results in a smaller amplitude (larger dip) of $T_{21}$ shown by figure \eqref{plot:s2_1a}.

In plot \eqref{plot:s1b}, lifetime of sterile neutrino is fixed to $6\times10^{26}$~sec and the values of $m_{\nu_s}$ varies from $2$~KeV (violet solid line) to $25$~KeV (yellow solid line). If one increases the sterile neutrino mass from 2~KeV (violet line) to 6~KeV (green line), the heating of IGM decreases significantly in the shaded region. It happens because $\rho_{\nu_s}=m_{\nu_s}n_{\nu_s}$, $n_{\nu_s}$ is the number density of sterile neutrinos. Therefore at a particular redshift, when one increases $m_{\nu_s}$ the number density of sterile neutrino decreases, and we get less photon injunction, produced from decaying sterile neutrinos, into the IGM. Hence, one gets less heating of IGM when the mass of sterile neutrino increases, and it results in a smaller amplitude (larger dip) of $T_{21}$ shown by figure \eqref{plot:s2_1b}.

If one considers the immediate and complete absorption of the photon energy into IGM,  then energy deposition efficiency, $f_{\rm abs}=1/2$ --- half of the total energy of sterile neutrino will be carried away by active neutrino \cite{Ripamonti:2006, Mapelli:2006}. The mass of the sterile neutrino in the equations \eqref{s4} and \eqref{s7}, enters through only $f_{\rm abs}$. Therefore, the energy deposition rate, equation \eqref{s6}, will depend only on  the lifetime of sterile neutrinos. This case has been depicted in figure \eqref{plot:s1c} for the different values of $\tau_{\nu_s}$. In this case, as expected, the heating of IGM increases more compared to the cases in figure \eqref{plot:s1a}. The corresponding profiles for 21 cm signal are shown in figure  \eqref{plot:s2_1c}. In figure \eqref{plot:s1c}, the gas temperature for $\tau_{\nu_s}=2\times10^{26}$~sec is higher than the CMB temperature in the shaded region-- (violet line). Therefore, we get a emission profile for $T_{21}$ in figure \eqref{plot:s2_1c} for $\tau_{\nu_s}=2\times10^{26}$~sec. For $\tau_{\nu_s}=6\times10^{26}$~sec, at redshift $\sim 17$, the gas temperature is comparable to the CMB temperature, therefore we do not see any absorption/emission in the 21 cm signal. Above the redshift $\sim17$, temperature of gas is lower than CMB, therefore, we see a small absorption in the profile. Below the redshift $\sim17$, the temperature of the gas is higher than CMB, therefore, we see a emission profile. For the case with $\tau_{\nu_s}=1\times10^{27}$~sec, in the shaded region, temperature of the gas is smaller than the CMB (cyan line), therefore, we get an absorption profile for the 21 cm signal--- figure \eqref{plot:s2_1c}.

\begin{figure}[hb!]
    \begin{center}
       {\includegraphics[width=4in,height=2.8in]{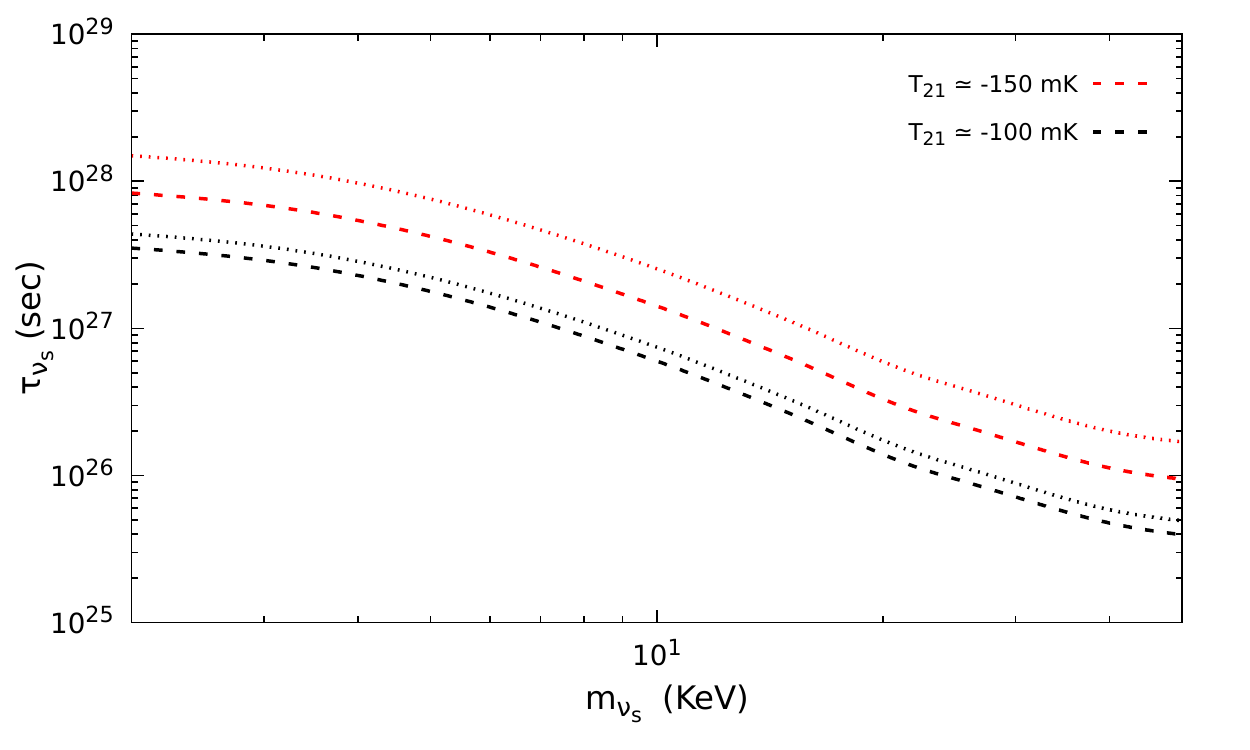}}       
    \end{center}
    \caption{The figure represents lower projected bounds on the lifetime of sterile neutrinos as a function of mass of sterile neutrinos by keeping 21~cm differential brightness temperature, $T_{21}\simeq-150$ and $-100$~mK. The dotted (dashed) line represents the case when energy transfer from CMB photons to gas is included (excluded) \cite{Venumadhav:2018}.}\label{plot:s2a}
\end{figure} 
\begin{figure}[]
	\begin{center}
	 {\includegraphics[width=6in,height=4in]{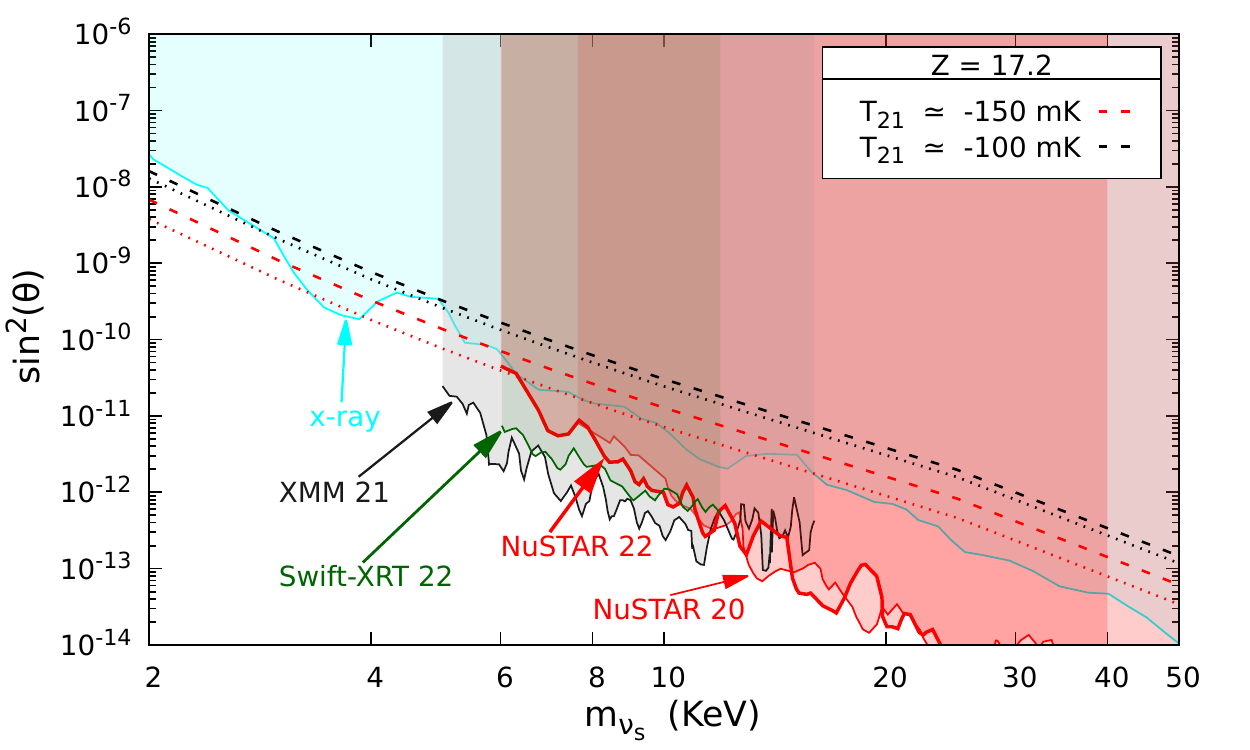}}
	\end{center}
	\caption{ The figure represents upper projected bounds on the mixing angle of sterile neutrinos with active neutrinos as a function of mass of sterile neutrinos by keeping 21~cm differential brightness temperature, $T_{21}\simeq-150$ and $-100$~mK. The dotted (dashed) line represents the case when energy transfer from CMB photons to gas is included (excluded) \cite{Venumadhav:2018}.  The shaded regions are excluded for corresponding observations. The x-ray constraint on mixing angle (cyan shaded region) has been taken from the Ref. \cite{ Boyarsky:2019}. The red shaded region depicts the upper bounds on $\sin^2(\theta)$ from NuSTAR observations \cite{Brandon:2020, NuStarGH_2022}. Here, we have also plotted the recently reported bounds (after publication of our article) on $\sin^2(\theta)$ by NuSTAR--- represented by NuSTAR 22 \cite{NuStarGH_2022} and by Swift-XRT--- represented by Swift-XRT 22 \cite{Swift-XRT_22}. The grey shaded region is excluded by  XMM-Newton \cite{foster:2021}.}\label{plot:s2b}
\end{figure} 

In figure \eqref{plot:s2a}, we plot the lower projected constraints on lifetime as a function of $m_{\nu_s}$ by requiring  $T_{21}$ such that it does not suppress the standard theoretical value of $T_{21}(z=17.2)\approx-220$~mK more than about a factor of 1/4 or 1/2. Considering $T_{21}<-150$~mK, will further strengthen our projected bounds. The red coloured curves depict the lower projected constraints on $\tau_{\nu_s}$ when $T_{21}\simeq-150$~mK, while the black coloured curves represent the lower projected constraint on $\tau_{\nu_s}$ when $T_{21}\simeq-100$~mK. To get the dashed line, we do not take into account the VDKZ18 heating of the gas. For the dotted line we consider  VDKZ18 heating of the gas. Inclusion of VDKZ18, gives more stringent projected constraint on $\tau_{\nu_s}$ as gas temperature rises due to the energy transfer from CMB photons mediated by Ly$\alpha$ photons. In figure \eqref{plot:s2b}, we obtained the  upper projected constraint on mixing angle of sterile neutrinos with active neutrinos as a function of mass. For reference, we have also plotted the x-ray constraint on mixing angle as function of $m_{\nu_s}$. The constraint is obtained by assuming solely radiative decay of sterile neutrinos. x-ray constraint comes from the fact that no such x-rays have been seen in observations \cite{Boyarsky:2019}. The red and black coloured curves depict the upper projected constraint on mixing angle when $T_{21}\simeq-150$~mK and -100~mK, respectively. To get the dashed curves, we do not take into account the VDKZ18 heating of the gas. For the dotted line we have included the VDKZ18 heating of the gas. Here, it is to noted that these bounds do not depend on dark-matter clustering. Therefore, the bounds are free of astrophysical parameters such as density profile or mass function of dark-matter halos, etc. To obtain these bounds, we do not consider any non-standard cooling mechanism to cool the IGM or any source of radio photons. The results in figure \eqref{plot:s2b} are comparable with the x-ray constraint for the higher mass of sterile neutrinos, while we get more stronger bounds for lower mass.


\section{Summary} 
We have constrained the sterile neutrino dark matter lifetime and mixing angle with active neutrino as a function of sterile neutrino mass, such that energy injection from radiative decay of sterile neutrino does not change the standard 21 cm absorption signal ($\sim-220$~mK) more than about a factor of 1/4 ($-150$~mK) or 1/2 ($-100$~mK) at the redshift, $z= 17.2\,$. We have considered the two scenarios to get the bounds: First, IGM evolution without the heat transfer from the background radiation to gas mediated by Ly$\alpha$ photons (VDKZ18 effect). Next, we have considered the VDKZ18 effect on the IGM gas. The following summarises our results for $T_{21} = -150$~mK\,: 

In the first scenario, the lower bound on the sterile neutrino lifetime varies from $8.3\times10^{27}$~sec to $9.4\times10^{25}$~sec by varying sterile neutrino mass from 2 KeV to 50 KeV.  The lifetime of sterile neutrino decrease when one increases the mass of the sterile neutrino.  It happens because $\rho_{\nu_s}=m_{\nu_s}n_{\nu_s}$. At a particular redshift, when one increases $m_{\nu_s}$, the $n_{\nu_s}$ decreases. Consecutively, one gets less radiative decay of sterile neutrinos. Therefore, we get more window to increase the gas temperature, i.e. we can decrease the lifetime of sterile neutrinos. The upper bound on the mixing angle ($\sin^2\theta$) varies from  $6.8\times10^{-9}$ to  $6.1\times10^{-14}$ by varying sterile neutrino mass from 2 KeV to 50 KeV.

In the second scenario, the lower bound on the sterile neutrino lifetime varies from $1.5\times10^{28}$ sec to  $1.7\times 10^{26}$ sec by varying sterile neutrino mass from 2 KeV to 50 KeV. While the upper bound on the mixing angle varies from $3.8\times10^{-9}$ to $3.42\times10^{-14}$ by varying sterile neutrino mass from 2 KeV to 50 KeV. 

We have also plotted the x-ray constraint to rule out some parameter space for mixing angle of the sterile neutrinos with active neutrinos \cite{Boyarsky:2019}. Although we have considered that sterile neutrinos account for all the dark matter in the Universe, sterile-neutrino may account for only a fraction of the dark matter abundance. In this scenario, the bounds on the sterile neutrino lifetime and mixing angle with active neutrino may modify.


\section{Additional study}
\subsection{Bounds in light of varying $T_{21}$ and redshift}

We have also studied the projected constraints on $\tau_{\nu_s}$ and $\sin^2(\theta)$ by varying the absorption amplitude of $T_{21}$ between $0$~mK (no signal) to $-200$~mK at $z=17$.  If we increase the value of $T_{21}$ above 0, it gives a emission signal instead of an absorption signal. Therefore, we restrict the maximum value  of  $T_{21}$ to 0. Also, we do not take the values of $T_{21}$ below $\sim-200$~mK, as the sterile neutrino term in the temperature evolution equation becomes insignificant compared to adiabatic and Compton scattering term. In the $\Lambda$CDM model without invoking any new physics, we get $T_{21}(z=17)=-220.215$ for the cosmological parameters $\Omega_{\rm m }=0.31$, $\Omega_{\rm b}=0.048$ and $h=0.68$\,. For the demonstration purpose of this, we take $m_{\nu_s}=2$~KeV. After inclusion of physics of decaying sterile neutrinos, we get $T_{21}(z=17)=-220.213$~mK for $\tau_{\nu_s}=4\times10^{32}$~sec. 
If we increase the value of $T_{21}$ only by $9.7\times10^{-2}$ percent (from $T_{21}=-220.213$~mK to $-220$~mK), the value of $\tau_{\nu_s}$ changes significantly from $4\times10^{32}$~sec to $3.78\times10^{30}$~sec--- decreases by more than a factor of hundred. Therefore, we do not consider bounds on the $\tau_{\nu_s}$ and $\sin^2(\theta)$ near the maximal absorptional value of $T_{21}$; and vary the value of $T_{21}$ from $-200$ to $0$~mK.

\begin{figure}[]
		\centering
		\subfloat[]
		{\includegraphics[width=2.9in,height=2.1in]{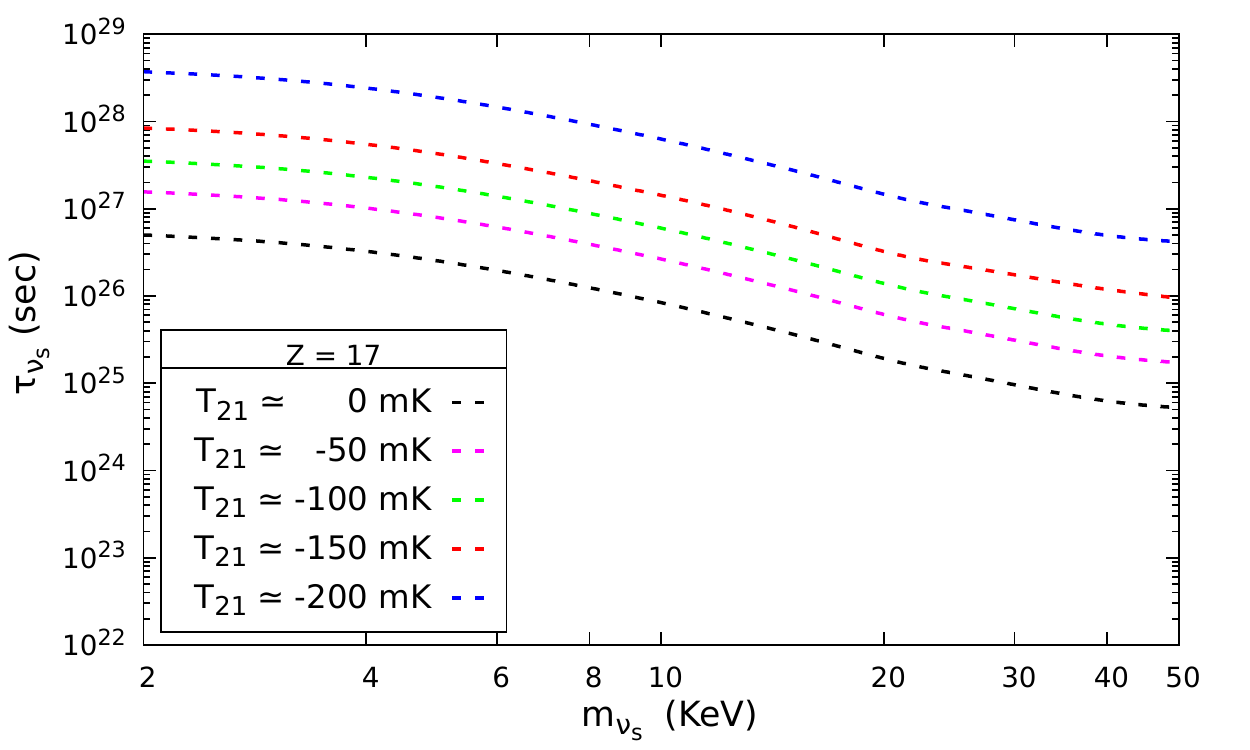}\label{ch2plot:1a}}
		\subfloat[] {\includegraphics[width=2.9in,height=2.1in]{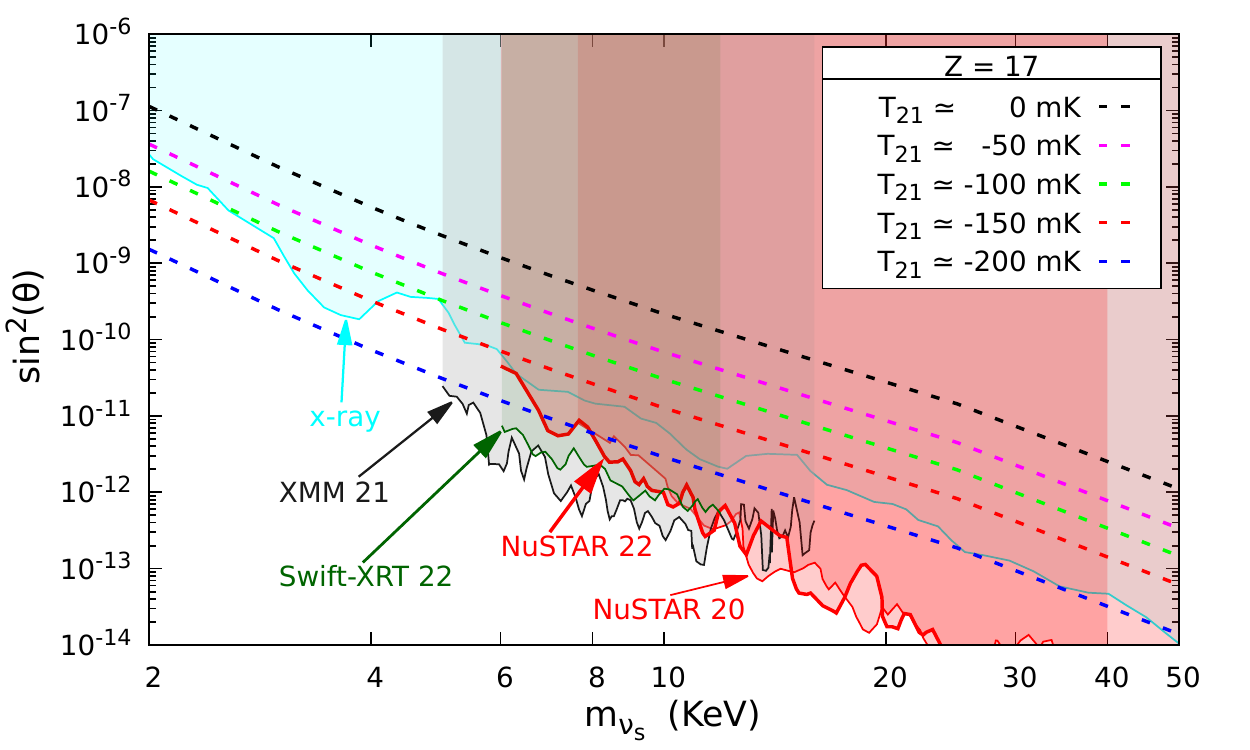}\label{ch2plot:1b}}
		\caption{Plot \eqref{ch2plot:1a}, shows lower projected bounds on the lifetime of sterile neutrinos as a function of mass, while plot \eqref{ch2plot:1b}, shows upper projected bounds on the mixing angle of sterile neutrinos with active neutrinos as a function of mass of sterile neutrinos for varying 21~cm differential brightness temperature ($T_{21}$) at $z=17$. In figure \eqref{ch2plot:1b}, the shaded regions are excluded for corresponding observations. The cyan shaded region represents the x-ray constraint \cite{Boyarsky:2019}. The red shaded region depicts the upper bounds on $\sin^2(\theta)$ from NuSTAR observations \cite{Brandon:2020, NuStarGH_2022}. Here, we have also included the recently reported bounds (after publication of our article) on $\sin^2(\theta)$ by NuSTAR--- represented by NuSTAR 22 \cite{NuStarGH_2022} and by Swift-XRT--- represented by Swift-XRT 22 \cite{Swift-XRT_22}. The grey shaded region is excluded by  XMM-Newton \cite{foster:2021}.}
		\label{ch2plot:1}
\end{figure}

In figure \eqref{ch2plot:1a}, we have plotted the lower projected bounds on the lifetime of sterile neutrinos as a function of mass for various values of $T_{21}$ at $z=17$. In figure \eqref{ch2plot:1b}, we have plotted the upper projected bounds on the mixing angle of sterile neutrinos with active neutrinos as a function of mass for various values of $T_{21}$ at $z=17$. NuSTAR 22 bound is reported in July 2022 \cite{NuStarGH_2022} and Swift-XRT 22 is reported in August 22 \cite{Swift-XRT_22}. The observational bounds indicate that the decay of sterile neutrinos will not significantly impact the thermal history of the Universe as the parameter space is excluded more stringently  by observations for a higher mass of sterile neutrinos. For example, XMM 21 excludes the values of $\sin^2(\theta)\gtrsim2\times10^{-12}$ for $m_{\nu_s}\simeq6$~KeV. If one wants to exclude this parameter space using 21-cm signal, it requires to consider $T_{21}<-200$~mK--- i.e. no significant modification in the thermal and ionization history of the Universe.

\begin{figure}[]
	\begin{center}
		\subfloat[]
		{\includegraphics[width=2.9in,height=2in]{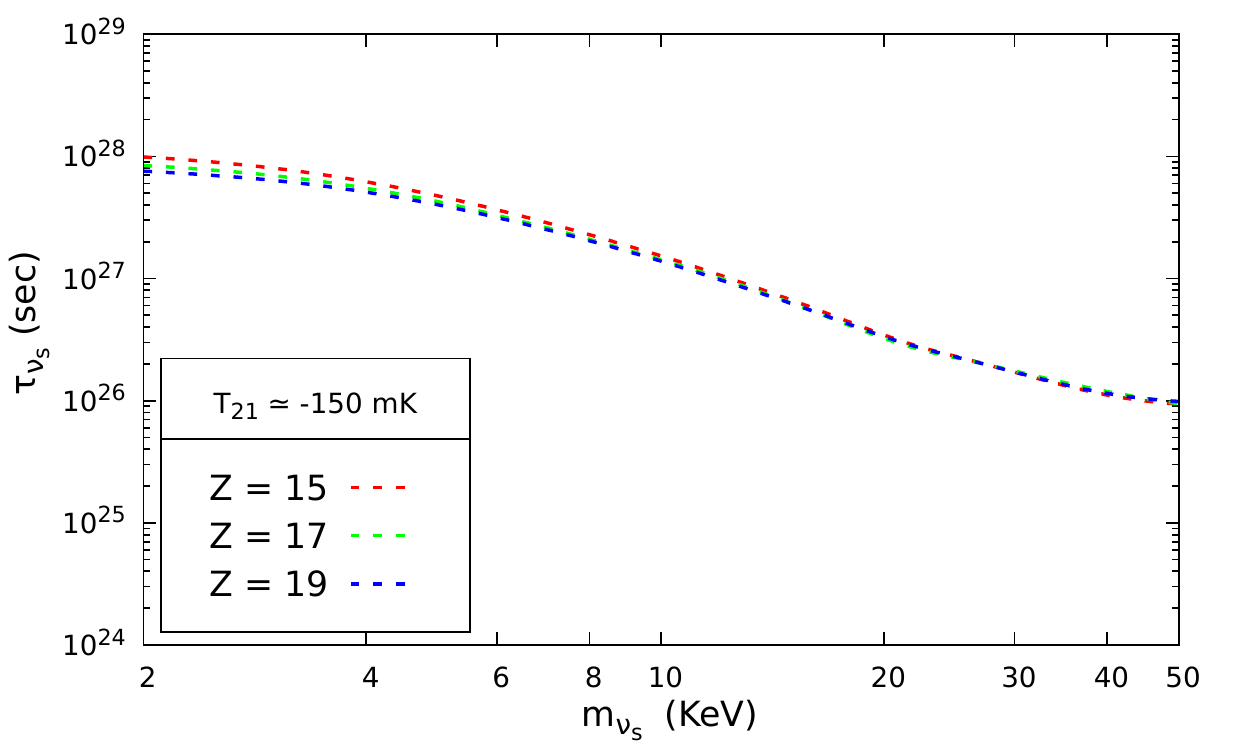}\label{ch2plot:2a}}
		\subfloat[] {\includegraphics[width=2.9in,height=2in]{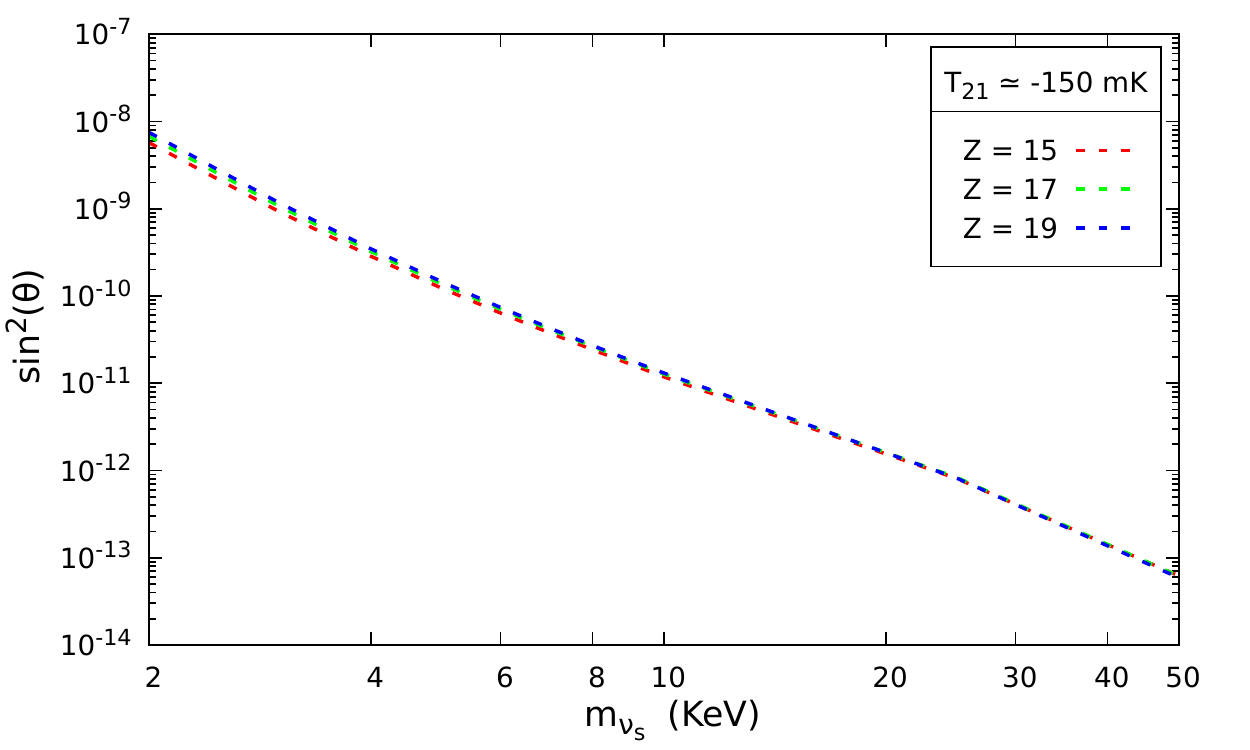}\label{ch2plot:2b}}
	\end{center}
	\caption{Plot \eqref{ch2plot:2a}, shows lower projected bounds on the lifetime of sterile neutrinos, while plot \eqref{ch2plot:2b}, shows upper projected bounds on the mixing angle of sterile neutrinos with active neutrinos by keeping  $T_{21}$ to $-150$~mK for different values of redshift. }
	\label{ch2plot:2}
\end{figure} 

\begin{figure}[]
\begin{center}
	\subfloat[]
	{\includegraphics[width=2.9in,height=2in]{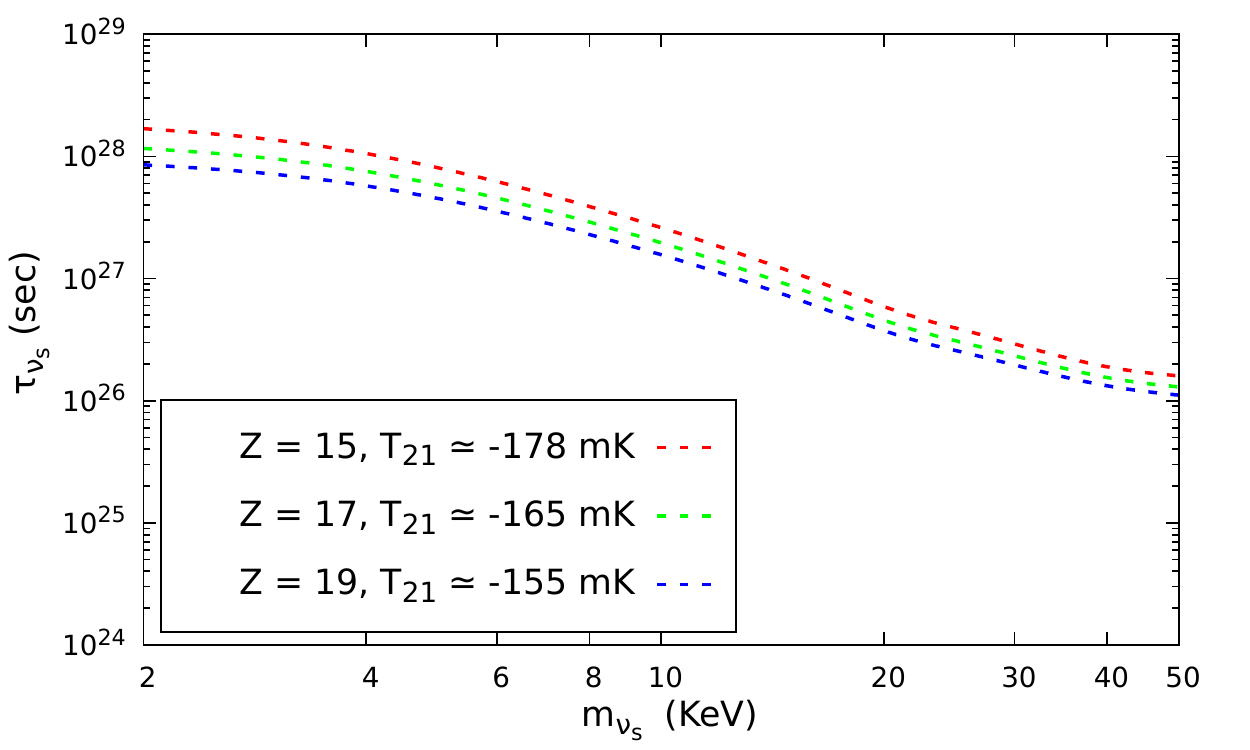}\label{ch2plot:3a}}
	\subfloat[] {\includegraphics[width=2.9in,height=2in]{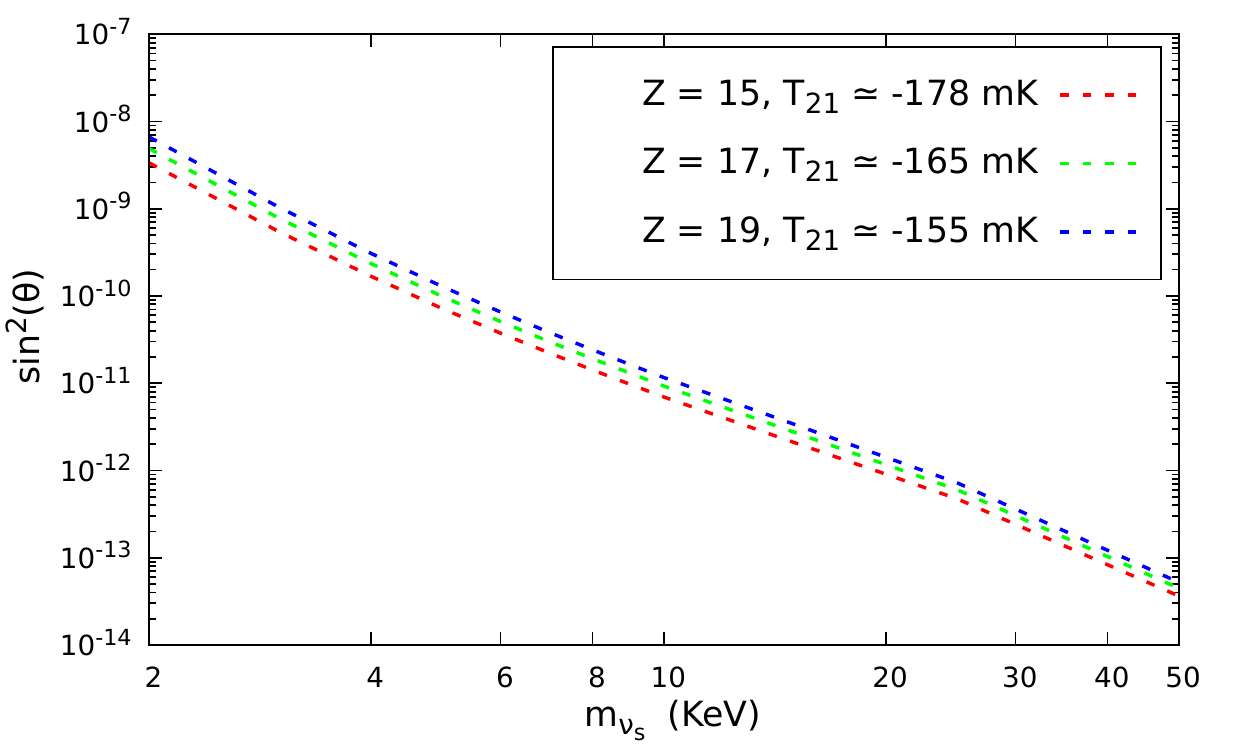}\label{ch2plot:3b}}
\end{center}
	\caption{Plot \eqref{ch2plot:2a}, shows lower projected bounds on the lifetime of sterile neutrinos, while plot \eqref{ch2plot:2b}, shows upper projected bounds on the mixing angle of sterile neutrinos with active neutrinos by keeping $T_{21}$ such that it does not change more than a factor of $1/4$ from the minimum possible amplitude based on $\Lambda$CDM model for corresponding values of redshift.}
	\label{ch2plot:3}
\end{figure} 

For further analysis with variation of redshift values, we have also added the plots for the case with different values of redshift keeping the value of $T_{21}$ constant--- presented in figure \eqref{ch2plot:2}. Here, we vary redshift between 15 to 19. As it is shown in figure \eqref{p_2a}, for fiducial models for Ly$\alpha$ coupling and x-ray heating, we can not take the spin temperature to be gas temperature above $z\sim17$ and also x-ray starts to dominate below $z\sim17$. Therefore, we restrict ourselves about redshift 17 and take a range from 15 to 19. Here, we do not see significant variation in the projected bounds of lifetime and mixing angle with variation of the values of redshift.

In figure \eqref{ch2plot:3}, we vary both the value of redshift and $T_{21}$, and plot the lower projected bounds on $\tau_{\nu_s}$ in figure \eqref{ch2plot:3a} and upper projected bounds on $\sin^2(\theta)$ in figure \eqref{ch2plot:3b}. Here, we choose the value of $T_{21}$, such that, it does not change more than a factor of $1/4$ from the minimum possible amplitude based on $\Lambda$CDM model. For the cosmological parameters, given above, we get the minimum possible amplitude of $T_{21}^{\rm Min}$ to be $-236.7$~mK at $z=15$, $-220.2$~mK at $z=17$ and $-206.1$~mK at $z=19$.



%
\clearpage
\pagestyle{empty}
\cleardoublepage
\pagestyle{fancy}
\begin{savequote}[75mm]
	``The universe doesn`t allow perfection."
	\qauthor{Stephen Hawking, \textit{A Brief History of Time}}
\end{savequote}

\chapter{Primordial Black Hole Dark Matter}\label{chap3}
\vspace{-1.5cm}

Primordial black holes have attracted much interest in recent years and have been a part of intense studies for more than five decades. The idea of the black hole goes back to the 18th century. In 1784, John Michell proposed that there could be such supermassive bodies that light could not pass them, or all light emitted would return towards them \cite{Michell:1784, Simon:1979, Colin:2009}. Later, in 1915 Albert Einstein developed the general theory of relativity. In 1916, Karl Schwarzschild found the solution of black holes by solving the Einstein field equations for a point mass \cite{Schwarzschild:1916}. Subsequently, in 1963, Roy Kerr found the solution of rotating black holes \cite{Kerr:1963}. In 1965, the more general solution of a rotating and charged black hole was found \cite{Newman:1965}. There is a possibility that a colossal number of black holes might have been formed in the very early Universe--- known as primordial black holes (PBHs). PBHs can be created by various mechanisms. It was first suggested by Zel’dovich and Novikov that the presence of initial inhomogeneities in the Universe can form PBHs \cite{Zeldovich:1967}. There is a possibility that for many regions in the space, gravitational energy of the initial density fluctuations can exceed the kinetic energy. These regions would have a gravitational collapse instead of the expanding with Universe creating collapsed objects with a minimum mass of $\sim10^{-5}$ g  \cite{Hawking:1971, Carr:1974H, Carr:1975}. There are various mechanisms that can produce inhomogeneities in the early Universe. For example, such density fluctuations can be generated due to the vacuum strings produced during the grand unification phase transition \cite{Vilenkin:1981}. Indeed, these fluctuations were present in the very early Universe, as evident from observations of structures in the Universe. The other explanations of PBHs formation include the collapse of cosmic string loops, collisions of bubbles, etc. The cosmic string loops can disappear in two ways: First, they can shrink into scalar and gauge particles. Second, some loops with specific initial shapes may disappear by collapsing in size below their Schwarzschild radius and form black holes \cite{Hawking:1989, Polnarev:1991, Hawking:1990, Garriga:1993}. In the article \cite{Ubi:1998}, the authors consider the formation of PBHs due to collapsing cosmic strings and argue that PBHs can significantly contribute to the dark matter density if their relic mass is larger than $10^3\,m_{\rm pl}$, here $m_{\rm pl}$ is the Planck mass. In another scenario, the collapse of the cusps neighbourhood of cosmic strings loops can also form a large number of spinning PBHs \cite{Jenkins:2020}. The collisions between the bubbles during various phase transitions in the Universe can also give rise to the formation of PBHs \cite{Hawking:1982, LA:1989, Jung:2021}. PBHs can also be produced in various inflation models \cite{Frampton:2010, Sebastien:2015, Inomata:2018}. 

Depending on the formation time $(t)$, PBHs can have a wide range of masses (in most of the cases roughly order of the Hubble horizon mass at the formation time) \cite{carr:2021, Carr:2010},
\begin{alignat}{2}
M_{\rm PBH}\, \sim\, 10^{15}\ \left[\frac{t}{10^{-23}\,\rm sec}\right]~{\rm g}. \label{eq3.1}
\end{alignat}
For example, PBHs with mass $10^{15}$~g might have formed at $t\sim10^{-23}$ after big-bang. In another example, PBHs formed during the QCD phase transition  ($t\sim10^{-5}$~sec)  might have a mass comparable to a solar mass \cite{Green:2014}. PBHs formed around neutrino decoupling  ($t\sim1$~sec) can have a mass about $10^5\ {\rm M_\odot}\,$.


\section{Primordial black holes as dark matter}

In the last decades, many particle-dark matter models have been proposed to explain the various astrophysical observations, as discussed in chapter \eqref{chap2}. The laboratory experiments for direct detection of dark matter have not observed any signature yet, for example, DarkSide-50, LUX, XENON1T, PandaX-II, CRESST, PICO, etc. \cite{DarkSide:2015, LUX:2017, XENON1T:2017, PandaX:2017, CRESST:2019, PICO:2019}. In this situation, it is desirable to look for alternative scenarios where dark matter may not be an elementary particle. As PBHs are massive, interact only gravitationally and are formed in the very early Universe, they can be considered as a potential candidate for non-particle dark matter. Recently, PBHs have gathered much attention in the scientific community after the black hole binary merger detection by Virgo and LIGO collaborations, and these events suggest that PBHs may constitute a fraction of dark matter  \cite{Bird:2016, Abbott:2016A, Abbott:2016B, Abbott:2016C, Abbott:2016D, Sasaki:2016}. PBHs having a mass below $\sim 10^{22}$~g can explain all the dark matter in the Universe as they are not ruled out by microlensing constraints \cite{Niikura:2019}. We will discuss other constraints on dark matter fractions in the form of PBHs with mass below  $\sim 10^{22}$~g later in the sections \eqref{chap3b} and \eqref{chap3Result}. One can explain the existence of dark matter in the form of PBHs without considering physics beyond the standard model (BSM) of particle physics by considering standard model Higgs fluctuations during inflation as instability can occur in Higgs potential at a scale  $\mathcal{O}(10^{11}~{\rm GeV})$ \cite{Espinosa:2018}. In the article \cite{Frampton:2010}, authors consider the double inflation model to explain the formation of PBHs between two inflations and argue that PBHs can be accounted for dark matter in the Universe. PBHs as missing matter or dark matter in the context of galaxy formation has been explored in old literature also \cite{Meszaros:1975, Chapline:1975}. Authors of the Ref. \cite{Sebastien:2015}, consider the formation of PBH dark matter due to the mild-waterfall phase of hybrid inflation and discuss how the tail of the mass distribution of PBHs can explain the origin for the supermassive black holes observed at galactic centres. These massive back holes can also provide the seed for present-day observed structures \cite{Sebastien:2015, Garc:2017}. A fraction/all of dark matter in the form of PBHs can produce the $r$-process nucleosynthesis--- a process that is responsible for producing about half of the heavier nuclei than iron, in the mass range $10^{-14}\,{\rm M}_\odot < M_{\rm PBH} < 10^{-8}\,{\rm M}_\odot$ \cite{Fuller:2017}.  Black holes can lose their mass by the emission of energetic particles due to Hawking evaporation \cite{Hawking:1974N}. For non-rotating and non-charged black holes formed in the very early Universe, their evaporation time scale can be given by \cite{carr:2021},
\begin{alignat}{2}
\tau(M_{\rm PBH})\, \sim\, \left[\frac{M_{\rm PBH}}{10^{15}\,\rm g}\right]^3~{\rm Gyr}. \label{eq3.2}
\end{alignat}
Therefore, PBHs having mass larger than $10^{15}$~g can survive the Hawking evaporation and account for present-day dark matter density \cite{Hawking:1975}. 

\subsection{Signature of Primordial Black Holes}

It is possible that a fraction of PBHs can grow to intermediate-mass black holes and explain the ultraluminous x-ray sources reported in various observations \cite{Sebastien:2015, Dewangan:2006, Madhusudhan:2006, Liu:2013}. There are several hints that indicate the presence of PBHs, such as dynamics and star clusters of ultra-faint-dwarf-galaxies, correlations between x-ray and infrared cosmic backgrounds, etc. (for a detailed review, see Ref. \cite{Sebastien:2018}). The presence of evaporating PBHs can explain the galactic/extra-galactic $\gamma$-ray background radiation \cite{Wright:1996, Lehoucq:2009, Carr:1976, Page:1976}, short-duration $\gamma$-ray bursts \cite{Cline:1997, Green:2001}, and reionization by injection of energetic photons and $e^{\pm}$ radiations into IGM \cite{Belotsky_2014, Belotsky:2015}. The emission of nucleons by evaporating PBHs can explain the observed baryon number density if more baryons are produced compared to antibaryons--- in a baryon-symmetric Universe \cite{Carr:1976}. Clustering between PBHs can provide the seeds for galaxy formation. PBHs evaporation can explain the observed point-like $\gamma$-ray sources \cite{Belotsky_2014}. The presence of massive PBHs can also serve as seeds for active galactic nuclei (AGN) \cite{Belotsky_2014}. 


\section{Existing bounds on Primordial Black Holes}\label{chap3b}

The fraction of dark matter in the form of PBHs ($f_{\rm PBH}\equiv\Omega_{\rm PBH}/\Omega_{\rm DM}$) is constrained from various astrophysical observations and theoretical predictions. Here, $\Omega_{\rm PBH}$ and $\Omega_{\rm DM}$ are the dimensionless density parameters for PBHs and dark matter, respectively. PBHs with mass smaller than $\sim\mathcal{O}(10^{15}~{\rm g})$ may have evaporated as of now and can be constrained from the impact on big bang nucleosynthesis by evaporated particles, background radiation etc. Higher mass PBHs can be constrained by the effect on large-scale structures, gravitational wave and lensing, and impact on thermal and ionization history of the IGM (for details, see the recent reviews \cite{carr:2021, Green:2021, Carr:2020} and the references cited therein). In the context of the 21 cm signal, the upper bound on the $f_{\rm PBH}$ can be found in Refs. \cite{Hektor:2018, Clark:2018, Mena:2019, Yang:2020, Halder_2021, Tashiro_2021, Yang_2020, Pablo:2021}. Angular momentum is a fundamental property of a black hole, and it can modify the Hawking evaporation drastically \cite{Chandrasekhar:1977, Taylor_1998, Arbey::2020, ArbeyAJ:2021}. In the case of rotating PBHs, authors of the Refs. \cite{Dasgupta:2020, Laha:2021} have reported the various types of bound on $f_{\rm PBH} $ as a function of PBHs mass and spin. Future collaboration, All-sky Medium Energy Gamma-ray Observatory (AMEGO)\footnote{\href{https://asd.gsfc.nasa.gov/amego/index.html}{https://asd.gsfc.nasa.gov/amego/index.html}} will be able to constrain some parameter space for the rotating  PBHs \cite{Ray:2021}. We discuss more bounds on the fraction of PBH dark matter in the result and discussion section \eqref{chap3Result}. In this chapter, we consider the rotating PBHs and constrain dark matter fraction in the form of PBHs as a function of their mass for various values of angular momentum in the light of global 21 cm signal \cite{Natwariya:2021PBH}.


\section{Impact on the thermal and ionization history}

During the cosmic dawn era, the evolution of the gas temperature and ionization fraction of the Universe are well-known \cite{Seager1999, Seager}. The addition of any exotic source of energy during the cosmic dawn era can significantly impact the ionization and thermal history of the Universe. Therefore, we can constrain the properties of such exotic sources from the observations during the cosmic dawn era. Evaporating PBHs can heat the gas and modify the free electron fraction in the IGM \cite{Laha:2021, Kim_2021}. Rotating PBHs can emit more particles into IGM and substantially affect the IGM evolution compared to non-rotating PBHs  \cite{Chandrasekhar:1977, Page:1976, Page::1976}. Therefore, it is important to study the properties of spinning PBHs. Black holes can get their spin depending on generation mechanisms, merger or accretion \cite{Kesden_2010, Cotner_2017, Harada_2021, Luca_2019, Luca_2020, Harada_2017, K_hnel_2020, flores:2021, Arbey_2020, He_2019, Cotner_2019}. PBHs with higher mass can have a lifetime larger/comparable than the age of the Universe. Therefore, they have enough time to accrete  mass and spin up \cite{Dong_2016}. In the present work, we consider the Hawking emission of PBHs into background radiations (photons and electron/positron) and  provide the projected constraints on the fraction of dark matter in the form of PBHs ($f_{\rm PBH}$) as a function of mass and spin. We analyse projected bounds on spinning PBHs such that 21 cm differential brightness temperature does not change by more than a factor of 1/4 from the $\Lambda$CDM model prediction ($|T_{21}|\sim220$~mK).

A rotating black hole with angular momentum $J_{\rm PBH}$ and having mass $M_{\rm PBH}$ can be defined with a rotation parameter, $a_*=J_{\rm PBH}/(G_{\rm N}\,M_{\rm PBH}^2)$ \cite{Page::1976}, where $G_{\rm N}$ is the gravitational constant. Rotating black hole with higher spin ($a_*\rightarrow1$) injects more energy into IGM and evaporates faster than non-rotating ones \cite{Chandrasekhar:1977, Taylor_1998, Arbey::2020, ArbeyAJ:2021}. Therefore, we expect that the bounds on $f_{\rm PBH}$ to be more stringent compared to non-rotating PBHs. The energy injection per unit volume per unit time due to $e^{\pm}$ and photons into IGM, for monochromatic mass distribution of PBHs, can be written as \cite{Laha:2021, mittal:2021},
\begin{alignat}{2}
		\Gamma_{\rm PBH}^{e^{\pm}}(z,a_*)&=2 \int\left[ f_c^e(E-m_e,z)\,(E-m_e)\left(\frac{d^2 N_{e}}{dt\,dE}\right)\, \right]\,n_{\rm PBH}\, dE\,,\label{eqt1}\\
		\Gamma_{\rm PBH}^{\gamma}(z,a_*)&=\int\left[\ f_c^\gamma(E,z)\,E\,\left(\frac{d^2 N_\gamma}{dt\,dE}\right)\ \right]\,n_{\rm PBH}\ dE\,.\label{eqt2}
\end{alignat}
Energy injection into IGM happens by three processes: heating, ionization, and excitation of the gas \cite{Slatyer:2016, Slatyer::2016, Liu_2020}. $f_c^i$ represents the energy deposition efficiency into IGM. Here, $c$ stands for above-mentioned three channels and $i\equiv$ (electron/ positron, photon) stands for different types of injected particles. The factor of 2 in equation \eqref{eqt1} accounts for the total contribution of electrons and positrons. $n_{\rm PBH}=f_{\rm PBH}\,(\rho_{\rm DM}/M_{\rm PBH})$ is the number density of the PBHs, and $\rho_{\rm DM}$ is the dark matter energy density. $d^2 N^i/(dt\,dE)\equiv d^2 N^i/(dt\,dE)\,\big(E,M_{\rm PBH}, a_*\big)$ represents the number of $i$ particles emitted by black hole per unit time per unit energy \cite{Page::1976, Mac:1990, Laha:2021, Arbey_2019},
\begin{alignat}{2}
\frac{d^2 N^i}{dt\,dE}=\frac{1}{2\,\pi}\ \sum_{\rm dof}\,\frac{\Gamma_i(E,M_{\rm PBH}, a_*)}{e^{E'/T_{\rm PBH}}\pm 1}\,,
\end{alignat}
here, $\Gamma_i$ is the greybody factor--- defines the probability of emitted particle $i$ from black hole to overcome its gravitational potential well. dof represents the degree of freedom \cite{Arbey_2019}. Moreover, $E$ is the total energy of emitted particle $i$ and $E'=E-n\Omega$. While, $n$ is the axial quantum number and $\Omega$ is the angular velocity at black hole horizon. We use the \texttt{BlackHawk} code\footnote{\href{https://blackhawk.hepforge.org/}{https://blackhawk.hepforge.org/}} to calculate the spectra due to photons, electrons and positrons; we take both the primary and secondary Hawking evaporation spectra into account--- i.e. emitted final particle $j$ per unit time and per unit energy  \cite{Arbey_2019, ArbeyJ:2021}
\begin{alignat}{2}
\frac{d^2 N^j}{dt\,dE}=\sum_i \frac{d^2 N^i}{dt\,dE''}\ \frac{dN^i_j}{dE}\ dE''\,,
\end{alignat}
here, $dN^i_j$ is the hadronization table accounts for the transformation of the primary spectra into secondary spectra \cite{Arbey_2019, ArbeyJ:2021, Hazma:2020}.

In the presence of Hawking radiation, the thermal evolution of the gas can be written as\cite{Chluba2015, Amico:2018},
\begin{alignat}{2}
\frac{dT_{\rm gas}}{dz}=2\,\frac{T_{\rm gas}}{1+z}+\frac{\Gamma_c}{(1+z)\,H}(T_{\rm gas}-T_{\rm CMB})-\frac{2\ \,\Gamma_{\rm PBH}\,}{3\,n_{\rm tot}(1+z)\,H}\,,\label{eqt3}
\end{alignat}
here, $\Gamma_{\rm PBH}=\Gamma_{\rm PBH}^{e^{\pm}}+\Gamma_{\rm PBH}^{\gamma}$ is the total energy injection per unit time and per unit volume into IGM. We consider the following numerical values of the cosmological parameters: $h=0.674$, $\Omega_{\rm M }=0.315$, $\Omega_{\rm b}=0.049$ and $T_{\rm CMB}|_{z=0}=2.725$~K \cite{Planck:2018,Fixsen_2009}. To compute the energy deposition efficiency, thermal and ionization history of the Universe, we use \texttt{DarkHistory}\footnote{\href{https://darkhistory.readthedocs.io/en/master/}{https://darkhistory.readthedocs.io/en/master/}} package with necessary modifications \cite{Liu_2020}. 


\section{Results and Discussion}\label{chap3Result}   
We take 21~cm differential brightness temperature such that it does not change, from its $\Lambda$CDM value ($\sim220$~mK), by more than a factor of 1/4 at redshift 17.2\,. We solve the coupled equations \eqref{eqt3} and (\ref{s4}--- replacing $\mathcal{E}$ with $\Gamma_{\rm PBH}$) for different mass, spin and fraction of PBH dark matter to get $x_{\rm HI}$ and $T_{\rm gas}$ at redshift $z=17.2\,$. To get any absorption signal in redshift range $15-20$, the gas temperature should be less than CMB temperature in shaded region--- redshift range from 15 to 20. By requiring $T_{21}\simeq-150$~mK at z=17.2, we constrain the parameter space of PBH dark matter. In the present chapter, we do not consider x-ray heating of the gas due to the uncertainty in the physics of the first stars--- as we discussed earlier. For a fix value of $T_{21}$ at a redshift, if one includes the x-ray heating of gas, our projected upper constraints on PBH dark matter fraction becomes stronger. Here, it is to be noted that the gas temperature may increase due to the energy transfer from the background radiation to the thermal motions of the gas mediated by Ly$\alpha$ radiation from the first stars \cite{Venumadhav:2018}. However, again due to the uncertainty in physics of the first star formation, we do not include this effect also. The inclusion of this effect will also further strengthen our  projected upper bounds on $f_{\rm PBH}$--- similar to discussed in chapter \eqref{chap2}. 


\begin{figure}[]
	\begin{center}
		{\includegraphics[width=4in,height=2.7in]{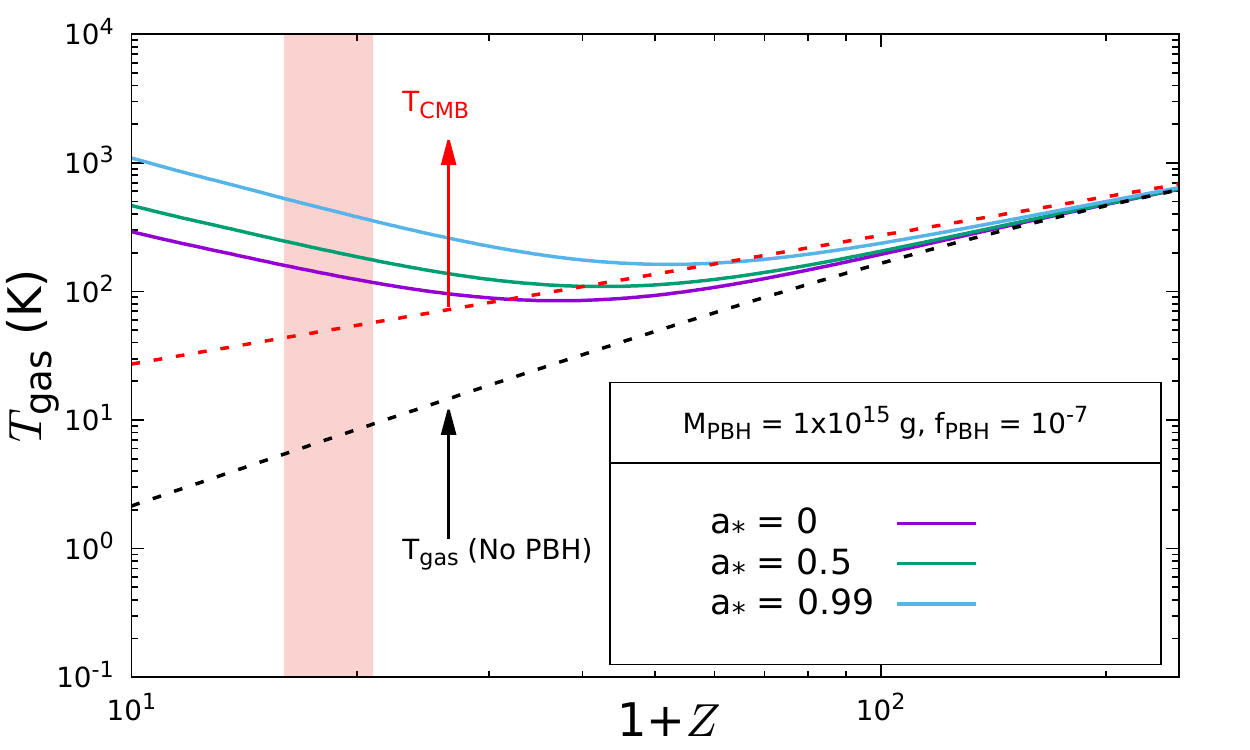}} 
	\end{center}
	\caption{The gas temperature evolution with redshift for evaporating primordial black hole. The red dashed lines represent the CMB temperature evolution. The black dashed lines depicts the $T_{\rm gas}$ when there is no PBHs. The shaded region corresponds to the redshift $15\leq z \leq 20$ (EDGES observed signal). In this figure, we consider PBHs mass and $f_{\rm PBH}$ to $1\times10^{15}$~g and $10^{-7}$, respectively, and vary the spin of PBHs.}\label{plot:t1a}
\end{figure}

\begin{figure}[]
	\begin{center}
		{\includegraphics[width=4in,height=2.7in]{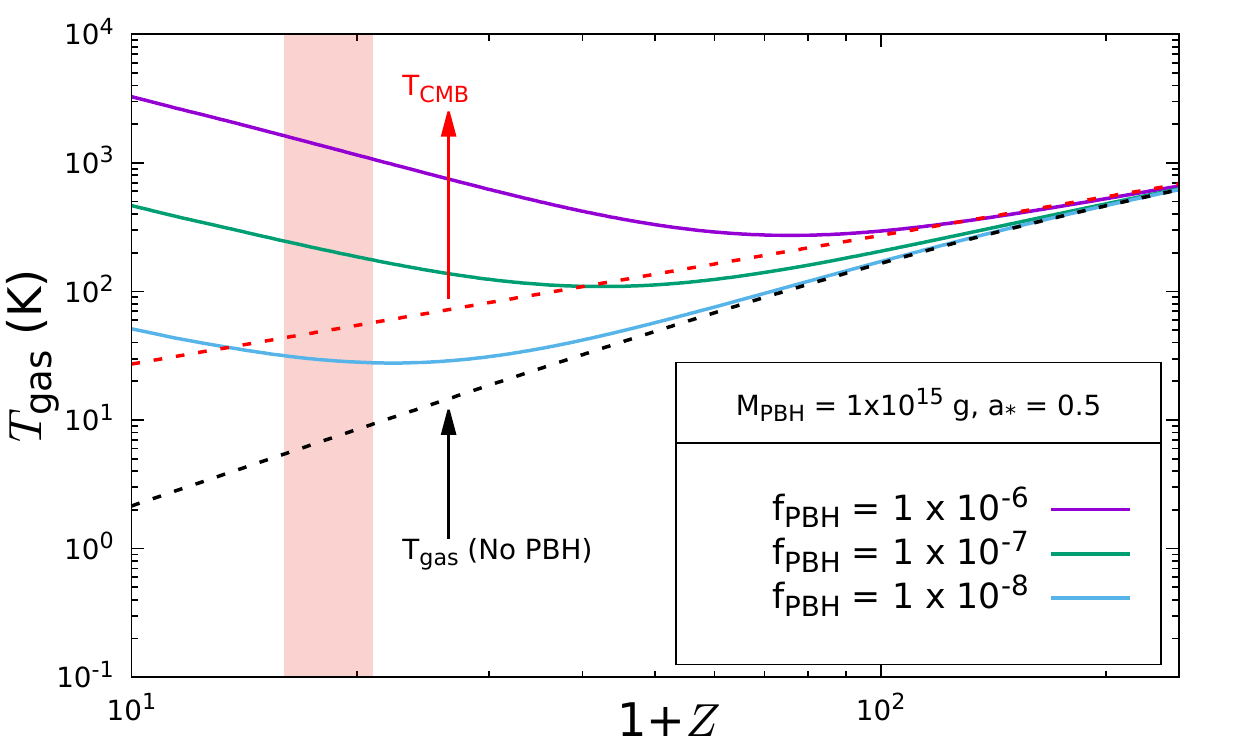}}  
	\end{center}
	\caption{ The caption is the same as in Figure \eqref{plot:t1a}, except here, we keep $M_{\rm PBH}=1\times10^{15}$~g and $a_*=0.5$ constant and vary $f_{\rm PBH}$.}\label{plot:t1b}
\end{figure}

\begin{figure}
	\begin{center}
		{\includegraphics[width=4in,height=2.7in]{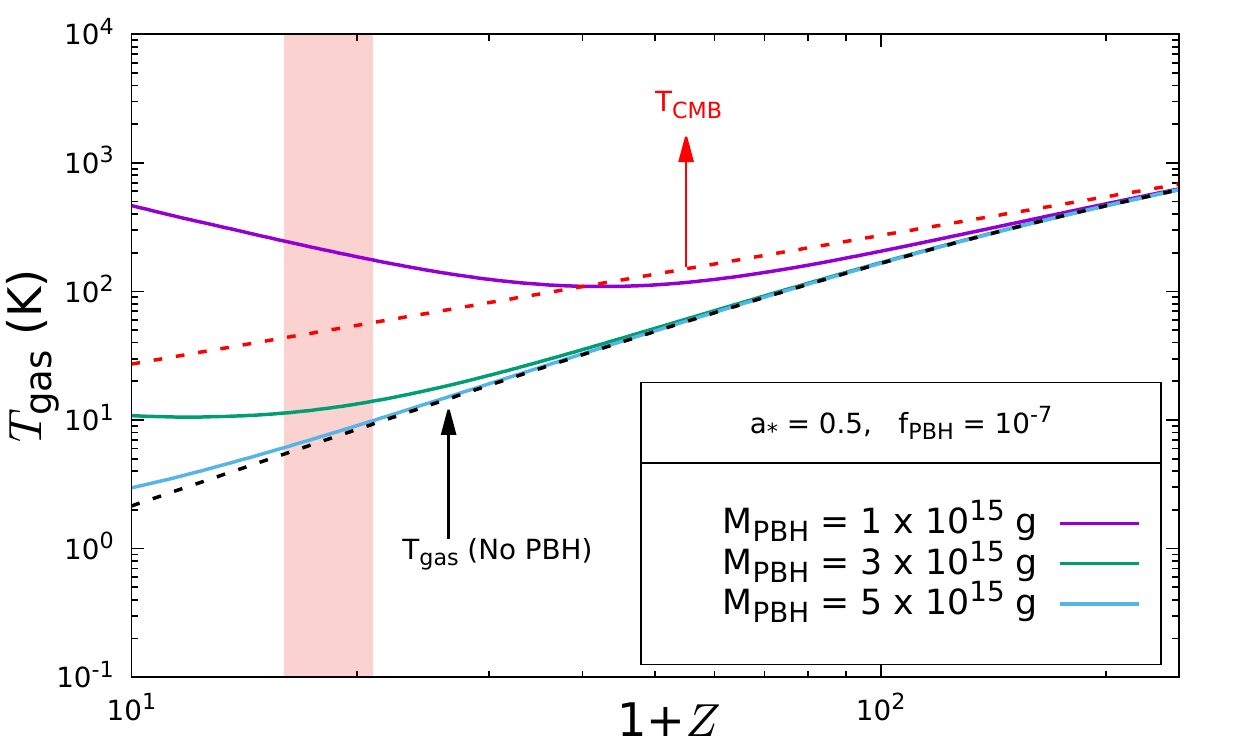}\label{plot:1c}} 
	\end{center}
	\caption{ The caption is the same as in Figure \eqref{plot:t1a}, except here, we vary the mass of PBHs and keep spin and $f_{\rm PBH}$ to $0.5$ and $10^{-7}$, respectively.}\label{plot:t1_2}
\end{figure}

%

In order to understand how spin, fraction and mass of PBH dark matter can affect the thermal evolution of the gas, we plot the figures \eqref{plot:t1a}, \eqref{plot:t1b} and \eqref{plot:t1_2}, respectively. The shaded region corresponds redshift range, $15-20\,$.  The red dashed curves in all plots depict the CMB temperature evolution, while the black dashed line represents the gas temperature when there are no evaporating PBHs. In Figure \eqref{plot:t1a}, we keep mass to $1\times10^{15}$~g and  $f_{\rm PBH}=10^{-7}$, and vary the spin of PBHs. As expected, when we increase the spin of PBHs, the gas temperature rises significantly in the shaded region. The solid violet curve represents the case when the spin of PBHs is 0. Increasing the spin to 0.5 (solid green line), the gas temperature increases. Further increasing $a_*$ to 0.99 (solid cyan line), the gas temperature rises further. In Figure \eqref{plot:t1b}, we keep  $M_{\rm PBH}=1\times10^{15}$~g, spin to 0.5 and vary $f_{\rm PBH}$. In this plot, as we increase the $f_{\rm PBH}$ from $10^{-8}$ (solid cyan line) to $10^{-6}$ (solid violet line), the IGM heating rises rapidly. If the gas temperature becomes larger than the CMB temperature in the shaded region, it can erase the 21 cm absorption signal; instead, it may give an emission signal. Therefore, at desired redshift (in our scenario $z = 17.2$), one has to keep $T_{\rm gas}< T_{\rm CMB}$ to get an absorption signal. Increasing $f_{\rm PBH}$, for a given mass, the number density of PBHs increases resulting in more energy injection into IGM by Hawking evaporation of PBHs. Therefore, $f_{\rm PBH}$ plays a significant role in deciding whether one gets an absorption profile or emission. In Figure \eqref{plot:t1_2},  we vary the mass of PBHs and keep spin and $f_{\rm PBH}$ constants to $0.5$ and $10^{-7}$, respectively. In this plot, as we increase the mass of PBHs from $1\times10^{15}$~g (solid violet line) to $5\times10^{15}$~g (solid cyan line), the gas temperature decreases. It happens for two reasons:  (i) Increasing the mass of PBHs leads to a decrease in the  total power contributions from Hawking evaporation of PBHs \cite{Mac:1990}. (ii)  Ignoring the integral dependency in equations \eqref{eqt1} and \eqref{eqt2}, $\Gamma_{\rm PBH}^{e^\pm}$ and $\Gamma_{\rm PBH}^{\gamma}$ are proportional to  $n_{\rm PBH}=f_{\rm PBH}\,(\rho_{\rm DM}/M_{\rm PBH})$. For a fixed dark-matter energy density and $f_{\rm PBH}$, the number density of PBHs increases by decreasing the black hole mass. Thus, energy injection into IGM per unit volume and time ($\Gamma_{\rm PBH}$) increases, and one gets more heating of the gas.\\

\vspace{0.15cm}
\noindent{\hrule height 0.08cm depth 0cm \relax} 
\begin{sidewaysfigure}
	\captionsetup{labelformat=empty}
	\begin{center}
		\subfloat[] {\includegraphics[width=4.1in,height=4.65in]{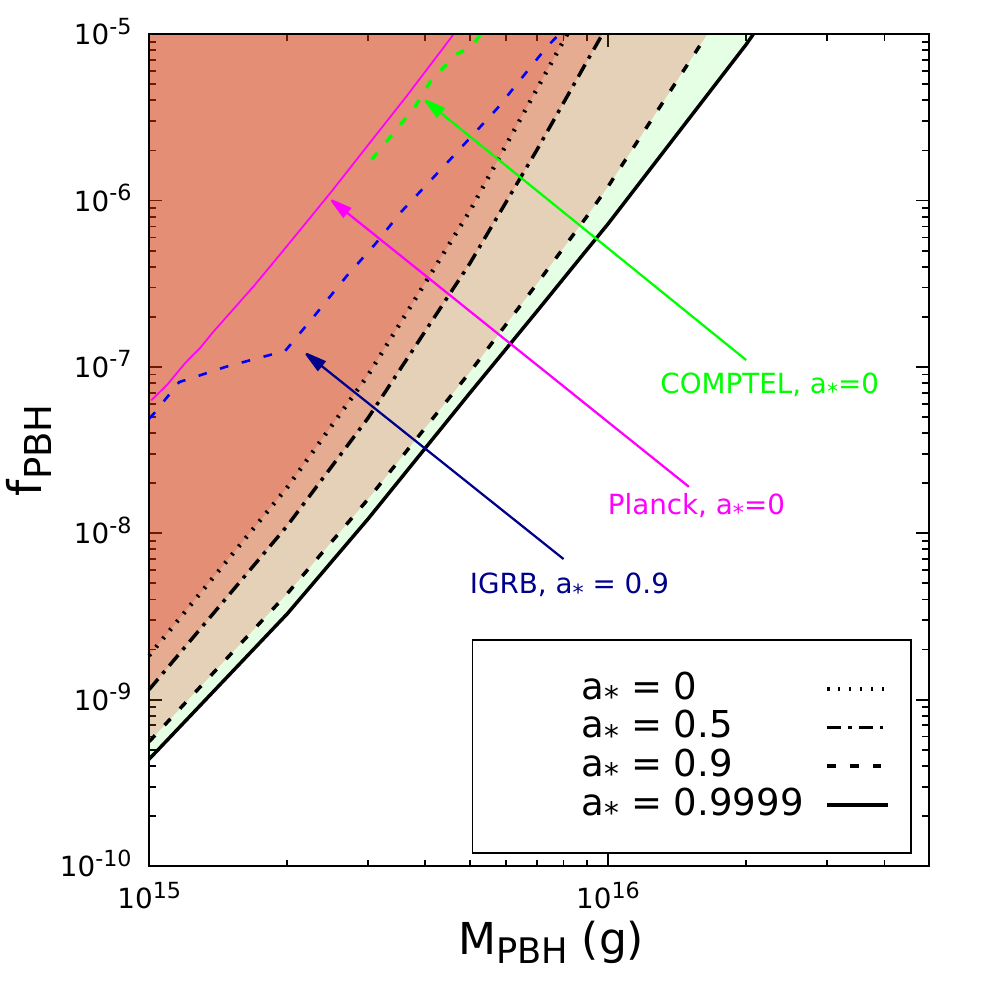}  \label{plot:t2a}}
		\subfloat[] { \includegraphics[width=4.1in,height=4.65in]{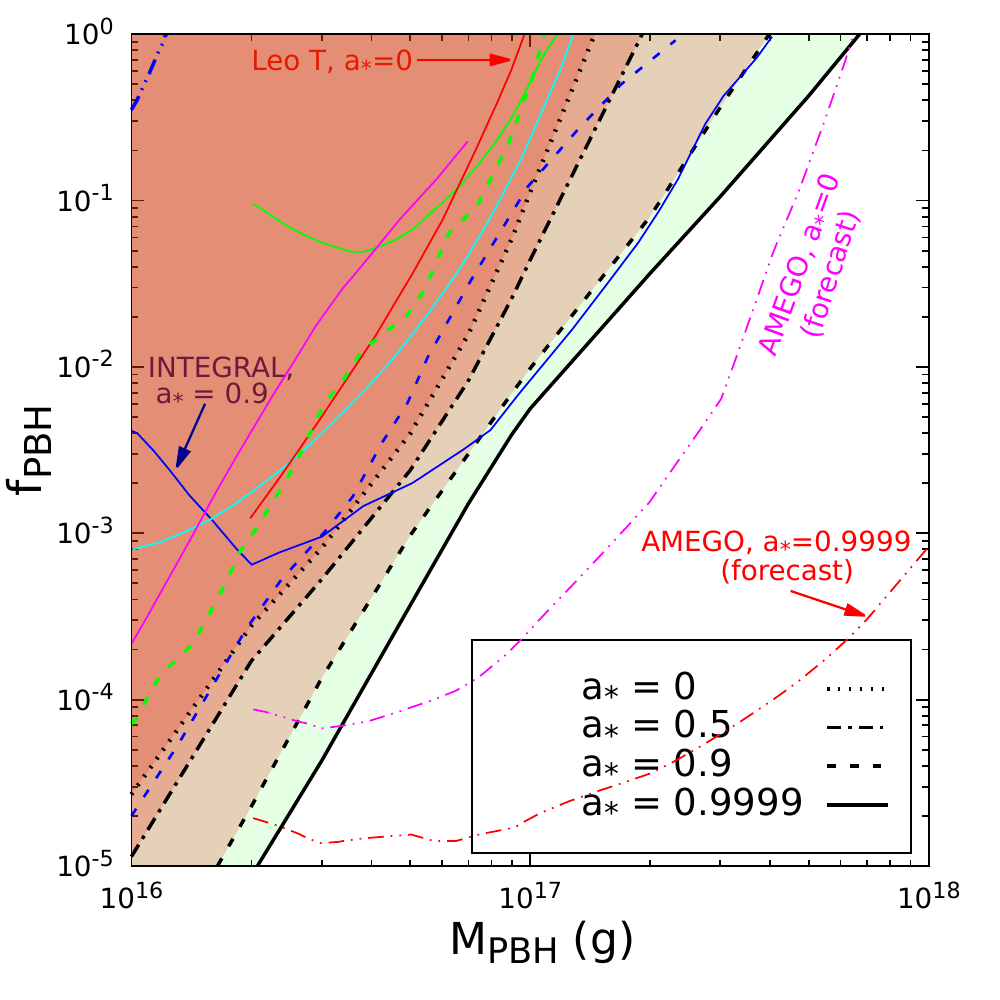}\label{plot:t2b}}  
	\end{center}
	\caption{}
\end{sidewaysfigure}
\clearpage
\begin{figure}[H]
	\captionsetup{labelformat=adja-page}
	\ContinuedFloat
	\caption{The projected upper bounds on the dark fraction of matter in the form PBHs ($f_{\rm PBH}=\Omega_{\rm PBH}/\Omega_{\rm DM}$) as a function of PBHs mass for different values of $a_*$. The shaded regions are excluded from our analysis for $f_{\rm PBH}$ when $a_*=0$ (dotted black line), 0.5 (dot-dashed black line), 0.9 (dashed black line) and 0.9999 (solid black line). The dashed blue curve depicts the upper constraint on $f_{\rm PBH}$ by observations of the diffuse Isotropic Gamma-Ray Background (IGRB) for $a_*=0.9$ \cite{Arbey::2020}. The double-dot-dashed blue curve represents the upper constraint on $f_{\rm PBH}$ from Diffuse Supernova Neutrino Background (DSNB) searches at Super-Kamiokande, while the solid blue line represents the INTErnational Gamma-Ray Astrophysical Laboratory (INTEGRAL) observation of 511~KeV $\gamma$-ray lines at Galactic centre constraint on $f_{\rm PBH}$ for $a_*=0.9$ \cite{Dasgupta:2020}.  The double-dot-dashed magenta (red) line represents the AMEGO forecast for $a_*= 0\ (a_*=0.9999)$ \cite{Ray:2021}. Near future, AMEGO collaboration will be able to probe the parameter-space above the magenta (red) double-dot-dashed curve for  $a_*= 0\ (a_*=0.9999)$. The solid green line stands for 95\% confidence level bound from INTEGRAL observation of Galactic gamma-ray flux for non-spinning PBHs \cite{Laha:2020}. Solid cyan curve depicts the upper bound from observing the 511 KeV $\gamma$-ray lines at the Galactic centre by assuming all the PBHs within a 3 Kpc radius of the Galactic centre for non-spinning PBHs \cite{Laha:2019}. The magenta solid line represents the Planck constraint \cite{Clark:2017}. The red solid line depicts the dwarf galaxy Leo T constraint \cite{Kim_2021} and the green dashed line shows the COMPTEL bound \cite{Coogan:2021} for non-spinning PBHs.}
	\label{plot:t2}
\end{figure}
\vspace{-0.31cm}
\noindent{\hrule height 0.08cm depth 0cm \relax} 
\vspace{1cm}

In Figure \eqref{plot:t2}, we plot the upper  projected bounds on the fraction of dark matter in the form of PBHs as a function of PBHs mass for different spins. Here, we have considered that 21 cm differential brightness temperature, $T_{21}$, remains $-150$~mK at redshift $z=17.2$. We vary the mass of PBHs from $10^{15}$~g to $10^{18}$~g. The shaded regions in both the plots are excluded for the corresponding PBH spins. The dashed blue curve represents the upper constraint on $f_{\rm PBH}$ by observations of the diffuse Isotropic Gamma-Ray Background (IGRB)  \cite{Arbey::2020}. The double-dot-dashed blue curve represents the upper constraint on $f_{\rm PBH}$ from Diffuse Supernova Neutrino Background (DSNB) searches at Super-Kamiokande, while the solid blue line represents the INTErnational Gamma-Ray Astrophysical Laboratory (INTEGRAL) observation of 511 KeV $\gamma$-ray line at Galactic centre constraint on $f_{\rm PBH}$ for $a_*=0.9$ \cite{Dasgupta:2020}.  For $a_*=0$, the observation at the Jiangmen Underground Neutrino Observatory (JUNO) will be able to place a 20 times stronger bound on the upper allowed value of $f_{\rm PBH}$ for $ M_{\rm PBH}=10^{15}$~g compared to Super-Kamiokande \cite{Wang:2021, Dasgupta:2020}. The double-dot-dashed magenta (red) line represents the AMEGO forecast for $a_*= 0\ (a_*=0.9999)$ \cite{Ray:2021}. In the near future, AMEGO collaboration will be able to probe the parameter-space above the magenta (red) double-dot-dashed curve for  $a_*= 0\ (a_*=0.9999)$. Solid green line stands for  95\% confidence level bound from INTEGRAL observation of Galactic $\gamma$-ray flux for non-spinning PBHs \cite{Laha:2020}. The solid cyan curve depicts the upper bound from the observation of 511 KeV $\gamma$-ray lines at the Galactic centre by assuming all the PBHs within a 3 Kpc radius of the Galactic centre for non-spinning PBHs \cite{Laha:2019}.  For the comparison, we have also plotted the bounds from Planck \cite{Clark:2017}, Leo T \cite{Kim_2021} and COMPTEL \cite{Coogan:2021} observations for non-spinning PBHs. In Figure \eqref{plot:t2a}, $f_{\rm PBH}$ varies from $1\times10^{-10}$ to $1\times10^{-5}$, while, in Figure \eqref{plot:t2b}, it varies from $1\times10^{-5}$ to its maximum allowed value 1 ($\Omega_{\rm PBH}=\Omega_{\rm DM}$). In Figure \eqref{plot:t2}, as we increase the value of spin from $0$ to its extremal value, $0.9999$, the upper bounds become more stringent. This is due to an increment in  evaporation of PBHs, and it results in more energy injection into the IGM \cite{Page::1976, Page:::1976, Page:1977}. As discussed earlier, increasing the mass of PBHs, energy injection into IGM decreases. Subsequently, one gets more window to increase the gas temperature or $f_{\rm PBH}$, and the upper bound becomes weaker. Therefore, in Figure \eqref{plot:t2}, the upper bound on $f_{\rm PBH}$ weakens as we increase the mass. Our upper projected constraint on $f_{\rm PBH}$ for $a_*=0.9$ is comparable to the INTEGRAL observation of 511~KeV $\gamma$-ray lines for PBHs mass larger than $\sim8\times10^{16}$ and becomes stronger for smaller PBH masses. Also, compared to IGRB \cite{Arbey::2020} and DSNB \cite{Dasgupta:2020}, our  projected bounds are stringent for the considered mass range of PBHs. We find the most robust lower projected constraint on the mass of PBHs, which is allowed to constitute the entire dark matter,  to $1.5\times10^{17}$~g, $1.9\times10^{17}$~g, $3.9\times10^{17}$~g and $6.7\times10^{17}$~g for PBH spins 0, 0.5, 0.9 and 0.9999, respectively. The lower bound on $M_{\rm PBH}$ for $\Omega_{\rm PBH}=\Omega_{\rm DM}$, for extremal spinning PBHs is nearly four times larger than non-spinning PBHs.


\section{Conclusions}
Spinning primordial black holes can substantially affect the ionization and thermal history of the Universe.  Subsequently, it can modify the 21 cm absorption signal in the cosmic dawn era by injecting energy due to Hawking evaporation. We study the upper  projected bounds on the fraction of dark matter in the form of PBHs as a function of mass and spin, considering that the 21 cm differential brightness temperature does not change by more than a factor of 1/4 from the theoretical prediction based on the $\Lambda$CDM framework.  Our  projected constraints are stringent compared to DSNB, INTEGRAL observation of the 511~KeV line, IGRB, Planck, Leo T and COMPTEL. In the near future, AMEGO collaboration will be able to probe some parameter space in our considered mass range of PBHs. In the present work, we have considered the monochromatic mass distribution of PBHs. The allowed parameter space can also be explored for different PBHs mass distributions such as log-normal, power-law, critical collapse, etc. \cite{Arbey_2019}.  Here, it is to be noted that we have not considered heating of IGM gas due to x-ray from the first stars in the vague of known physics of the first stars.  For a fix value of $T_{21}$ at a redshift, if one includes the x-ray heating of the gas, the projected bounds becomes stronger.


\newpage

\section{Additional study}

\begin{figure}[b!]
	\subfloat[]
	{\includegraphics[width=2.9in,height=2in]{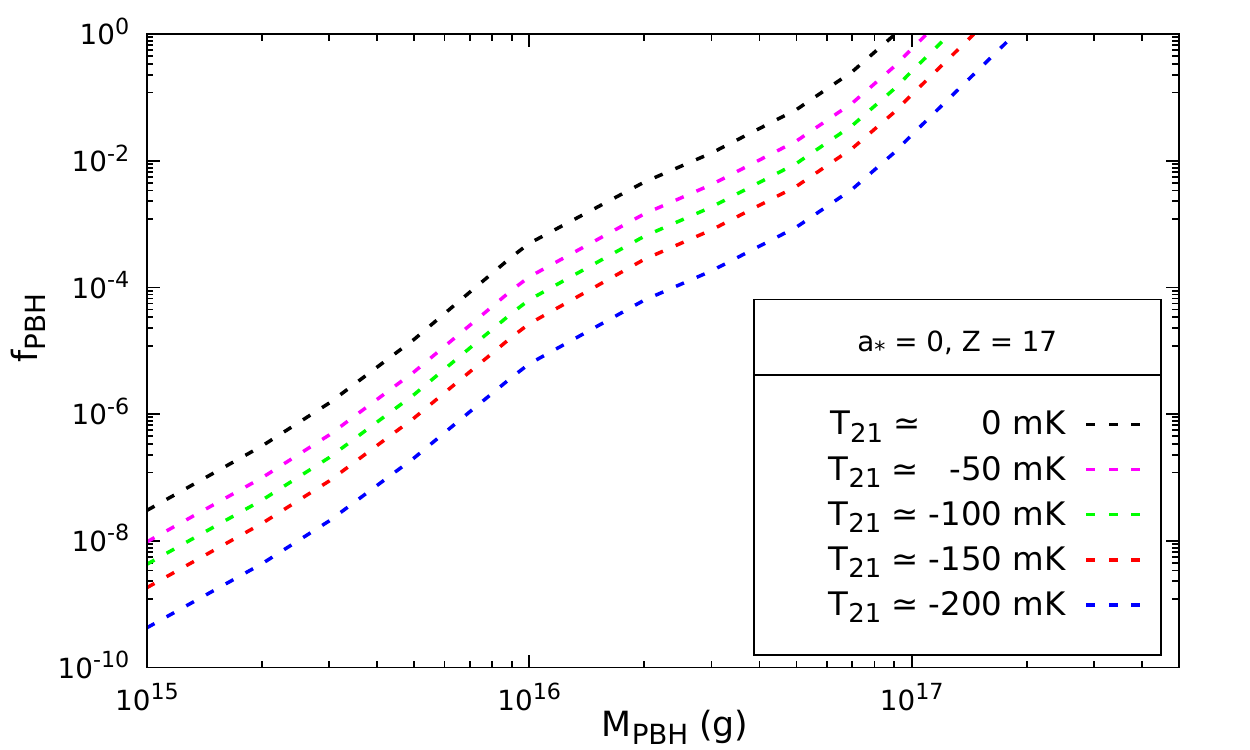}\label{ch3plot:4a}}
	\subfloat[] {\includegraphics[width=2.9in,height=2in]{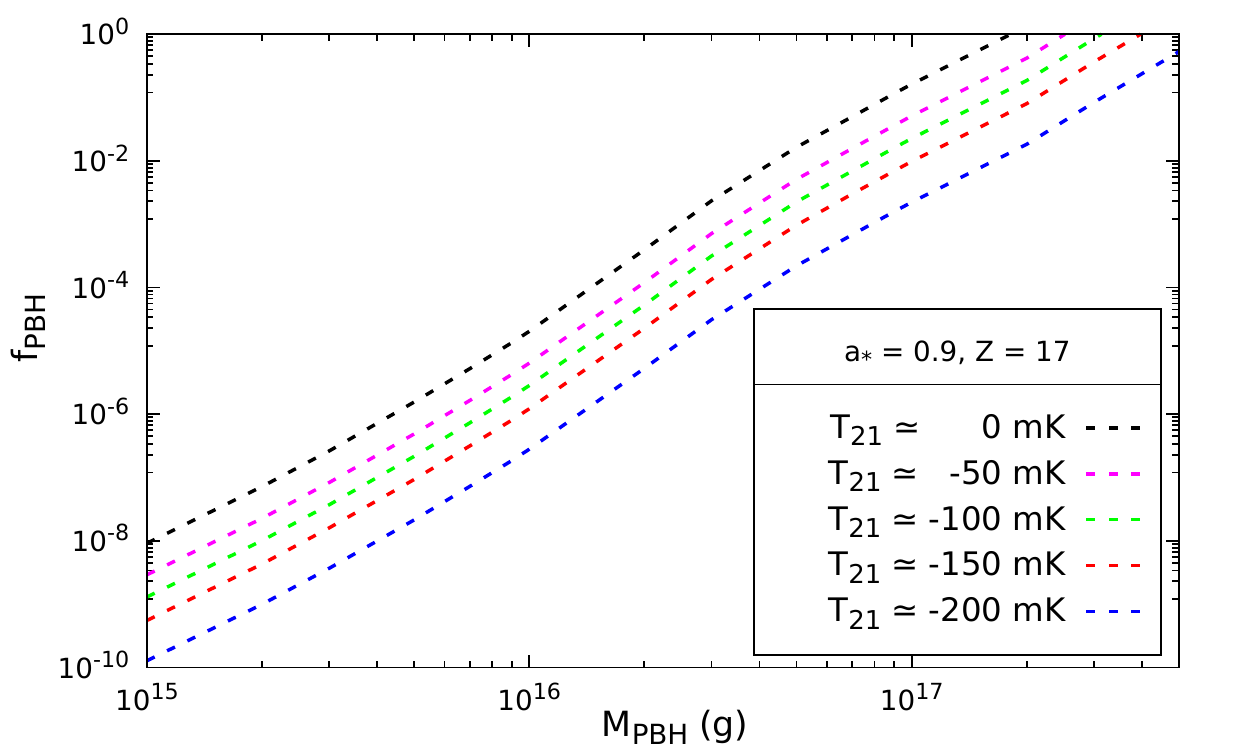}\label{ch3plot:4b}}
	\caption{ The plots represent the  upper projected bounds on the fraction of dark matter in the form of primordial black holes ($f_{\rm PBH}$) as a function of mass of PBHs ($M_{\rm PBH}$) for varying 21~cm differential brightness temperature ($T_{21}$) at $z=17$. Figure \eqref{ch3plot:4a} represents the case when spin of PBHs: $a_*=0$, while, figure \eqref{ch3plot:4b} represents the case with $a_*=0.9\,$.}
	\label{ch3plot:4}
\end{figure}

\begin{figure}[]
	\subfloat[]
	{\includegraphics[width=2.9in,height=2in]{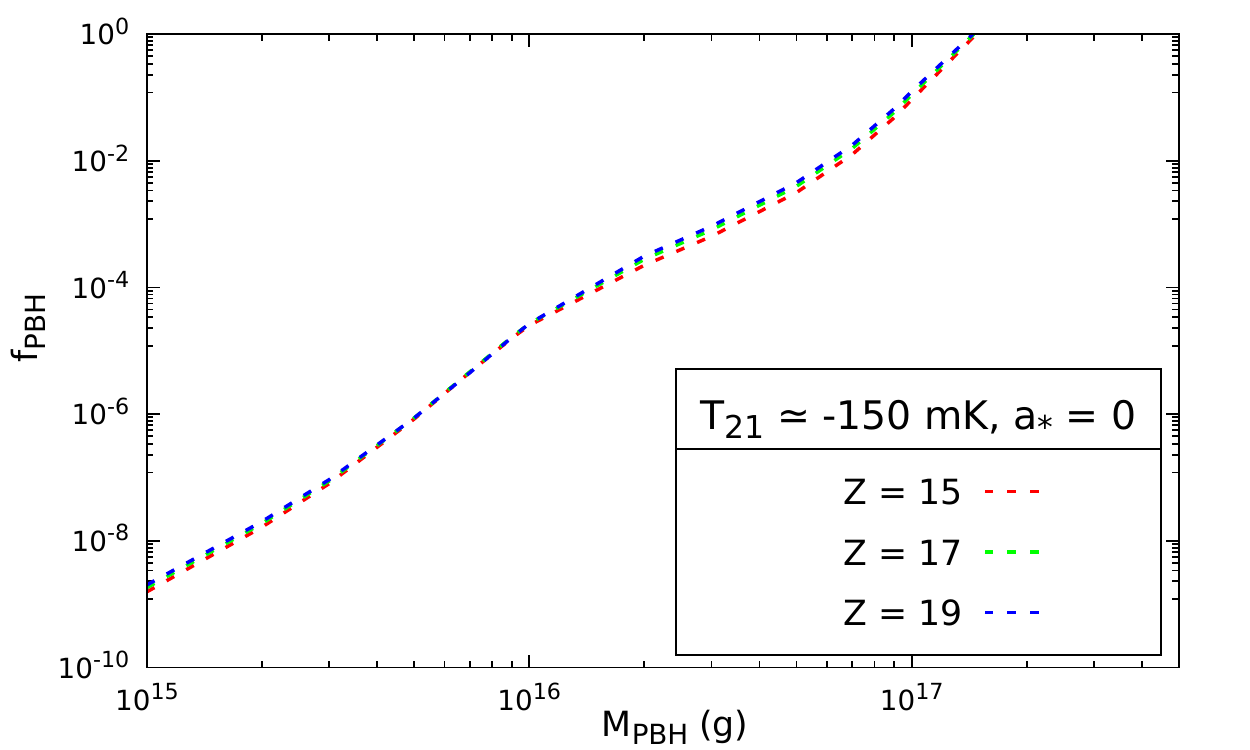}\label{ch3plot:5a}}
	\subfloat[] {\includegraphics[width=2.9in,height=2in]{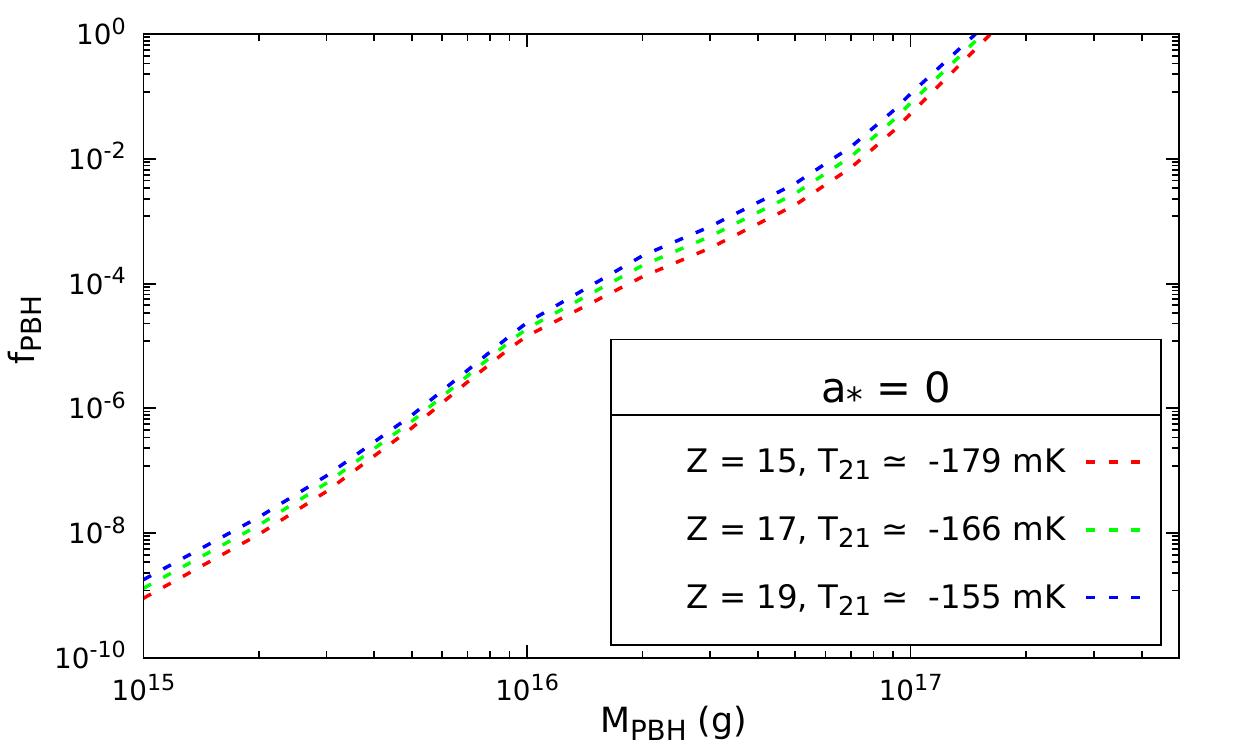}\label{ch3plot:5b}}
	\caption{ Figure \eqref{ch3plot:5a} represents upper projected bounds on $f_{\rm PBH}$ when $T_{21}\simeq-150$~mK for different values of redshift ($z$). Figure \eqref{ch3plot:5b} represents upper bounds on $f_{\rm PBH}$ when $T_{21}$ does not change more than a factor of $1/4$ from the minimum possible amplitude based on $\Lambda$CDM model for corresponding values of redshift ($T_{21}^{\rm Min}(z=15)\simeq-238$~mK, $T_{21}^{\rm Min}(z=17)\simeq-221.2$~mK and $T_{21}^{\rm Min}(z=19)\simeq-207$~mK). Here, the cosmological parameters are: $h=0.674$, $\Omega_{\rm M }=0.315$, $\Omega_{\rm b}=0.049$ \citep{Planck:2018}. Both figures obtained for $a_*=0$.}
	\label{ch3plot:5}
\end{figure}

\subsection{Bounds in light of varying $T_{21}$ and redshift}

We also study  upper projected bounds on the fraction of the dark matter in the from of PBHs by varying the amplitude of 21~cm differential brightness temperature and redshift. In figure \eqref{ch3plot:4}, we have plotted upper bounds on the fraction of dark matter in the form of primordial black holes as a function of mass for various values of $T_{21}$ at $z=17$. To understand that how bounds change on $f_{\rm PBH}$ with $T_{21}$, we consider two scenarios for the spin of PBHs: $a_*=0$ (figure \ref{ch3plot:4a}) and  $a_*=0.9$ (figure \ref{ch3plot:4b}). In both of the plots, we notice that when we change the value of $T_{21}$ from $-200$~mK to $-150$~mK the bound relaxes with a factor of $\sim 4.3\,$. By changing the $T_{21}$ from $-150$~mK to $-100$~mK bounds relax by a factor of $\sim2.3$, and, going from $-100$~mK to $-50$~mK the bounds relax by a factor of $\sim2.25$. By changing the $T_{21}$ from $-50$~mK to $0$, bounds relax by a factor of $\sim3.1\,$. This similar pattern also occurs for the case of sterile neutrinos and the factors remain same. Therefore, we can also find the constraints on $f_{\rm PBH}$ for other values of spin when the bound on $f_{\rm PBH}$ is given for any value of $T_{21}\in \{0,-50,-100,-150,-200\}$~mK.

In figure \eqref{ch3plot:5a}, similarly to sterile neutrino case, we see that bounds do not change significantly for a fix value of $T_{21}$ at different values of redshift. In figure  \eqref{ch3plot:5b}, upper bounds on $f_{\rm PBH}$ are obtained such that  $T_{21}$ does not change more than a factor of $1/4$ from the minimum possible amplitude based on $\Lambda$CDM model for corresponding values of redshift. In this chapter,  we consider the following numerical values of the cosmological parameters: $h=0.674$, $\Omega_{\rm M }=0.315$, $\Omega_{\rm b}=0.049$ and $T_{\rm CMB}|_{z=0}=2.725$~K \citep{Planck:2018,Fixsen_2009}. Therefore, we get minimum possible value of $T_{21}$ based on $\Lambda$CDM model to $-238$~mK at $z=15$, $-221.2$~mK at $z=17$ and $-207$~mK at $z=19$.

\clearpage
\pagestyle{empty}
\cleardoublepage
\pagestyle{fancy}
\begin{savequote}[75mm]
	``Astronomy and Pure Mathematics are the magnetic poles toward which the compass of my mind ever turns."
	\qauthor{Carl Friedrich Gauss, \textit{In Letter to Bolyai (30 Jun 1803)}}
\end{savequote}

\chapter[PMFs \& Excess Radio Background]{Primordial Magnetic Fields and Excess Radio Background}\label{chap4}
\vspace{-1.5cm}
%
%
%

Observations suggest that the magnetic fields are ubiquitous in the Universe--- from the length scale of planets and stars to the cluster of galaxies \cite{Haverkorn:2008, Fletcher:2011, Carilli:2002, Brandenburg20051}. Fermi\footnote{\href{https://fermi.gsfc.nasa.gov/}{https://fermi.gsfc.nasa.gov/}} and High Energy Stereoscopic System (HESS)\footnote{\href{https://www.mpi-hd.mpg.de/hfm/HESS/}{https://www.mpi-hd.mpg.de/hfm/HESS/}} gamma-ray observation suggests that even voids could host magnetic fields with strength $\mathcal{O}(10^{-16}~{\rm G})$ with a typical coherent scale of Mpc \cite{Neronov:2010, Vovk:2012}. Magnetic fields can also play a significant role in reionization, relic electron density and structure formation \cite{Sethi:2004pe}. The presence of magnetic fields can substantially affect the evolution and dynamics of structures in the Universe as they can contribute to the total pressure against gravitational collapse. This could modify the total matter power spectrum on small scales, $\lesssim 1 $~Mpc \cite{Kim:1994zh, Beck:1996, Sethi:2004pe, Tashiro:2006uv, Beck:2015, Back:2015R}. The presence of magnetic fields during recombination can also have important consequences, such as it could have lead to the collapse of gas clouds after recombination, formation of first pre-galactic stars, quasars \cite{Hogan:1983cj}. The Earth has a magnetic field of the order of $\mathcal{O}({\rm G})$, and it is sustained for years by some dynamo mechanism. Similarly, other astronomical objects near to Earth, such as Sun and other solar system planets, also show the presence of magnetic fields \cite{Weiss:2002}. Our home galaxy Milky Way, other spiral galaxies and their interstellar medium (ISM) contain magnetic fields with the strength $\mathcal{O}(\mu{\rm G})$ \cite{Beck:1996, Beck:2000, Weiss:2002, Back:2015R, Beck:2020}. Moreover, galaxy clusters, intercluster medium, filaments, IGM, etc., also show the magnetic fields \cite{Carilli:2002, Vogt:2005, Neronov:2010, Taylor:2011, Vernstrom:2021}. These magnetic fields are likely to be seeded by primordial magnetic fields (PMFs). These PMFs might have originated in the very early Universe, and subsequently amplified in the small scale structures by some mechanisms \cite{Neronov:2010, Subramanian:2019}.

\section{Generation of primordial magnetic fields}

The origin and evolution of PMFs is one of the outstanding problems of modern cosmology (Ref. \cite{Subramanian:2016, Subramanian:2019} and references cited therein). It would be very difficult to explain the magnetic fields in the voids and high redshift galaxies with only late-time astrophysical processes without magnetic fields from the very early Universe. Therefore, these magnetic fields indeed may have a primordial origin \cite{Grasso:2000wj, Neronov:2010, Subramanian:2015lua, Bertone:2006}.  There are several theoretical models that can generate the magnetic field in the early Universe with a large coherent scale. The two scenarios to generate PMFs that are vastly discussed in the literature are phase transitions in the early Universe and various models of inflation (for details, see the recent review \cite{Subramanian:2019}). In the Ref. \cite{turner:1988mw}, the authors discuss how the inflation model can generate large scale, $\sim\mathcal{O}({\rm Mpc})$, magnetic fields. The generated magnetic fields have a small strength. To amplify the field, one has to break the conformal invariance of the electromagnetic field. The authors consider three mechanisms to break the conformal invariance: Coupling of the photon to the axions, gravitational field and massless-charged-nonconformally invariant scalar field \cite{turner:1988mw}. Authors of the Ref. \cite{Ratra:1992}, extend the inflation model by introducing the coupling between the Maxwell field and the scalar field ($\Phi$) responsible for inflation ($\propto e^{\alpha\Phi} F_{\mu\nu} F^{\mu\nu}$), here, $F_{\mu\nu}$ is the electromagnetic field tensor. This scenario can generate magnetic fields with a present-day strength up to nG with the coherence scale of a few Mpc depending on the parameter $\alpha$. A similar mechanism to generate the magnetic fields during inflation is based on the superstring cosmology \cite{Lemoine:1995, Gasperini:1995vg}. The Lagrangian, similar to considered by \cite{Ratra:1992} with $\alpha=-1$, naturally arises from the effective action in low-energy string theory. Here, inflation is driven by the kinetic part of the dilaton scalar field--- $\Phi'$. Whereas in the article \cite{Ratra:1992}, it is driven by the false vacuum scalar field potential---  which is too steep for producing the slow-roll inflation \cite{Lemoine:1995, Grasso:2000wj}. In the article \cite{Demozzi:2009}, authors argue that the back reaction of generated magnetic fields via inflation can spoil the inflation. Considering the backreaction, the authors put an upper bound on the present-day strength of magnetic fields to $10^{-32}$~G on the Mpc scale. This strength seems too small for galactic dynamos to amplify to explain the observed magnetic fields. In the recent article \cite{Talebian:2020}, it is shown that this issue can be circumvented for some parameter space. The authors find that magnetic fields with a present-day strength of $\sim10^{-13}$~G with a scale of Mpc can be generated while keeping the backreactions under control. Magnetic fields can also arise during electroweak \cite{Baym:1996} and quantum-chromo-dynamics \cite{Quashnock:1989sl} phase transitions. Other mechanisms include cosmic strings \cite{Hill:1988, Vachaspati:1991V}, primordial plasma vorticity \cite{Harrison:1970}, etc. In this chapter, we obtain the upper bounds on present-day strength of PMFs for various values of spectral index in the light of EDGES\footnote{Recently, the EDGES signal has been questioned in many articles. We discuss this point in chapter \eqref{chap6}.} observation and excess radio background observed by the ARCADE 2 \& LWA 1 observation \cite{Natwariya:2021}. Here, we obtain the bounds on PMFs in both the presence and absence of heating effects due to first stars.



\section{Existing bounds on primordial magnetic fields}

The present-day strength, spectral index and coherence scale of PMFs depends on their generation mechanisms. Therefore, the constraints on PMFs can give a hint of the early Universe physics. Recently in the Ref. \cite{Jedamzik:2020L}, authors show that PMFs can be used as a remedy to resolve the Hubble tension between different observations. The present-day amplitude of PMFs is constrained from the BBN, formation of structures and temperature anisotropies \& polarization of CMB \cite{ Trivedi:2012ssp, Trivedi:2013wqa, Sethi:2004pe}. Authors of the Ref. \cite{Minoda:2018gxj}, put an upper constraint to $\sim 10^{-10}$~G on 1~Mpc scale by considering $T_{\rm gas}\lesssim T_{\rm CMB}$ (i.e. $T_{21}\lesssim0$) so that, PMFs can not erase the $T_{21}$ absorptional signal in the redshift range $15\lesssim z\lesssim20$. Planck 2015 results put individual upper constraints of the $\mathcal{O}({\rm nG})$ for different cosmological scenarios on 1~Mpc scale \cite{Planck:2016}. The authors of the Ref. \cite{Natwariya:2020}, in the context of EDGES observation, put an upper and lower constraint on the PMFs to be $6\times 10^{-3}~{\rm nG} $ and $5\times 10^{-4}~{\rm nG}$ respectively.  Also, the lower bound on the present-day strength of PMFs found in Refs. \cite{Ellis:2019MMVA, Fermi_LAT:2018AB, Tavecchio:2010GFB}. Further, in the Ref. \cite{Neronov:2010}, authors put a lower bound on the strength of intergalactic magnetic fields of the order of $3\times10^{-16}$~G using Fermi observations of TeV blazars. Authors of the reference \cite{Cheng:1996vn}, report upper bound of $2\times10^9$~G at the end of BBN. Presence of PMFs can modify the present-day relic abundance of He$^4$ and other light elements. Therefore, magnetic fields can be constrained by observations of light element abundances \cite{Matese:1969cj, Greenstein:1969, Tashiro:2005ua, Sethi:2004pe, Choudhury:2015P}. The authors of the Ref. \cite{Jedamzik:2019}, put an upper bound of $47$~pG for scale-invariant PMFs by comparing CMB anisotropies, reported by the Wilkinson Microwave Anisotropy Probe (WMAP) and Planck, with calculated  CMB  anisotropies. 
%
%
%
%
%

\section{Evolution of PMFs after recombination}

The generation of the magnetic fields in the early Universe for the various cosmological scenarios has been studied in the earlier literature (for example see Refs. \cite{Quashnock:1989sl, Grasso:2000wj, Subramanian:2010, Pandey:2020, Ellis:2019MMVA}). It is to be noted that decaying magnetic fields has been studied in several literatures. In these works, the authors consider the decay of the PMFs by ambipolar diffusion and turbulent decay  \cite{Sethi:2004pe, Chluba2015, Bhatt2019pac, Minoda:2018gxj, Bera:2020}. Ambipolar diffusion of magnetic fields is important in neutral medium as it is inversely proportional to free-electron fraction ($x_e$) and $x_e\sim 10^{-4}$ after redshift $z\lesssim100$ \cite{Chluba2015, Peebles:1968ja, Sethi:2004pe}. The presence of PMFs can induce the Lorentz force in the gas. The force exerts only on free electrons and ions leaving the neutral components unaffected. This can result in creating a velocity difference between charged and neutral components. The velocity difference can enhance the collision frequency in the gas, resulting in a dissipation of magnetic energy into the gas--- known as the ambipolar diffusion of magnetic fields \cite{Shu:1992fh}. After the recombination ($z\sim1100$), the radiative viscosity of fluid dramatically decreases, and velocity perturbations are no longer damped. Therefore, the tangled magnetic fields having length scale smaller than the magnetic Jeans length can dissipate via another mode--- turbulent decay \cite{Sethi:2004pe, Chluba2015, Schleicher:2008aa}. Magnetic heating of the gas due to the turbulent decay decreases with redshift but later when ionization fraction decreases, heating increases due to ambipolar diffusion \cite{Chluba2015, Sethi:2004pe}. We further discuss about the ambipolar and turbulent decay in section \eqref{PMFI}. Decaying PMFs can inject magnetic energy into the thermal energy of the IGM and heat the gas above $6.7$~K at $z=17$, and even it can erase the EDGES absorption signal \cite{Minoda:2018gxj, Sethi:2004pe, Chluba2015}. Still, the EDGES absorption signal can be explained by considering the possible early excess of radio radiation \cite{Feng2018}.

\section{Background excess radio radiation}

The Absolute Radiometer for Cosmology, Astrophysics and Diffuse Emission (ARCADE 2) collaboration\footnote{\href{https://asd.gsfc.nasa.gov/archive/arcade/}{https://asd.gsfc.nasa.gov/archive/arcade/}}, a double-nulled balloon-borne instrument with seven radiometers, measured the absolute sky temperature in a frequency range of  $3-90$~GHz. The observation reported excess radio radiation in a frequency range of $3-10$~GHz \cite{Fixsen2011},
\begin{alignat}{2}
T(\nu)=T_0+T_r\,(\nu/\nu_0)^\beta\,,\label{radex}
\end{alignat}
here, ARCADE 2 observation fitted the parameters as: $T_0=2.731 \pm 0.004$~K, $\beta=-2.6$, $T_r=21.1 \pm 3.0$~K and $\nu_0=310$~MHz. By combining ARCADE 2 with the Low-frequency data \cite{Roger:1999, Maeda:1999, Haslam:1981, Reich:1986} and Far Infrared Absolute Spectrophotometer (FIRAS) data \cite{Fixsen_2002}, the parameters can be fitted as: $T_0=2.725 \pm 0.001$~K, $\beta=-2.599 \pm 0.036$, $T_r=24.1 \pm 2.1$~K and $\nu_0=310$~MHz in a frequency range of $22$~MHz$-10$~GHz \cite{Fixsen2011}. This is measured at present-day ($z=0$). The radiation temperature maps with redshift as: $\propto(1+z)$, we can multiply $T(\nu)$ with ($1+z$) for past \cite{Feng2018, Fialkov:2019, Reis:2020, Yang:2018, Mondal:2020,banet:2020},
\begin{alignat}{2}
T(z)=T_0\,(1+z)\,\left[\ 1+\frac{T_r}{T_0}\ \left(\frac{78}{310}\right)^\beta\times\,\left(\frac{\nu}{78~{\rm MHz}}\right)^\beta\ \right]\,.\label{ARC}
\end{alignat}
This radiation is several times larger than the observed radio counts due to the known Galactic and extragalactic radio processes and sources, such as star-forming galaxies, AGN-driven sources--- quasars and radio galaxies, etc. \cite{Singal_2018, Singal_2022}. The presence of early excess radiation can not be completely ruled out at the time of cosmic dawn. For example, in the redshift range $z\approx30$ to 16, accretion onto the first intermediate-mass black holes can produce a radio radiation \citep{Ewall-Wice2018}. Accreting supermassive black holes  \cite{Biermann:2014} or supernovae \cite{Jana:2018} can also produce radio background due to synchrotron emission at the time of cosmic down by accelerated electrons in the presence of the magnetic field. The enhancement in the background radiation is also supported by the first station of the Long Wavelength Array (LWA  1)\footnote{\href{https://leo.phys.unm.edu/~lwa/} {https://leo.phys.unm.edu/~lwa/}} in frequency range $40-80$~MHz. The excess observed by LWA 1 can also be fitted by the same model given by equation \eqref{radex}. After the inclusion of LWA 1 data with ARCADE 2 \cite{Fixsen2011} and Low-frequency data \cite{Roger:1999, Maeda:1999, Haslam:1981, Reich:1986}, the parameters change as: $T_0=2.722\pm0.022$~K, $\beta=-2.58\pm0.05$ and $T_r=30.4\pm2.6$~K at $\nu_0=310$~MHz \cite{Dowell2018, Dowell:2017}. For the observation of 21-cm signal, we can write: $\nu=1420.4/(1+z)$~MHz. In the equation \eqref{ARC}, the factor of $(\nu/78~{\rm MHz})^\beta$ can be defined as a fraction of excess radio background, $A_r$. Depending on the origin, $A_r$ can have different values--- we discuss about this more in next sections. Therefore, in the final analysis, we vary the value of excess radiation fraction.

\subsection{Excess radiation during the cosmic dawn}\label{GMFER}
In this work, we use the EDGES signal in the presence of excess radio radiation to constrain the strength of PMFs.  Some of the processes which we have discussed responsible for the excess radio background can occur at earlier redshift ($z\sim 17$) \cite{Ewall-Wice2018, Biermann:2014, Jana:2018}. Also, one of the interesting proposals in the Ref. \cite{Feng2018} is to argue that such a possibility can exist at the time of cosmic dawn, and it can help to explain the EDGES signal. Here, authors show that the EDGES absorption signal can be explained by having only 10 percent of the observed radio background by ARCADE 2. In  Ref. \cite{Lawson:2019, Lawson:2013}, the authors claim that thermal emission from the axion quark nugget dark matter model can explain the EDGES signal, and it can also contribute a fraction of the radiation excess observed by ARCADE 2. At present, there exist several theoretical models to explain this excess at the time of cosmic dawn.  The stimulated emission from Bose stars can give a large contribution to the radio background and explain the EDGES and ARCADE 2 observations \cite{Levkov:2020}. 
The radio emission from accreting Pop III black holes can produce the EDGES like signal by increasing background radiation temperature \cite{Mebane:2020}. In other scenarios,  the  EDGES anomaly can be explained by axion-photon conversion in the presence of intergalactic magnetic fields \cite{Moroi2018} or by radiative decays of standard model neutrino induced by magnetic fields \cite{AristizabalSierra2018}. Radio excess can also be explained by the cusp region of superconducting cosmic strings \cite{Brandenberger:2019}. In ref. \cite{Chianese:2019}, authors consider radiative decays of relic neutrino and show that it can potentially explain the ARCADE 2 excess together with the EDGES signal. Depending on the origin, the excess fraction of radio radiation can have a different value. We discuss the constraints on excess radiation later.  Considering the above possibilities of having early excess radiation, we believe that it is important to analyze constraints on the primordial magnetic field in the presence of such radiation.




\subsection{Phenomenological model for excess radiation}

As discussed in subsection \eqref{GMFER}, the possibility of an excess radio radiation background over the CMBR can not be denied. For the excess radio background, we consider the phenomenological model following the Ref. \cite{Fialkov:2019, Reis:2020, Yang:2018, Mondal:2020,banet:2020}. Here, Authors consider a uniform redshift-independent synchrotron-like radiation, motivated by the ARCADE 2 and LWA 1 observations. This model can explain the EDGES anomaly in addition to enhancement of cosmic down power spectrum.  Accordingly, from equation \eqref{ARC} and following the Refs. \cite{Fialkov:2019, Reis:2020, Yang:2018, Mondal:2020,banet:2020},
\begin{alignat}{2}
T_R=T_0\,(1+z)\,\left[1+A_r\,\left(\frac{\nu_{\rm obs}}{78~{\rm MHz}}\right)^\beta\ \right] \,,\label{eq3}
\end{alignat}
where, $T_0 =2.725$~K is the present day CMB temperature and $\beta=-2.6$ is the spectral index. Here, $A_r$ is the amplitude defined relative to the CMB at reference frequency of 78~MHz. For the 21~cm signal $\nu_{\rm obs}$ is  $1420.4/(1+z)$~MHz. Authors of the Ref. \cite{Fialkov:2019}, put a limit on the excess radiation background to $1.9<A_r<418$ at reference frequency of 78~MHz by considering the effect of an uniform radiation excess on the 21~cm signal from the cosmic dawn, dark ages and reionization. Authors consider a synchrotron-like spectrum with spectral index $-2.6\,$. The case with $A_r\sim418$ corresponds to the LWA 1 limit on $A_r$ at the reference frequency of 78~MHz \cite{Dowell2018, Fialkov:2019}. The stringent constraint on excess radiation comes from the Low-Frequency Array (LOFAR) to  $A_r<182$ (95 percent CL) and $A_r<259$ (99 percent CL) for a spectral index of $-2.6\,$ \cite{Mondal:2020}.


\section{Impact on the thermal and ionization history due to primordial magnetic fields}\label{PMFI}

In the presence of decaying magnetic fields, the gas temperature can increase. $T_{\rm gas}$ can even increase above the background radiation and can erase the 21 cm absorption signal  reported by EDGES \cite{Sethi:2004pe, Schleicher:2008aa, Chluba2015, Minoda:2018gxj}. Therefore, present-day PMFs strength can be constrained by the EDGES observation in the presence of excess radiation reported by ARCADE 2 and LWA 1 \cite{Bowman:2018yin, Feng2018, Fixsen2011, Kogut:2011, Dowell2018, Fialkov:2019}. In the presence of turbulent decay and ambipolar diffusion, the thermal evolution of the gas with the redshift can be written as \cite{Shu:1992fh, Sethi:2008eq, Schleicher:2008aa,Sethi:2004pe, Chluba2015},
\begin{alignat}{2}
\frac{dT_{\rm gas}}{dz}=2\,\frac{T_{\rm gas}}{1+z}&+\frac{\Gamma_c}{(1+z)\,H}(T_{\rm gas}-T_{\rm CMB})\nonumber\\
&-\frac{2}{3\,n_{\rm tot}(1+z)\,H}(\Gamma_{\rm turb}+\Gamma_{\rm ambi})\,,\label{eq6}
\end{alignat}
Here, $f_{\rm He}=0.079$ and $T_{\rm CMB}=T_0\,(1+z)$ is the cosmic microwave background (CMB) temperature. At early times, $T_{\rm gas}$ remains in equilibrium with CMB temperature due to Compton scattering. However, the gas temperature will not be strongly affected by the comparatively small amount of energy in the non-thermal radio radiation. Therefore, $T_{\rm gas}$ and $T_\alpha$ can be assumed independent of the excess radiation \cite{Feng2018}. The change in the free electron fraction ($x_e$) with redshift is given by equation \eqref{s4} with $\mathcal{E}=0\,$. Heating rate per unit volume due to the ambipolar diffusion ($\Gamma_{\rm ambi}$) and turbulence decay ($\Gamma_{\rm turb}$) is given by \cite{Sethi:2004pe, Chluba2015},
\begin{alignat}{2}
&\Gamma_{\rm ambi}=\frac{(1-x_e)}{\gamma\, x_e\, (M_{\rm H}\,N_b)^2}\ \frac{|(\bm \nabla\times\bm B)\times\bm B|^2}{16\,\pi^2}\,,\label{eq9}\\
&\Gamma_{\rm turb}=\frac{1.5\ m\ \left[\ln(1+t_i/t_d)\right]^m}{\left[\ln(1+t_i/t_d)+1.5\ln\{(1+z_i)/(1+z)\}\right]^{m+1}}H\,E_B\,,\label{eq10}
\end{alignat}
here, $m=2(n_B+3)/(n_B+5)$,  $z_i=1088$ is the redshift when heating starts due the magnetic fields (recombination epoch), $\gamma=1.9\times 10^{14}\,(T_{\rm gas}/{\rm K})^{0.375}{\rm cm}^3/{\rm g}/{\rm s}$ is the coupling coefficient, $M_{\rm H}$ is the mass of hydrogen atom and $N_b$ is the number density of baryons. $t_d=1/\big(k_d\,V_A(k_d,z)\big)$ is the decay time for the turbulence.  For matter dominated era, $t_i=2/\big(3\,H(z_i)\big)$ and $V_A(k_d,z)=B(k_d,z)/\big(4\,\pi\,\rho_b(z)\,\big)^{1/2}\,$ is the Alfv\'{e}n wave velocity. $B(k_d,z)$ is the magnetic field strength smoothed over the scale of $k_d$ at redshift $z$. $k_d$ is constrained by the damping wavenumber of Alfv\'{e}n wave. PMFs with wavenumber ($k$) larger than $k_d$, are strongly damped by the radiative-viscosity \cite{Sethi:2004pe, Schleicher:2008aa, Jedamzik1998, Kunze_2014, Subramanian:1997gi, Mack:2002}. Moreover, $E_B=B^2/(8\pi)$ is the magnetic field energy density,
\begin{equation}
\frac{dE_B}{dz}=4\,\frac{E_B}{1+z}+\frac{1}{H\ (1+z)}\ (\,\Gamma_{\rm turb}+\Gamma_{\rm ambi}\,)\,.\label{11}
\end{equation}
Here, we assume that PMFs are isotropic and homogeneous Gaussian random magnetic field, whose power spectrum is given by the following equation \cite{Landau:1987, Sethi:2004pe, Tashiro:2006uv, Minoda:2018gxj}
\begin{equation}
\langle \tilde{{\bf B}}_i ({\bf k}) \, \tilde{{\bf B}}^*_j ({\bf q})\rangle = \frac{(2\pi)^3}{2}\delta_D^3({\bf k}-{\bf q})\left(\delta_{ij}-\frac{k_i k_j}{k^2}\right)P_B(k)\,,
\end{equation}
here, $P_B(k)$ is the magnetic power spectrum, $k=|{\bf k}|$ is the comoving wave number and $\delta_D$ is the Dirac delta function. Here, we consider a power-law spectrum of the magnetic fields in the Fourier space for $k< k_d$ \cite{Minoda:2018gxj},
\begin{alignat}{2}
P_B(k)=\frac{(2\pi)^2}{\Gamma\big[(n_B+3)/2\big]}\ B_0^2\ \left(\frac{k}{\rm Mpc^{-1}}\right)^{n_B}~{\rm Mpc^3}\label{PB}\,.
\end{alignat} 
Here, $n_B$ is the spectral index. In particular, $n_B=2$ for white noise \cite{Hogan:1983cj}, $n_B=4$ for the Batchelor spectrum \cite{Durrer:2003ja} and $n_B=-2.9$ for nearly scale invariant spectrum \cite{Sethi:2004pe}. As discussed above, magnetic fields are strongly damped by the large radiative-viscosity for wavenumber larger than $k_d$ before recombination. Therefore, we consider a sharp cut-off for power spectrum of PMFs: $P_B(k)= 0$ for $k\ge k_d$ \cite{Minoda:2018gxj}. Following the Ref. \cite{Minoda:2018gxj}, we take the time evolution of the Alfv\'{e}n wave damping scale: $k_d(z)=k_{d,\rm i}\,f(z)$ and $f(z_i)=1$. Here, $k_{d,\rm i}$ is the damping wavenumber at recombination epoch,
\begin{alignat}{2}
k_{d,\rm i}=2\pi~{\rm Mpc^{-1}}\, \Bigg[1.32\times 10^{-3} \left(\frac{B_0}{\rm nG}\right)^2\,\left(\frac{0.02}{\Omega_{\rm b}h^2}\right)\,\left(\frac{\Omega_{\rm m}h^2}{0.15}\right)^{1/2}\Bigg]^{-\frac{1}{n_B+5}}\label{kdi} \,.
\end{alignat} 
Here, to smooth the  magnetic field amplitude over the inverse length scale of $k_{d,\rm i}\,$, we choose the Gaussian window function in Fourier space ($k$) as \cite{Caprini:2004,Minoda:2018gxj,Planck:2016},
\begin{alignat}{2}
B_{k_{d,\rm i}}^2=\int_{0}^{\infty}\frac{d^3k}{(2\pi)^3}\ {\rm e}^{-k^2\big(\frac{2\pi}{k_{d,\rm i}}\big)^2}\,P_B(k)=B_0^2\left[\frac{k_{d,\rm i}}{2\pi~{\rm Mpc^{-1}}}\right]^{n_B+3}\label{Bkd}\,.
\end{alignat}
The magnetic field strength smoothed over the scale of 1~Mpc, $$B_{1~\rm Mpc}^2=\int\, (dk/2\pi)^3\ \exp[-(k/{\rm Mpc^{-1}})^2]\ P_B(k)=B_0^2\,.$$ Lorentz force and the magnetic energy density can be calculated as \cite{Minoda:2018gxj},
\begin{equation}
|(\bm \nabla\times\bm B)\times\bm B|^2=\int_{k_1,k_2}k_1^2\ P_B(k_1)\ P_B(k_2)\ f^{2n_B+8}(z)\ (1+z)^{10}\,,\label{LF}
\end{equation}
here $\int_{k_1,k_2}[\cdots]=\int \int d^3k_1/(2\pi)^3 \times d^3k_2/(2\pi)^3\,[\cdots]$, and
\begin{equation}
E_B=\frac{1}{8\pi}\,\int\frac{d^3k}{(2\pi)^3}\  P_B(k)\ f^{n_B+3}(z)\ (1+z)^{4}\,.\label{EB}
\end{equation}
We can get the redshift evolution of the function $f(z)$, by substituting equation \eqref{EB} in equation \eqref{11}.
\section{Impact on the thermal and ionization history due to background radiation}\label{Heat_result} 
Heating of IGM due to background radio radiation during cosmic dawn era has been discussed in chapter \eqref{chap2}. After inclusion of heating due to excess radio radiation and x-ray,  the equation \eqref{eq6}  will modify,
\begin{alignat}{2}
\frac{dT_{\rm gas}}{dz}=\frac{dT_{\rm gas}}{dz}\Bigg|_{[{\rm eq. \eqref{eq6}}]}+\frac{dT_{\rm gas}}{dz}\Bigg|_{\rm x-ray}-\frac{\Gamma_{R}}{(1+z)\,(1+f_{He}+x_e)}\,,\label{eq12}
\end{alignat}
where, ${dT_{\rm gas}}/{dz}\big|_{[{\rm eq. \eqref{eq6}}]}$ stands for the gas temperature evolution represented in equation \eqref{eq6}. To include the x-ray heating of the gas, we consider the $tanh$ parameterization \cite{Kovetz2018,Mirocha:2015G,Harker:2015M}. In the presence of x-ray radiation, the ionization fraction evolution will also change.  For the present case, we consider the fiducial model, for x-ray heating and ionization fraction evolution, motivated by Ref.  \cite{Kovetz2018}. The heating effects of both the VDKZ18 (the last term in equation \ref{eq12}--- discussed in chapter \ref{chap2}) and x-ray are shown in plots (\ref{p_1a}, \ref{p_1b}, \ref{p_2a}, \ref{p_3b} \& \ref{p_4b}).


\section{Result and discussion}\label{S_result}
We consider the following values for the cosmological parameters: $\Omega_{\rm m }=0.31$, $\Omega_{\rm b}=0.048$, $h=0.68$, $\sigma_8=0.82$ and $n_s=0.97$ \cite{Planck:2018}. To study the gas temperature evolution with redshift in the presence of primordial magnetic field dissipation, we solve the coupled equations (\ref{s4} with $\mathcal{E}=0$), \eqref{eq6} and \eqref{11}. To get the Lorentz force term in equation \eqref{eq9}, we solve the equation  \eqref{LF}. Similarly, to get the magnetic field energy density in equation \eqref{eq10}, we solve the equation \eqref{EB}. To get the evolution of the $f(z)$ with redshift, $df(z)/dz$, we substitute equation \eqref{EB} in equation \eqref{11} with initial condition $f(z_i)=1\,$. To obtain upper constraint on PMFs strength, we solve the equation \eqref{t21f} with equations \eqref{eq6}, (\ref{s4} with $\mathcal{E}=0$) and \eqref{11} for $T_{21}\simeq-300$~mK or $-500$~mK by varying $B_0$, $n_B$ and $A_r$. For infinite Ly$\alpha$ coupling $T_S\simeq T_{\rm gas}$, therefore, $T_S$ solely depends on the gas temperature. While, for finite Ly$\alpha$ coupling, $T_S$ depends on both the gas and background radiation temperature.   

\begin{figure}
    \begin{center}
        {\includegraphics[width=4.5in,height=3in]{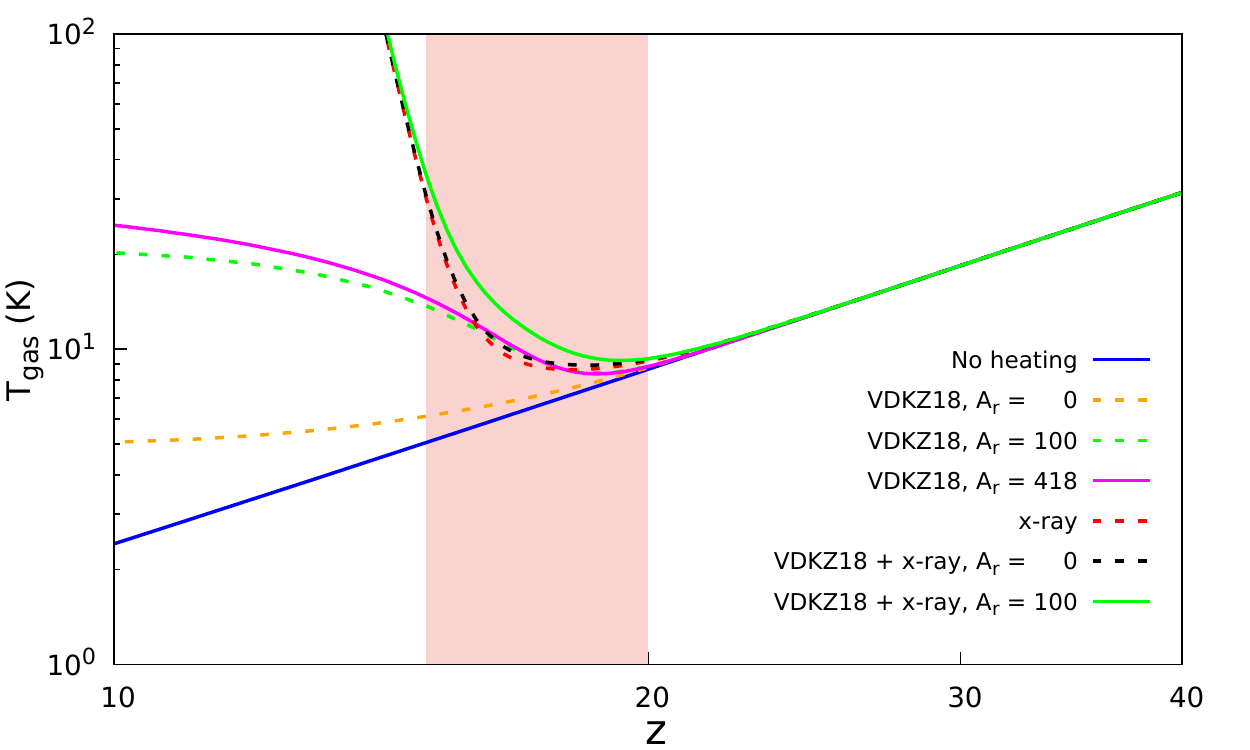}} 
    \end{center}
    \caption{The gas temperature evolution with redshift. The solid blue lines represent the case when there is no x-ray, VDKZ18 or magnetic heating. VDKZ18 corresponds to the heat transfer from the background radiation to gas mediated by Ly$\alpha$. The shaded region represents the EDGES observation redshift range, $15\leq z \leq 20$ . In this figure, we consider only VDKZ18 and x-ray heating with excess radiation ($A_r$).}\label{p_1a}
\end{figure}

\begin{figure}
    \begin{center}
        {\includegraphics[width=4.5in,height=3in]{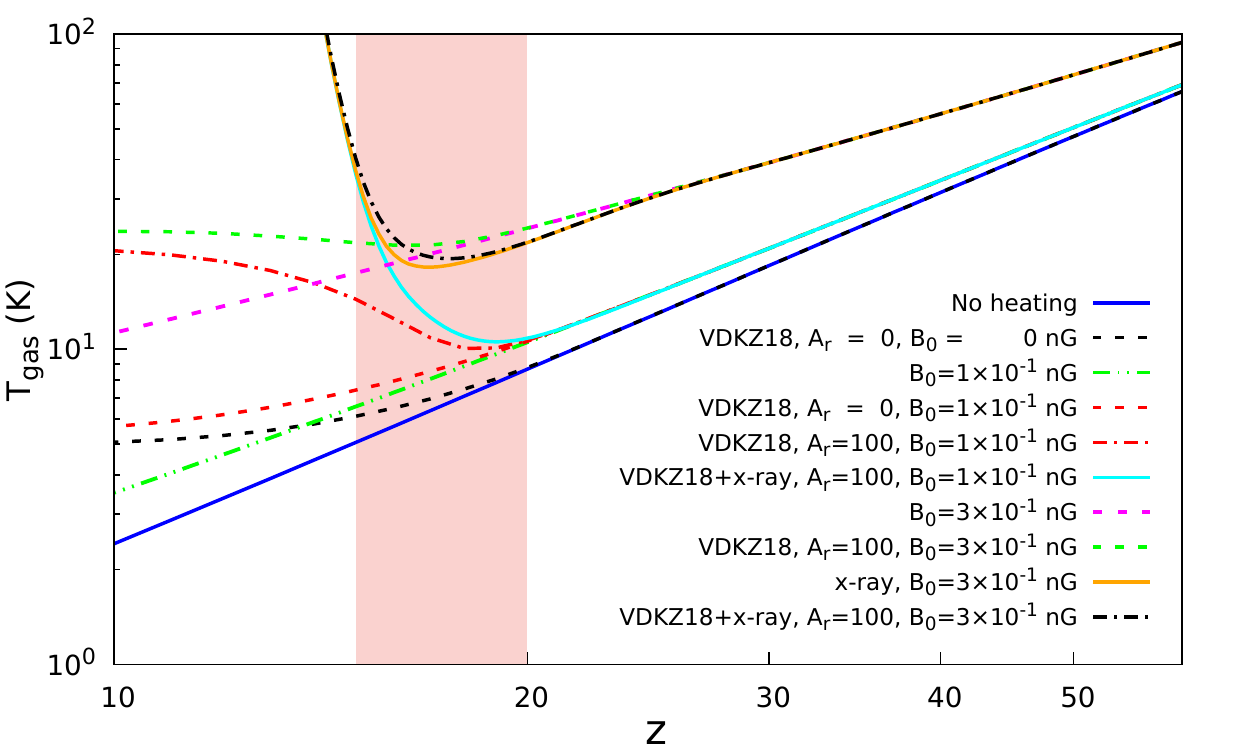}} 
    \end{center}
    \caption{The caption is same as in figure \eqref{p_1a}, except here, we include different combination of VDKZ18, x-ray and magnetic heating, and spectral index is fixed to $-2.99\,$.}\label{p_1b}
\end{figure}
\begin{figure}
    \begin{center}
        {\includegraphics[width=4.5in,height=3in]{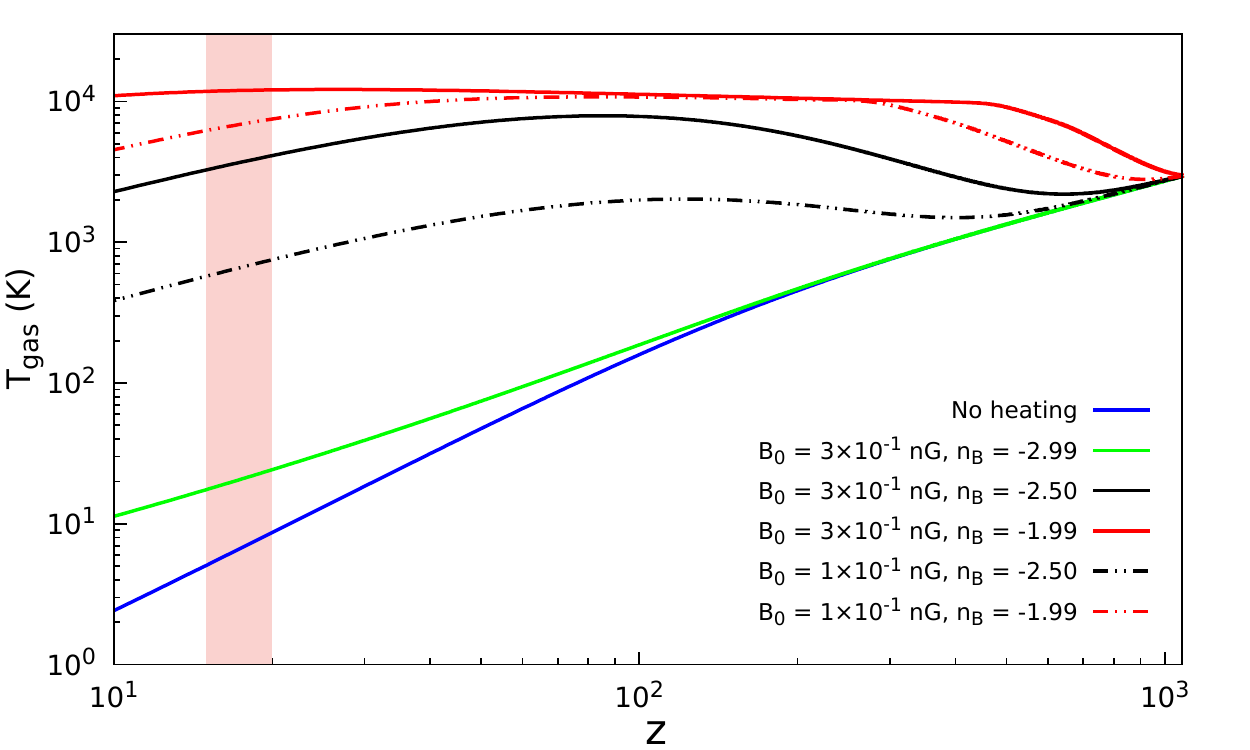}} 
    \end{center}
    \caption{The caption is same as in figure \eqref{p_1a}, except here, we vary the spectral index and plot magnetic heating of the gas.}\label{p_1c}
\end{figure}
%
In figures \eqref{p_1a}, \eqref{p_1b} \& \eqref{p_1c}, we plot the gas temperature vs. redshift for different values of present-day strengths of PMFs ($B_0$) and excess radio background fraction ($A_r$). The solid blue lines represent the case when there is no heating of the IGM gas, i.e. no x-ray, VDKZ18 or magnetic heating. The pink shaded band in the figure shows the EDGES redshift range, $15\leq z \leq 20$, for the 21~cm absorption signal. In plot \eqref{p_1a}, we consider only VDKZ18 and x-ray heating. The orange dashed line describes the heating due to  VDKZ18 only while keeping $A_r=0$.  Next, we increase the value of $A_r$ from 0 to 100.  This case is described by the dashed-green line in plot \eqref{p_1a}, which shows a significant rise in the gas temperature due to the excess radiation fraction. Further, if one increases the $A_r$  to its LWA 1 limit, i.e. $A_r= 418$, the gas temperature does not change significantly from $A_r =100$ case, as shown by the solid magenta curve. It happens because $\Gamma_{R}\propto (T_R/T_S-1)\sim T_R/T_S$, equation \eqref{seq13}. As we increase $A_r$, $T_R/T_S$ increases slowly. For example, at $z=17$, $T_R/T_S$ is $6.5$ for $A_r=0$, $51.4$ for $A_r=100$ and $54.9$ for $A_r=418$. Here, we can see that, even increasing $A_r$ to $\sim4$ times (100 to 418), $T_R/T_S$ increases by only $6.8$ percent.   Therefore, increasing further $A_r$ will not affect gas temperature significantly. To analyse the role of x-ray heating, we have first considered the heating due to x-ray only, depicted by the red dashed line. The inclusion of VDKZ18 for $A_r=0$ further increases the gas temperature slightly, as shown by the black dashed line. In this case of inclusion of x-ray heating, if we increase the value of $A_r$ to 100, there is a significant increase in the gas temperature as shown by the solid green line. We find the contribution due to x-ray heating dominates for redshift values $z\lesssim15$. 


In plot  \eqref{p_1a}, we compare the contribution of VDKZ18 and x-ray heating. In plot \eqref{p_1b}, we compare the contributions of VDKZ18, x-ray and magnetic heating while keeping the spectral-index, $n_B=-2.99$ for a nearly scale-invariant magnetic field spectrum.  While in figure \eqref{p_1c}, we vary the magnetic spectral index ($n_B$) and plot the magnetic heating of the gas.


In plot \eqref{p_1b}, we have included the effect of primordial magnetic fields on the IGM gas evolution.  The solid blue line represents the case when there is no heating, and the dashed-black curve shows the case of VDKZ18 with no magnetic fields and x-ray for $A_r=0$. The double dot-dashed green curve represents the case when there is only the magnetic heating with a magnetic field strength of $B_0=1\times10^{-1}$nG.  Next, we include the case of VDKZ18 for $A_r=0$ in the pure magnetic heating scenario, as shown by the red dashed curve. Now, if we increase  $A_r$ from 0 to 100, the gas temperature rises significantly in the shaded region as shown by the dash-dotted red curve in figure \eqref{p_1b}. Now the further addition of x-ray heating is shown by the cyan plot, which shows significant heating in the shaded region.  Next, for more analysis, we increase the magnetic field strength from $B_0=1\times10^{-1}$~nG to $B_0=3\times10^{-1}$~nG and study cases with VDKZ18 and x-ray as before. The magenta dashed line depicts the case with only magnetic heating. The green dashed line shows the case of VDKZ18 with $A_r=100$. The orange curve shows the case with magnetic and x-ray heating only. Here, as expected, the gas temperature decreases after the inclusion of the x-ray effect with the magnetic fields. It happens because the ionization fraction increases by x-ray radiation. Ambipolar diffusion evolves as $\Gamma_{\rm ambi}\propto(1-x_e)/x_e$; therefore, as ionization fraction increases, ambipolar diffusion of the magnetic field decreases. Thus, the heating due to magnetic fields also decreases.  Therefore, including the x-ray contribution with the magnetic field decreases the magnetic field diffusion. Hence, the gas temperature decreases (this effect also occurs for $B_0=1\times10^{-1}$~nG, but it is not visible in the plot). The black dot-dashed line includes all the three effects: magnetic and x-ray heating together with VDKZ18 for $A_r=100$ and $B_0=3\times10^{-1}$~nG. Here, the addition of the VDKZ18 heating for $A_r=100$ increases the gas temperature above the solid orange line. It is also lower than the magenta dashed line because of the inclusion of the x-ray contribution.  At the smaller redshift, x-ray heating dominates over all other heating mechanisms, and all lines merge. 


In figure \eqref{p_1c}, we plot the magnetic heating of the gas for the different spectral index ($n_B$) and $B_0$.  The solid lines, except the blue one, represent the magnetic heating for $B_0=3\times10^{-1}$~nG, while double dot-dashed lines are for $B_0=1\times10^{-1}$~nG. Increasing the spectral index, the magnetic heating due to ambipolar diffusion and turbulent decay increases as $\Gamma_{\rm ambi}\propto \big(1/\Gamma[(n_B+3)/2]\big)^2$ and  $\Gamma_{\rm turb}\propto 1/\Gamma[(n_B+3)/2]$ (by ignoring the logarithmic and integral dependencies). For example, if one changes $n_B$ from its value $-2.99$ to  $-1$ then $1/\Gamma[(n_B+3)/2]$ changes from $5\times10^{-3}$ to  1. Therefore, by increasing $n_B$ from $-2.99$ to $-1$,  magnetic heating enhances significantly. To get $T_{21}$ (equation \ref{t21f}) around $-500$~mK or $-300$~mK at $z=17.2$, one needs to ensure that even by increasing  $n_B$, that the factor $x_{\rm HI}\left(1-{T_R}/{T_S}\right)$  remains same. Thus from equations \eqref{eq9}, \eqref{eq10} and \eqref{PB} when we increase $n_B$, we have to decrease $B_0$ so that the magnetic heating contribution to the gas remains the same. Therefore, by increasing $n_B$,  the upper bound on $B_0$ will become more stringent. Here, we also include the collisional ionization of the gas in equation \eqref{s4}, as this term is important only when gas temperature is $\gtrsim1.58\times10^5$~K. Otherwise this term is exponentially suppressed as $\propto \exp[-(13.6~{\rm eV})/T_{\rm gas}]$     \cite{Sethi:2004pe, Asselin:1988, Shiraishi:2014}. In plot \eqref{p_1c}, the gas temperature rises by increasing $B_0$, as more magnetic energy is getting injected into thermal energy of the gas via $\Gamma_{\rm ambi}\propto E_B^2$ and $\Gamma_{\rm turb}\propto E_B$. However, for redshift $z\lesssim 100$, the gas temperature starts decreasing as the cooling effect due to expansion of the Universe become dominant, as can be seen in equations \eqref{eq6} \& \eqref{11} (it also depends on the strength and spectral index of the magnetic field). Since,  with the expansion of the Universe, magnetic energy density ($E_B$) also dilutes, the contributions from $\Gamma_{\rm ambi}$ and $\Gamma_{\rm turb}$ decreases as can be seen from equations \eqref{eq9}, \eqref{eq10} and \eqref{11}.

\begin{figure}
    \begin{center}
        {\includegraphics[width=4.5in,height=3in]{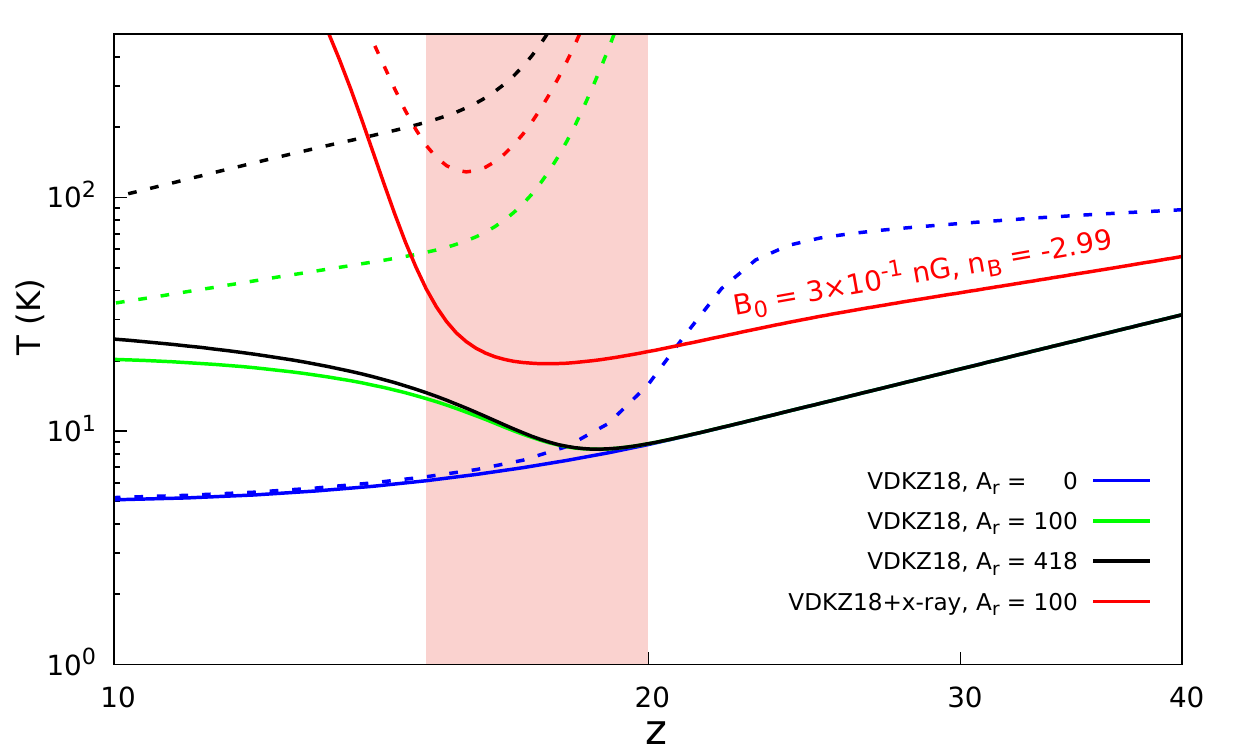}} 
    \end{center}
    \caption{This figure shows the gas (solid lines) and spin (dashed lines) temperature evolution, The shaded region corresponds to the redshift $15\leq z \leq 20\,$--- the redshift range for EDGES reported signal.} \label{p_2a}
\end{figure}
\begin{figure}
    \begin{center}
        {\includegraphics[width=4.5in,height=3in]{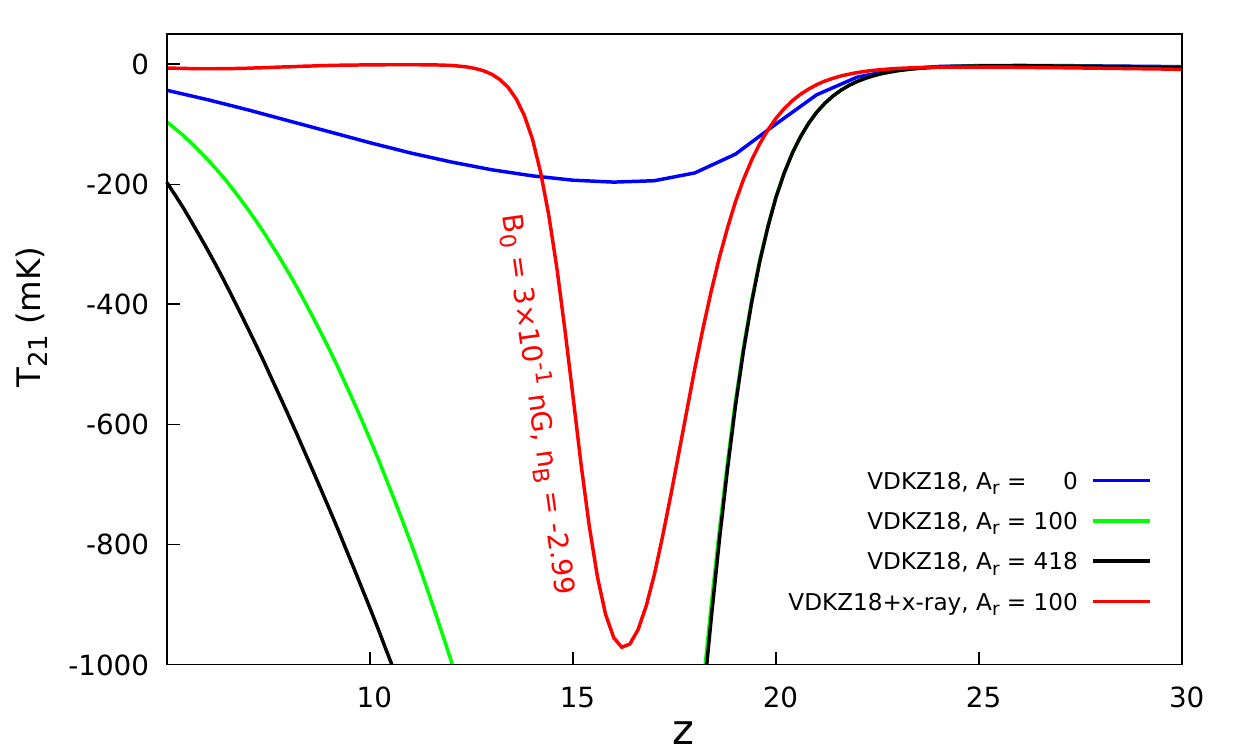}} 
    \end{center}
    \caption{This figure shows the 21 cm differential brightness temperature with redshift for same cases in plot \eqref{p_2a}.} \label{p_2b}
\end{figure}
In figure \eqref{p_2a}, we plot the spin (dashed lines) and gas (solid lines) temperature.  For $A_r=0$, i.e. $T_R=T_{\rm CMB}$,  we get $T_{\rm gas}\simeq T_S$ as seen by the overlapping  dashed and solid  blue lines in the shaded region. $x_\alpha$ and $x_c$ are $\propto 1/T_R\,$ as can be seen from equations \eqref{xalph} and \eqref{xc}. Therefore, the coupling between the gas and spin temperature decreases by increasing $A_r$. As discussed before, increasing the value of $A_r$ above $\sim 100$, the spin temperature increases, but the increment in gas temperature becomes insignificant, and the $T_R/T_S$ ratio increases slowly. Therefore, as $x_\alpha$ and $x_c$ decreases, the difference between the gas and spin temperature increases, as shown in the plot \eqref{p_2a}. Increasing the values of $A_r$ from $100$ (green lines) to $418$ (black lines), the difference between gas and spin temperatures increases. Figure \eqref{p_2b}, shows the plots for 21 cm differential brightness temperature vs. redshift,  for all the cases discussed in plot \eqref{p_2a}. As we increase the $A_r$ from 0 to 100 the $|T_{21}|$ increases. By increasing $A_r$  from 100 to 418, values of $T_{21}$ does not change significantly. Further, including x-ray heating and magnetic heating (for $B_0=3\times10^{-1}$~nG and $n_B=-2.99$) the gas temperature rises and $|T_{21}|$ decreases.

\begin{figure}
    \begin{center}
        {\includegraphics[width=5.5in,height=3.5in]{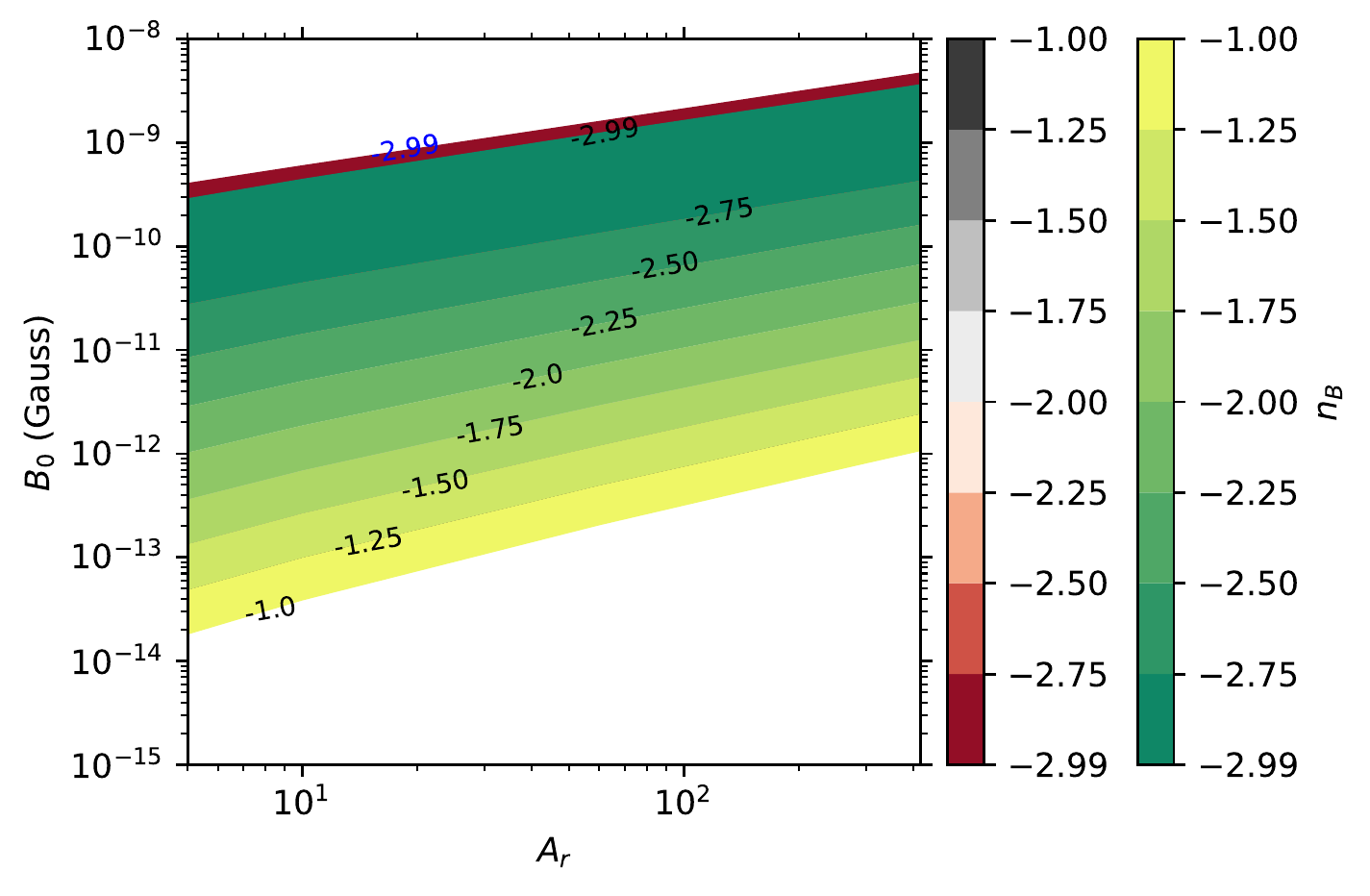}} 
    \end{center}
    \caption{In this figure, we study upper bounds on present-day magnetic field strength ($B_0$) with excess radiation fraction $(A_r)$ for different values of the spectral index, $n_B$. The green-yellow and red-grey colour schemes represent the cases when $T_{21}|_{z=17.2}\simeq-500$~mK and $-300$~mK, respectively. For $T_{21}|_{z=17.2}\simeq-300$~mK case the value of $n_B$ written with blue coloured text , while for $-500$~mK  case it is written with black coloured text. Here, we consider $T_{\rm S}\simeq T_{\rm gas}$ and do not take into account the x-ray and VDKZ18 effects.}\label{p_3a}
\end{figure}

\begin{figure}
    \begin{center}
        {\includegraphics[width=5.5in,height=3.5in]{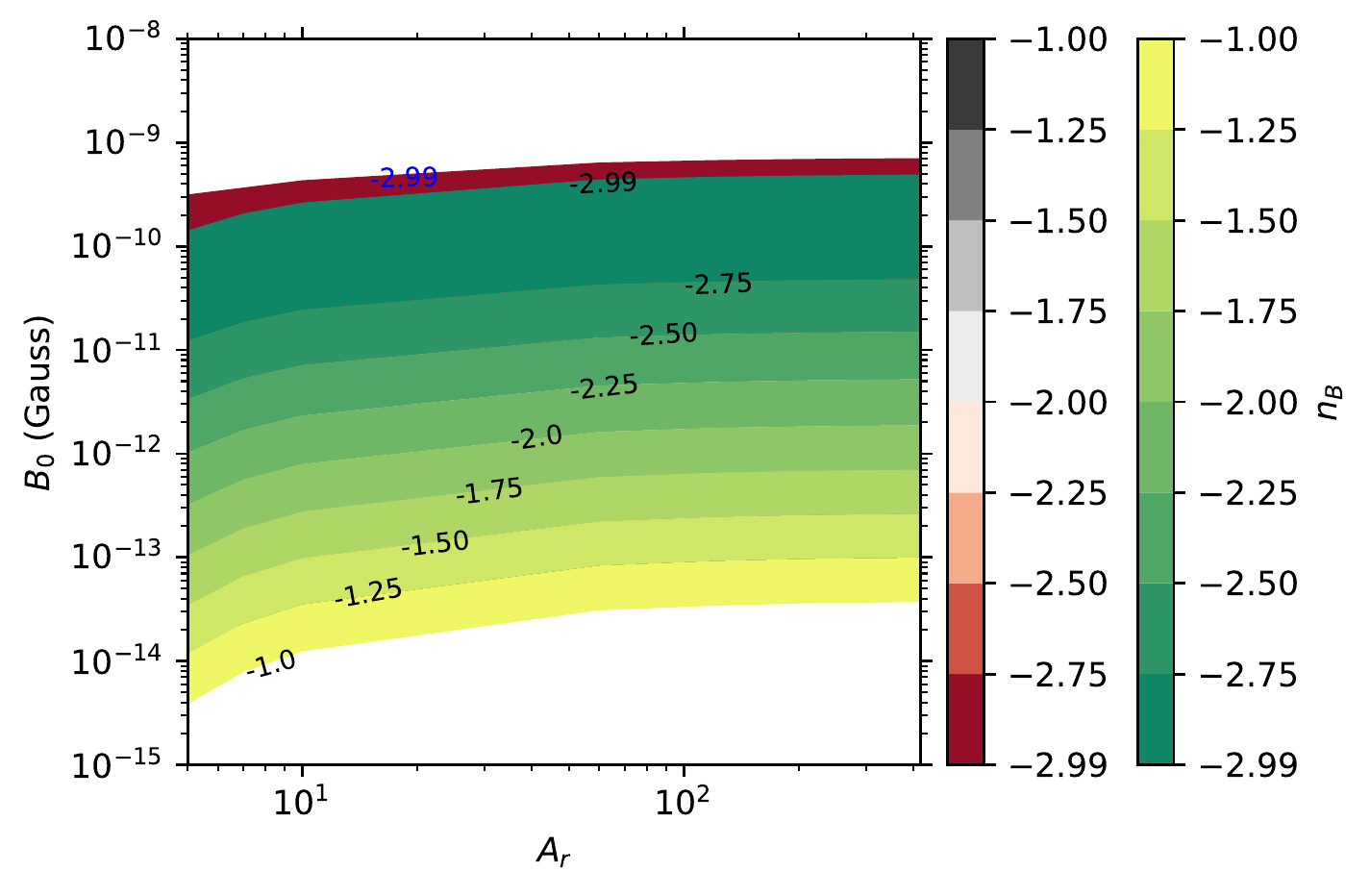}} 
    \end{center}
    \caption{The caption is same as in figure \eqref{p_3a}, except here, we consider the effects of VDKZ18 and x-ray heating on the gas due to first stars after $z\lesssim 35$ and consider finite Ly$\alpha$ coupling.}\label{p_3b}
\end{figure}
In figures \eqref{p_3a} and \eqref{p_3b}, we plot the maximally allowed values of $B_0$ versus radiation excess ($A_r$) for different spectral indexes. The colour-bars represent the variation of the magnetic field spectral index. In the plots, the spectral index varies from its nearly scale-invariant value (-2.99) to -1. Here, we consider both the EDGES best fit and upper constraint on the 21 cm absorption signal for constraining $B_0$. The green-yellow colour scheme represents the case with  $T_{21}|_{z=17.2}$ $\simeq-500$~mK, while the red-grey colour scheme represents the case with $T_{21}|_{z=17.2}\simeq-300$~mK. Numerical values of $n_B$ for the different colour bands are written with different colour. For $T_{21}|_{z=17.2}\simeq-300$~mK case the value of $n_B$ written with blue coloured text , while for $T_{21}|_{z=17.2}\simeq-500$~mK  case it is written with black coloured text. The colour-bars are common for both the plots.


In figure \eqref{p_3a}, we consider infinite Ly$\alpha$ coupling ($x_\alpha\gg x_c,\,1$), i.e. $T_S\simeq T_{\rm gas}$.  Here, we do not consider the x-ray and VDKZ18  effects on the  gas and thus the 21 cm signal $T_{21}\propto(1-T_R/T_{\rm gas})$. As we increase $A_r$, the amplitude of $|T_{21}|$ increases, and we get more window to increase the gas temperature. In this plot, we consider heating only due to the decaying magnetohydrodynamics. Therefore, we can increase $B_0$ as we increase $A_r$. As discussed earlier, by decreasing $n_B$, the amplitude of the magnetic field power spectrum also decreases, resulting in less magnetic energy dissipation into the gas kinetic energy. Thus by reducing values of $n_B$ from -1 to -2.99, we get more window to increase $B_0$. Next, when one increases  $T_{21}$ from -500~mK to -300~mK,  the allowed value of $B_0$ also increases. This is shown by the red-grey colour scheme in figures \eqref{p_3a} and \eqref{p_3b}.  In figure \eqref{p_3b}, we consider the effects of VDKZ18 and x-ray on IGM gas evolution due to first stars after $z\lesssim 35$ and consider finite Ly$\alpha$ coupling. As discussed earlier, $T_{\rm gas}\neq T_S$ for $A_r>0$ and the difference between gas and spin temperature increases as $A_r$ increases. Thus, in the presence of first star's effects, the upper bound on the present-day strength of PMFs modifies. Following the Refs. \cite{Kovetz2018, Mirocha:2015G, Harker:2015M}, we consider WF coupling coefficient, $x_\alpha = 2A_\alpha(z) \times (T_0/T_R)$. Here, $A_\alpha(z) = A_\alpha (1 + \tanh[(z_{\alpha0}-z)/\Delta z_\alpha])$, the step height $A_\alpha=100$, pivot redshift $z_{\alpha0}=17$ and duration $\Delta z_\alpha=2$. The collisional coupling coefficient, $x_{c} = T_{10}/T_R\times (N_H\,k_{10}^{HH})/A_{10}$. After the inclusion of x-ray and VDKZ18 heating effects, the gas temperature remains $>10$~K. Therefore, we can take $k_{10}^{HH}\,\approx 3.1$ $\times10^{-11}$ $(T_{\rm gas}/{\rm K})^{0.357}$  $\exp(-32~{\rm K}/T_{\rm gas})$ ${\rm cm^3/sec}$ for ${\rm 10~K}<T_{\rm gas}<{\rm 10^3~K}$. As illustrated in plot \eqref{p_1a}, \eqref{p_1b}, \eqref{p_2a} \& \eqref{p_2b}, increasing excess radiation fraction $A_r$ above $\sim 100$, the $T_R/T_S$ remains nearly constant and this also mean that $T_{21}$ remain unchanged. Consequently one can not increase the value of $B_0$ and one gets nearly flat profile for $B_0$ for $A_r\gtrsim100$ in figure \eqref{p_3b}. 


\begin{figure}
    \begin{center}
        {\includegraphics[width=5.5in,height=3.5in]{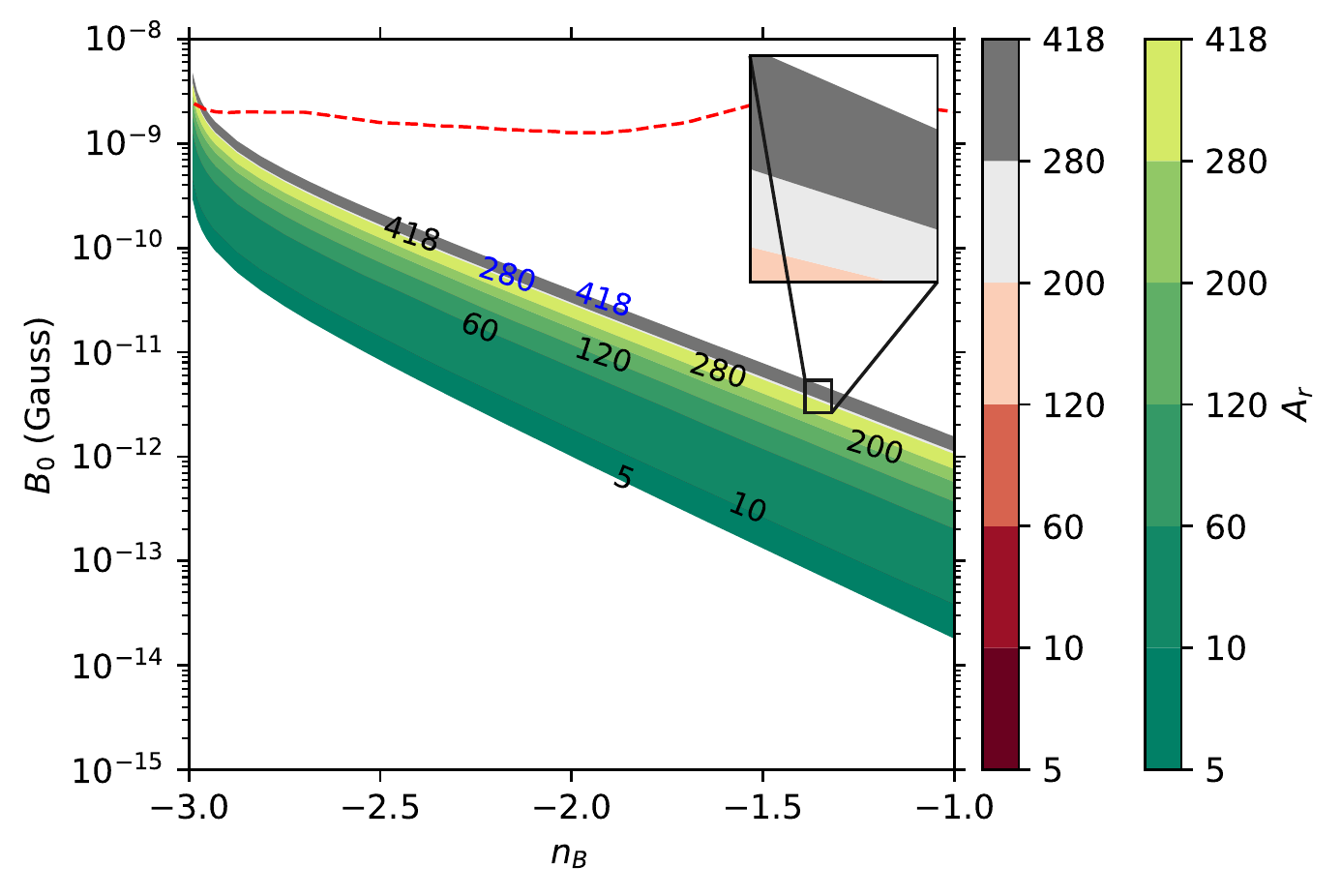}} 
    \end{center}
    \caption{In this figure, we study upper bounds on the present-day magnetic field strength ($B_0$) with spectral index ($n_B$) for different values of excess radiation fraction $(A_r)$. The green-yellow and  red-grey colour schemes represent the cases when $T_{21}|_{z=17.2}\simeq-500$~mK and $-300$~mK, respectively. For $T_{21}|_{z=17.2}\simeq-300$~mK case the value of $n_B$ written with blue coloured text , while for $-500$~mK  case it is written with black coloured text. The red dashed line depicts the Planck 2015 upper constraint on the present-day magnetic field strength \cite{Planck:2016, Minoda:2018gxj}. Here, we consider $T_{\rm S}\simeq T_{\rm gas}$ and do not take into account the x-ray and VDKZ18 effects.} \label{p_4a}
\end{figure}
\begin{figure}
    \begin{center}
        {\includegraphics[width=5.5in,height=3.5in]{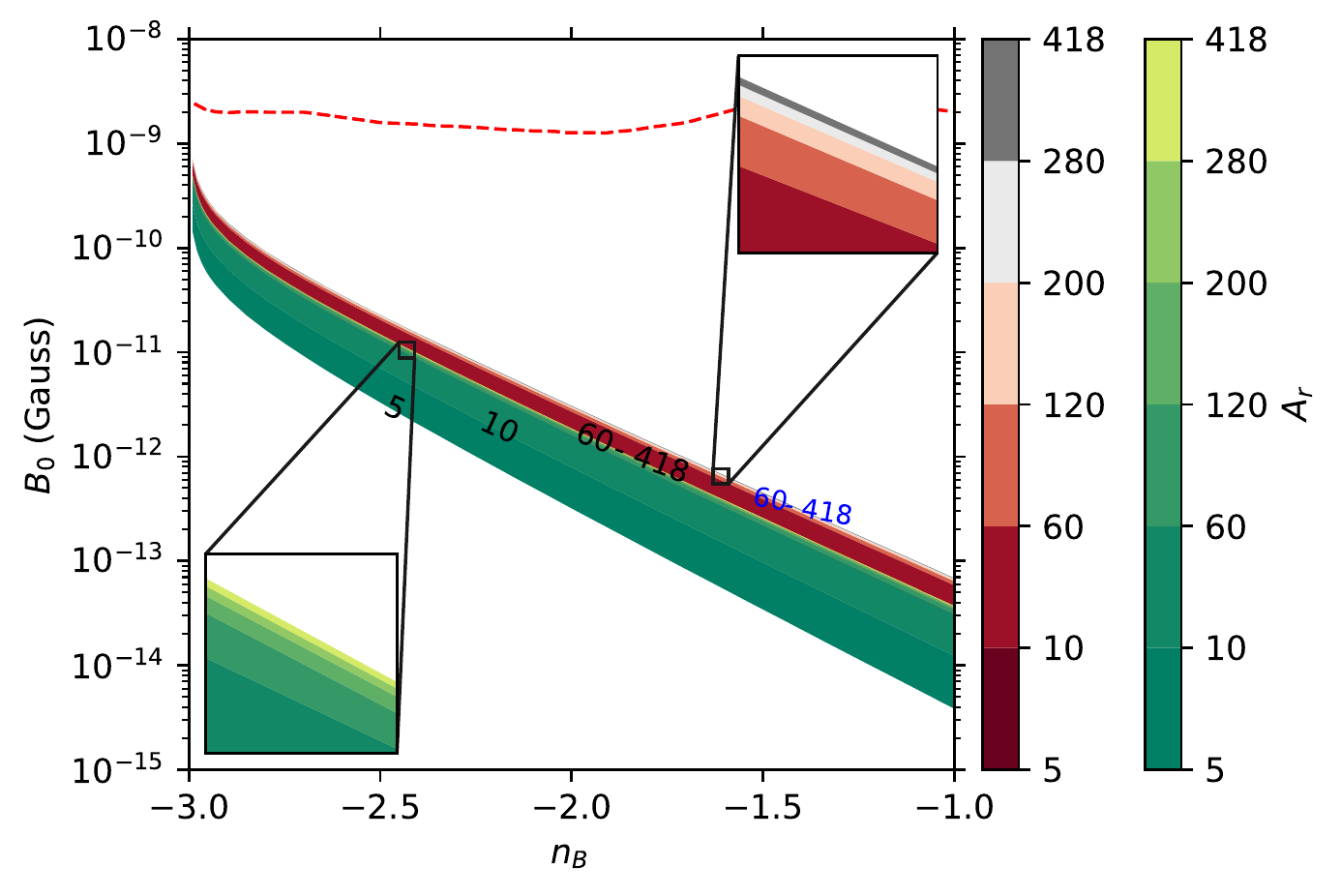}} 
    \end{center}
    \caption{The caption is same as in figure \eqref{p_4a}, except here, we consider the heating effects of VDKZ18 and x-ray on IGM gas due to first stars after $z\lesssim 35$ and consider finite Ly$\alpha$ coupling. The colour-bars are common for both plots.}\label{p_4b}
\end{figure}


In figures \eqref{p_4a} \& \eqref{p_4b}, we plot the maximally allowed values of $B_0$ vs $n_B$ for various values of $A_r$. The colour-bars represent the variation in $A_r$. In the plots, $A_r$ varies from 5 to LWA 1 limit $\sim 418$. We consider both the EDGES best fit and upper constraint on 21 cm absorption signal for constraining $B_0$. The green-yellow scheme represent the case with  $T_{21}|_{z=17.2}\simeq-500$~mK, while the red-grey colour scheme represent the case  $T_{21}|_{z=17.2}\simeq-300$~mK. Numerical values of $A_r$ for the different colour bands are written in different colours. For $T_{21}|_{z=17.2}\simeq-300$~mK case the value of $A_r$ written with blue coloured text , while for $T_{21}|_{z=17.2}\simeq-500$~mK  case it is written with black coloured text. The spectral index ranges from -2.99 to -1.  The red dashed line represents the Planck 2015 upper constraint on the present-day magnetic field strength with spectral index in both plots. This constraint has been taken from Refs. \cite{Planck:2016, Minoda:2018gxj}.


In plot \eqref{p_4a}, we consider $T_{\rm S}\simeq T_{\rm gas}$ and we do not take into account the x-ray and VDKZ18 effects on IGM gas evolution. The zoomed inset in the figure shows the contour plot when $T_{21}|_{z=17.2}\simeq-300$~mK. Here, considering $T_{21}|_{z=17.2}\simeq-300$~mK,  for $n_B<-2.98$ the  $A_r\gtrsim 200$ is excluded similarly for $n_B<-2.96$ the $A_r\gtrsim 280$ is excluded by Planck 2015 upper constraint on $B_0$. Likewise, for $T_{21}|_{z=17.2}\simeq-500$~mK, for $n_B<-2.97$ the  $A_r\gtrsim 280$ is excluded. For spectral index -2.9 and excess radiation fraction 418, we get the upper constraint on $B_0$ to be $\sim 1$~nG and   $1.3$~nG by requiring $T_{21}|_{z=17.2}\simeq-500$~mK (EDGES best fit constraint) and $-300$~mK (EDGES upper constraint), respectively. While for $n_B=-1$,  these bound change to $1.1\times10^{-3}$~nG and $1.6\times10^{-3}$~nG for $T_{21}|_{z=17.2}\simeq-500$~mK and $-300$~mK, respectively. In plot \eqref{p_4b}, we include both the VDKZ18 and x-ray effect and consider finite Ly$\alpha$ coupling. As discusses earlier, for $A_r\gtrsim100$, $T_R/T_S$ ratio remain nearly constant. Thus, in the plot \eqref{p_4b}, we can see that for $A_r\gtrsim100$, the upper bound on $B_0$ is not changing significantly--- the plots are merged for $A_r\gtrsim100$. These plots have been shown by the zoomed inset. The right upper zoomed inset is shown for $T_{21}\simeq-300$~mK, while left lower zoomed inset is shown for green-yellow contour plots when $T_{21}\simeq-500$~mK. Hence, further increasing $A_r>100$ will not change significantly the upper bound on $B_0$. As illustrated in figure \eqref{p_2a}, $T_S> T_{\rm gas}$ for $A_r >0$, and $T_{21}\propto (1-T_R/T_S)$. Therefore, to get $T_{21}\simeq-300$~mK or $-500$~mK, we need to lower $B_0$ compared to previous scenario--- figure \eqref{p_4a}. Hence, we get the more stringent upper bound on present-day magnetic field strength in figure \eqref{p_4b}. For spectral index -2.9 and excess radiation fraction 418, we get the upper constraint on $B_0$ to be $\lesssim 1.7\times10^{-1}$~nG and   $1.2\times10^{-1}$~nG by requiring $T_{21}|_{z=17.2}\simeq-300$~mK and $-500$~mK, respectively. For $n_B=-1$, we get $B_0\lesssim6.9\times10^{-5}$~nG and $3.7\times10^{-5}$~nG by requiring EDGES upper and best fit constraint on 21~cm differential brightness temperature. Decreasing the values of $A_r$, the upper constraint on $B_0$ becomes more stringent. For example, when $A_r=5$, we get upper bound on present day magnetic field strength to be $\lesssim1.4\times10^{-1}$~nG for spectral index -2.99, and for spectral index $n_B=-1$ we get $B_0\lesssim3.8\times10^{-6}$~nG by requiring EDGES best fit constraint on $T_{21}$. The upper bounds are also well below the Planck 2015 constraint \cite{Planck:2016}.


\section{Conclusions}
In the present work, we study the upper constraint on the strength of the primordial magnetic fields for different spectral index using the bound of EDGES observation on $T_{21} $, in the presence of uniform redshift-independent synchrotron like radiation reported by ARCADE 2 and LWA 1 \cite{Fixsen2011, Feng2018, Dowell2018, Fialkov:2019}. We have considered excess radiation fraction up to the LWA 1 limit (i.e. $A_r\sim418$) at the reference frequency of 78~MHz  \cite{Dowell2018, Fialkov:2019}. To get the upper constraint on $B_0$, we have used both the EDGES upper and best-fit constraints on $T_{21}$.  We have considered two scenarios: First, infinite Ly$\alpha$ coupling (i.e. $x_\alpha\gg x_c, 1$) without the effects of x-ray and VDKZ18 on IGM gas evolution. In another scenario, we consider the finite  Ly$\alpha$ coupling with x-ray and VDKZ18 effects. The following summarises our results for $T_{21}=-500$~mK:

In the first scenario, for $A_r=418$, we get $B_0\lesssim3.7$~nG for spectral index -2.99, while for $n_B=-1$ we get $B_0\lesssim1.1\times10^{-3}$~nG. When $A_r=5$, upper constraint on present-day magnetic field strength varies from $B_0\lesssim 2.9\times10^{-1}$~nG to $1.8\times10^{-5}$~nG by varying $n_B$ from -2.99 to -1, respectively. 

In the second scenario, the upper bounds on $B_0$ will modify \cite{Venumadhav:2018, Kovetz2018}. For $A_r=418$, we get the upper constraint on magnetic field to be $B_0(n_B=-2.99)\lesssim4.9\times10^{-1}$~nG and $B_0(n_B=-1)\lesssim3.7\times10^{-5}$~nG. While for $A_r=5$, we get upper bound on present day magnetic field strength to be $\lesssim1.4\times10^{-1}$~nG for spectral index -2.99, and for spectral index -1 we get $B_0\lesssim3.8\times10^{-6}$~nG.

We would like to note that these upper bounds on $B_0$ that we have reported here are also consistent with the Planck observations \cite{Planck:2016, Planck:2014}.

\clearpage
\pagestyle{empty}
\cleardoublepage
\pagestyle{fancy}
\begin{savequote}[75mm]
	``Who sees the future? Let us have free scope for all directions of research"
	\qauthor{Ludwig Eduard Boltzmann, \textit{``Lectures on Gas Theory" translated by Stephen G. Brush}}
\end{savequote}

\chapter[PMFs \& Baryon-Dark matter Interaction] {Primordial Magnetic Fields and  Baryon-Dark matter Interaction}\label{chap5}
\vspace{-1.5cm}

In the previous chapter \eqref{chap4}, we have analysed the upper bound on present-day strength of PMFs in the light of EDGES observation and excess radio background reported by ARCADE 2 and LWA 1 observations \cite{Natwariya:2021}. As discussed earlier in chapter \eqref{chap1}, to explain EDGES observation one requires that either the background radio radiation should be grater than $\sim$104~K in the absence of any non-standard mechanism for the evolution of the gas temperature or the gas temperature should be less than $3.2$~K for the standard evolution of CMB temperature at the centre of the ``U" profile for the best fitting amplitude \cite{Bowman:2018yin}. The first possibility has been investigated by authors of the Ref. \cite{Moroi:2018vci, FRASER2018159, Pospelov:2018, Liu:2019H}. In the second scenario, IGM gas can be cooled by emitting the photons between the Ly-limit to Ly-$\gamma$ wavelengths \cite{Chuzhoy_2007,Chuzhoy2006}. There are very few mechanisms to cool the gas. Since the dark matter is colder than the gas, effective cooling of the gas can be obtained by elastic scattering between the dark matter and baryon particles \cite{Barkana:2018nd, Barkana:2018lgd, Tashiro:2014tsa}. A new kind of interaction between dark matter and baryons was proposed by the authors of reference \cite{Munoz:2015bk, Barkana:2018lgd} to explain the EDGES absorptional signal. The authors consider a non-standard ``Coulomb-like" interaction: $\sigma= \hat{\sigma}\ v^{-4}$; $v$ is the relative velocity between the dark matter and baryons and $\hat{\sigma}$ is the strength of baryon-dark matter interaction cross-section \cite{Barkana:2018lgd, Tashiro:2014tsa, Dvorkin:2013cea, Barkana:2018nd, Munoz:2015bk}. Here, the interaction between dark matter and baryons does not depend on whether the baryons are free or bound within atoms \cite{Barkana:2018lgd}. The cooling of the gas, by transferring energy to the dark matter, is tightly constrained because of constraints on the dark matter mass and cross-section by cosmological and astrophysical observations \cite{Barkana:2018nd, Barkana:2018lgd, Berlin2018, Creque-Sarbinowski:2019mcm}. In the present chapter, we reanalyse the constraints on PMFs in the presence of baryon-dark matter interaction proposed by the authors of reference \cite{Barkana:2018lgd}. In the presence of baryon-dark matter interaction the bounds on magnetic field, baryon dark matter cross-section strength ($\hat \sigma$) and dark matter mass ($M_{\rm DM}$) can strongly influence each other. This requires to rework the bounds on $\hat \sigma\,$, $M_{\rm DM}$ and $B_0$ which can explain the observed absorption signal by EDGES collaboration. The upper limit on the magnetic field strength can modify in presence of baryon-dark matter interaction cross-section. In the presence of a strong magnetic field, a large baryon-dark matter interaction cross-section is required to balance magnetic heating of gas to explain the EDGES signal as compared to a weak magnetic field. Subsequently, the strong magnetic-fields can even erase the 21~cm signal--- this gives an upper bound on the strength of magnetic-fields, dark matter mass and baryon-dark matter cross-section.

In order to explain the EDGES absorption signal, the gas temperature needs to be cooler than the $\Lambda$CDM prediction. During the Cosmic dawn era, the Universe was at its coldest phase, and the relative velocity between the dark matter and baryon was very small, $\mathcal{O}(10^{-6})$. Also, the temperature of the dark matter was colder than the baryon temperature during this period, so an interaction of the baryon with dark matter can cool the gas temperature. Since the relative velocity is small, scattering cross section of the type $\sigma= \hat{\sigma}\ v^{-4}$ can enhance the interaction rate and cool the gas sufficiently to explain EDGES absorption dip \cite{Tashiro:2014tsa, Dvorkin:2013cea, Barkana:2018lgd}. In this chapter, we consider magnetic heating of the gas and dark matter via ambipolar and turbulent decay. Here, we take cosmological parameters $\Omega_b$, $\Omega_m$, and $h$ as $\Omega_b=0.04859$, $\Omega_m=0.315$ and $h=0.68$ \cite{Planck:2018}.

\section{Baryon-dark matter interaction in presence of magnetic fields}
\label{gas_dm_mag}

In this section, we discuss the effects of magnetic fields on the gas temperature in the presence of baryon-dark matter interaction. The gas temperature evolves as discussed in the chapter \eqref{chap4}, except here, the cooling rate ($dQ_{\rm gas}/dt$) will add due to the energy transfer from gas to dark matter \cite{Munoz:2015bk, Chluba2015},
\begin{alignat}{2}
\frac{dT_{\rm gas}}{dz} & =  2\,\frac{T_{\rm gas}}{1+z} + \frac{\Gamma_{C}}{(1+z)H} (T_{\rm gas}-T_{\rm CMB}) \nonumber \\ 
&\qquad-\frac{2}{3\,n_{\rm tot}(1+z)\,H}(\Gamma_{\rm turb}+\Gamma_{\rm ambi})+\frac{2}{3\,(1+z)\,H}\,\frac{d{Q_{\rm gas}}}{dt}\,.\label{eq:baryon_temp}
\end{alignat}
The cooling rate ($dQ_{\rm gas}/dt$) depends on the temperature difference and relative velocity between dark matter and baryons,
\begin{equation}
\frac{d{Q_{\rm gas}}}{dt} = \frac{2\, M_b\,\rho_{\rm DM}\,\hat{\sigma}\,e^{-r^2/2}}{\sqrt{2\, \pi}\,(M_{b} + M_{\rm DM})^2 \,u_{\mathrm{th}}^3} \bigg(T_{\rm gas}-T_{\rm DM}\bigg)-\mu\,\frac{\rho_{\rm DM} }{\rho_{\rm M}}\, v\, D(v)~,\label{eq:heat_tranfer}
\end{equation}
here,  $M_b\approx M_{\rm H}$ is the baryon mass and can be taken as mass of hydrogen atom. $\rho_{\rm DM}$ and $\rho_{\rm M}$ are the dark matter and total matter energy density, respectively. Moreover, $r=v/u_{\rm th}$, $v$ is the relative motion between baryons and dark matter while $u_{\rm th}^2=T_{\rm gas}/M_b+T_{\rm DM}/M_{\rm DM}\,$. Here, $T_{\rm DM}$ is the dark matter temperature and $\mu=M_b\,M_{\rm DM}/(M_b+M_{\rm DM})$ is the reduced mass. The first term in equation \eqref{eq:heat_tranfer}, arises due to the temperature difference between dark matter and gas. As $T_{\rm DM} < T_{\rm gas}$, the first term is positive. It implies that the energy of gas is being transferred to dark matter with time. The second term in equation \eqref{eq:heat_tranfer}, comes due to the friction   between two fluids caused by velocity difference--- drag term, and it is given by $\mu\,(\rho_{\rm DM}/\rho_{\rm M})\,v\,D(v)$, 
\vspace{-0.2cm}
\begin{alignat}{2}
D(v) \equiv  \frac{\rho_{\rm M}\,\hat{\sigma}}{M_{b} + M_{\rm DM}}\ \frac{1}{v^{2}}\ F(r) \label{eq:drag1}\, ,
\end{alignat}
here, $r = {v}/{u_{\mathrm{th} }}$ and the function $F(r)$ is defined as,
\vspace{-0.2cm}
\begin{alignat}{2}
F(r) \equiv \mathrm{erf}\left(\frac{r}{\sqrt{2}}\right) - \sqrt{\frac{2}{\pi}}\ r \ e^{-r^{2}/2}\,,
\end{alignat}
here, ${\rm erf()}$ is the Gauss error function. When the relative velocity between dark matter and baryons is zero, i.e. $r=0$, one gets $F(0)=0$. In this case there will not be any drag heating of gas and dark matter. As $r\rightarrow\infty$, $F(r)\rightarrow1$. For any value of $r\geq0$, one finds that $F(r)\geq0$. Therefore, the last term in equation \eqref{eq:heat_tranfer} always remains negative. It implies that the energy of gas always increases due to the drag. In equation \eqref{eq:heat_tranfer}, one can check that the heating gets maximize due to drag as $M_{\rm DM}\rightarrow M_b$. The dark matter temperature evolution can be written as, 
\begin{alignat}{2}
\frac{dT_{\rm DM}}{dz} = 2\,\frac{T_{\rm DM}}{(1+z)} +  \frac{2}{3\,(1+z)\,H}\,\frac{d{Q_{\rm DM}}}{dt}\label{eq:dm_temp}\,, 
\end{alignat}
here, first term represents the cooling of the dark matter due to expansion of the Universe. Heat transfer rate for dark matter $\left({d{Q_{\rm DM}}}/{dt}\right)$ can be obtained by interchanging $b \leftrightarrow {\rm DM}$ and $T_{\rm gas}\leftrightarrow T_{\rm DM}$ in equation \eqref{eq:heat_tranfer}. As drag term \eqref{eq:drag1} remains symmetric under the transformation $b \leftrightarrow {\rm DM}$, it heats the dark matter also. We can also check that total energy density of the system is conserved \cite{Munoz:2015bk},
\begin{alignat}{2}
N_{\rm DM}\,\frac{dQ_{\rm DM}}{dt}+N_{b}\,\frac{dQ_{\rm gas}}{dt}-\frac{\rho_{\rm DM}\, \rho_b}{\rho_{\rm M}}\ v\ D(v)=0\,,
\end{alignat}
here, $N_{\rm DM}$ and $N_b$ are number density of dark matter and baryons. As the relative motion between dark matter and baryons is damped due to friction between both fluids and expansion of the Universe, one can write the evolution of relative motion as,
\begin{alignat}{2}
\frac{dv}{dz}& =  \frac{v}{1+z} + \frac{D(v)}{(1+z)\,H}\,.\label{eq:vel}
\end{alignat}
Temperature evolutions of the gas and dark matter require free electron fraction. It  is given by equation \eqref{s4} with $\mathcal{E}=0\,$. As it has been confirmed in Ref. \cite{Minoda:2018gxj}, that cooling due to effects like Ly$\alpha$ emission, Bremsstrahlung and recombination does not have that much effects on the dynamics of the gas and dark matter, therefore, we have not considered these effects in the present work.

\section{Results and Discussion}\label{sec:res-dis}

\begin{figure}
    \begin{center}
        {\includegraphics[width=4.5in,height=3in]{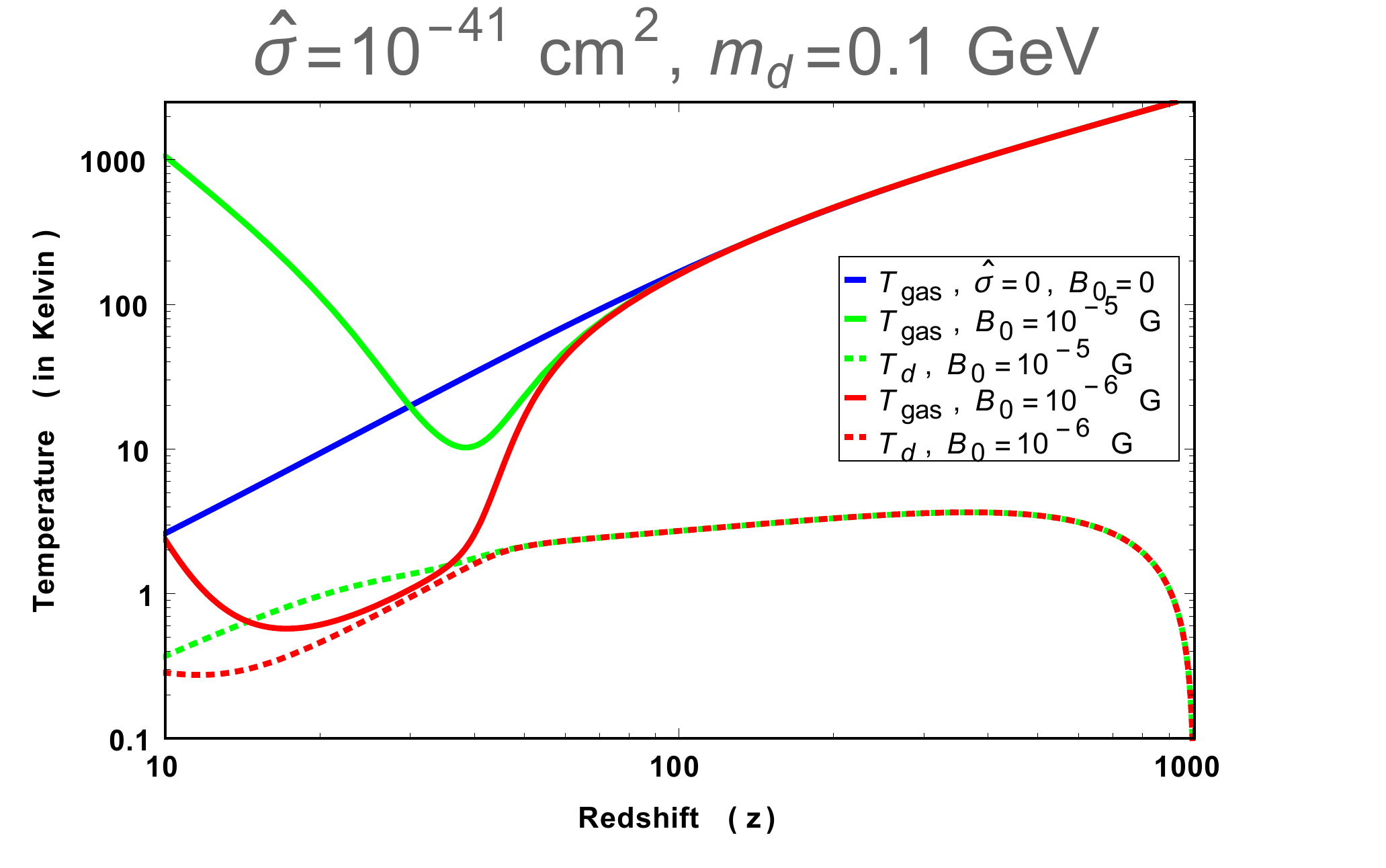}} 
    \end{center}
    \caption{This figure shows the temperature evolutions of baryon and dark matter in the presence of PMFs and baryon-dark matter interaction. Blue line corresponds to temperature evolution of gas in the absence of both magnetic heating and baryon-dark matter interaction. The red (green) solid lines represents the variation of the gas temperature and the dotted red (green) line shows the variation of the dark matter temperature in presence of PMFs and the baryon-dark matter interaction. In this plot we vary the strength of PMFs, and keep $\hat{\sigma}$ \& dark matter mass constant to $10^{-41}~{\rm cm^2}$ \& $10^{-1}$~GeV, respectively. In all figures, notation for the mass of dark matter is written with $m_d$. While in the text, it is written as $M_{\rm DM}$.}\label{fig:Temp-crossmass}
\end{figure}

\begin{figure}
    \begin{center}
        {\includegraphics[width=4.5in,height=3in]{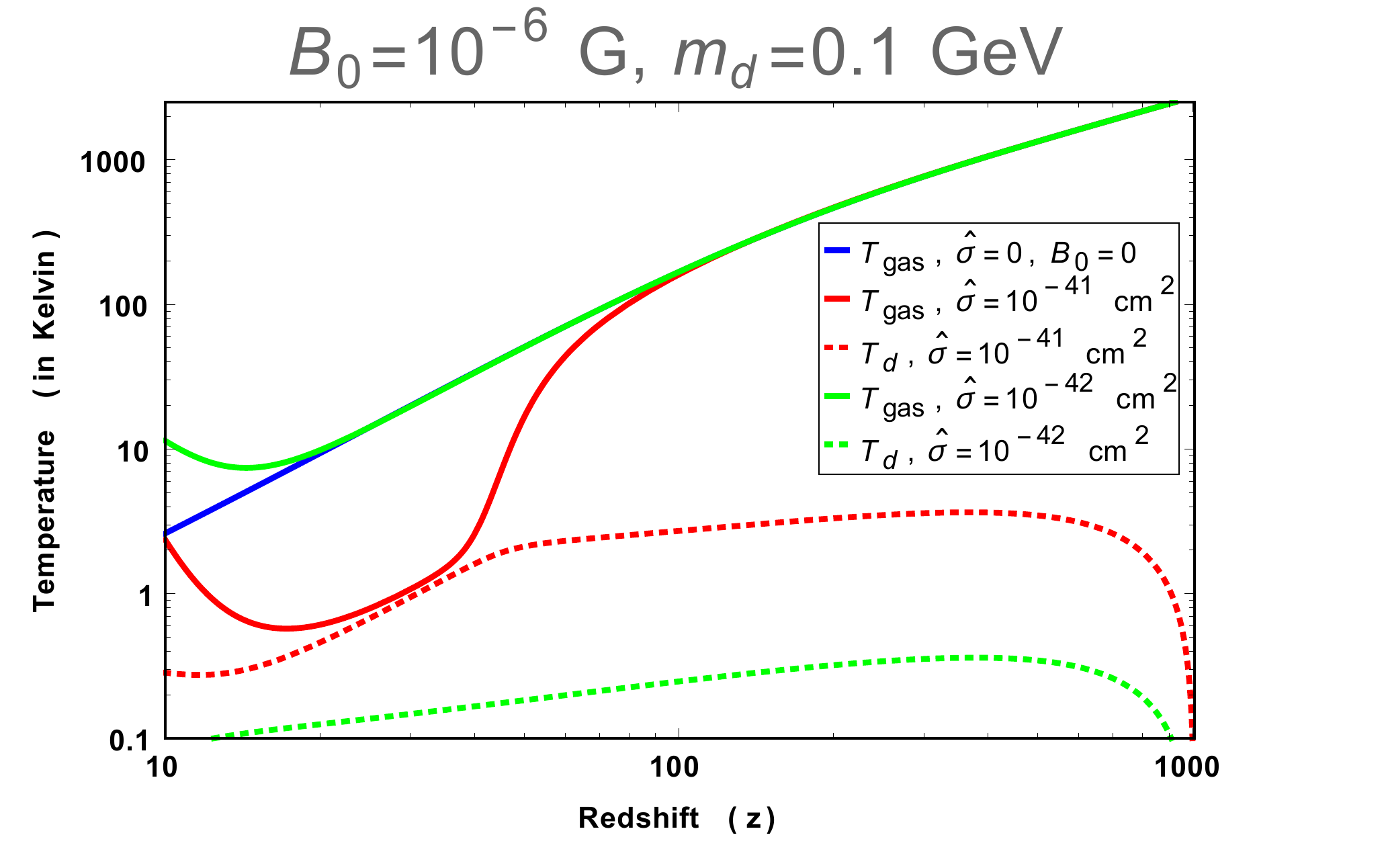}} 
    \end{center}
    \caption{The caption is same as in figure \eqref{fig:Temp-crossmass}, except here, we only vary the strength of baryon-dark matter cross-section, and keep  $B_0$ \& dark matter mass constant to $10^{-6}~{\rm G}$ \& $10^{-1}$~GeV, respectively.}\label{fig:Temp-cross-mass}
\end{figure}

\begin{figure}
    \begin{center}
        {\includegraphics[width=4.5in,height=3in]{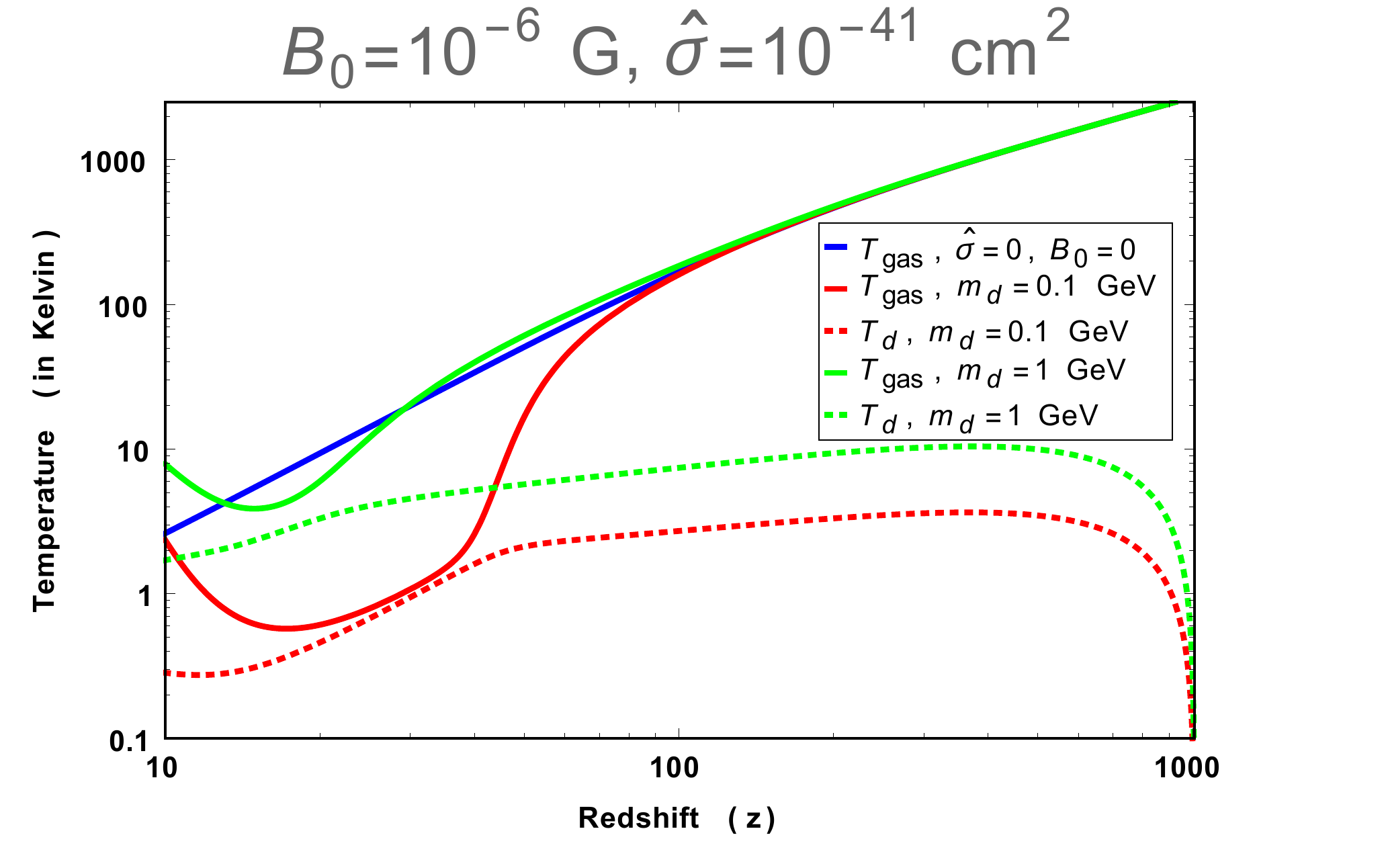}} 
    \end{center}
    \caption{The caption is same as in figure \eqref{fig:Temp-crossmass}, except here, we only vary the dark matter mass, and keep  $B_0$ \& $\hat{\sigma}$ constant to $10^{-6}~{\rm G}$ \& $10^{-41}~{\rm cm^2}$, respectively. }\label{fig:Tem-B-cross}
\end{figure}

Solving coupled equations (\ref{s4} with $\mathcal{E}=0\,$, \ref{11}, \ref{eq:baryon_temp}, \ref{eq:dm_temp} and \ref{eq:vel}) with initial conditions $T_{\rm gas}(1010)\simeq T_{\rm CMB}(1010)$, $T_{\rm DM}(1010)\sim 0$~K, $x_e(1010)=0.057$ and $B(z)=B_0\, (1+z)^2|_{z=1010}$ is the initial magnetic field strength, we get the temperature evolution of the dark matter and gas for different dark matter masses, strength of baryon-dark matter interaction cross-sections and magnetic field's strengths. Figures \eqref{fig:Temp-crossmass}, \eqref{fig:Temp-cross-mass} and \eqref{fig:Tem-B-cross}  show the evolution of the gas and dark matter temperature with redshift ($z$). The solid blue line in all these figures correspond to gas temperature when both the magnetic field and baryon-dark matter interaction are zero. In this case, gas temperature falls as $T_{\rm gas} \propto (1+z)^2$ after $z\sim200$ and reaches 6.8 K at $z=17$.

In figure \eqref{fig:Temp-crossmass}, temperature evolution of the gas and dark matter  is given for different strength of PMFs at constant  $\hat{\sigma} = 10^{-41} ~ {\rm cm}^2$ and $M_{\rm DM}=10^{-1}$~GeV. For both the cases $B_0 = 10^{-5}$~G and $10^{-6}$~G, gas temperature falls down due to Hubble expansion and baryon-dark matter interaction till $z \sim 30$ and $\sim 20$, respectively, then temperature rises due to magnetic heating. We note that, $T_{\rm DM}$ also increases due to the energy transfer from gas to dark matter depending on $\hat{\sigma}$ and $M_{\rm DM}$. Larger the strength of magnetic fields, earlier the heating begins. For example, heating  for the case with $B_0=10^{-5}$~G starts earlier compared to the case with $B_0=10^{-6}$~G in figure \eqref{fig:Temp-crossmass}. Although $T_{\rm DM}$ at $z \sim 1010$ is taken to be zero, it increases due to the energy transfer from baryons to dark matter. By increasing $B_0$, magnetic-heating of the gas rises, subsequently, the value of $T_{\rm DM}$ also rises. It can be seen in figure \eqref{fig:Temp-crossmass}, temperature of dark matter for $B_0=10^{-5}$~G is larger compared to $B_0=10^{-6}$~G.

Figure \eqref{fig:Temp-cross-mass} shows the temperature evolution of gas and dark matter for different strength of baryon-dark matter interaction cross-section when $B_0=10^{-6}$~G and $M_{\rm DM}=10^{-1}$~GeV are fixed. Larger the $\hat\sigma$, more heat transfers from gas to dark matter and cools the gas efficiently. For the green lines $\hat{\sigma}=10^{-42}~{\rm cm^2}$. As we increase $\hat{\sigma}$ to $10^{-41}~{\rm cm^2}$, the gas temperature decreases--- shown by red solid line. It decreases because the energy transfer from gas to dark matter becomes more efficient by increasing interaction between dark matter and baryons. It results in more heating of dark matter--- shown by red dashed line.

For $B_0=10^{-6}$ G and $\hat{\sigma} = 10^{-41} ~{\rm cm}^2$, temperature evolution for different dark matter mass is shown in Figure \eqref{fig:Tem-B-cross}. As we increase the dark matter mass from $10^{-1}$ GeV to $1$~GeV, temperature of both the dark matter and gas increases, and it becomes more efficient for large dark matter mass \cite{Munoz:2015bk}.  This drag heating is important when mass of dark matter is around $\sim1$~GeV \cite{Munoz:2015bk}. When $M_{\rm DM}$ approaches to $1$~GeV, in addition to magnetic heating of the gas, the heating due to drag term  also becomes effective . Therefore, the  gas temperature for $M_{\rm DM}=1$~GeV is higher than $M_{\rm DM}=10^{-1}$~GeV.
%

\subsection{Correlation between dark matter mass and baryon-dark matter cross section}\label{dm-cross}

\begin{figure}
    \begin{center}
        {\includegraphics[width=4.5in,height=3in]{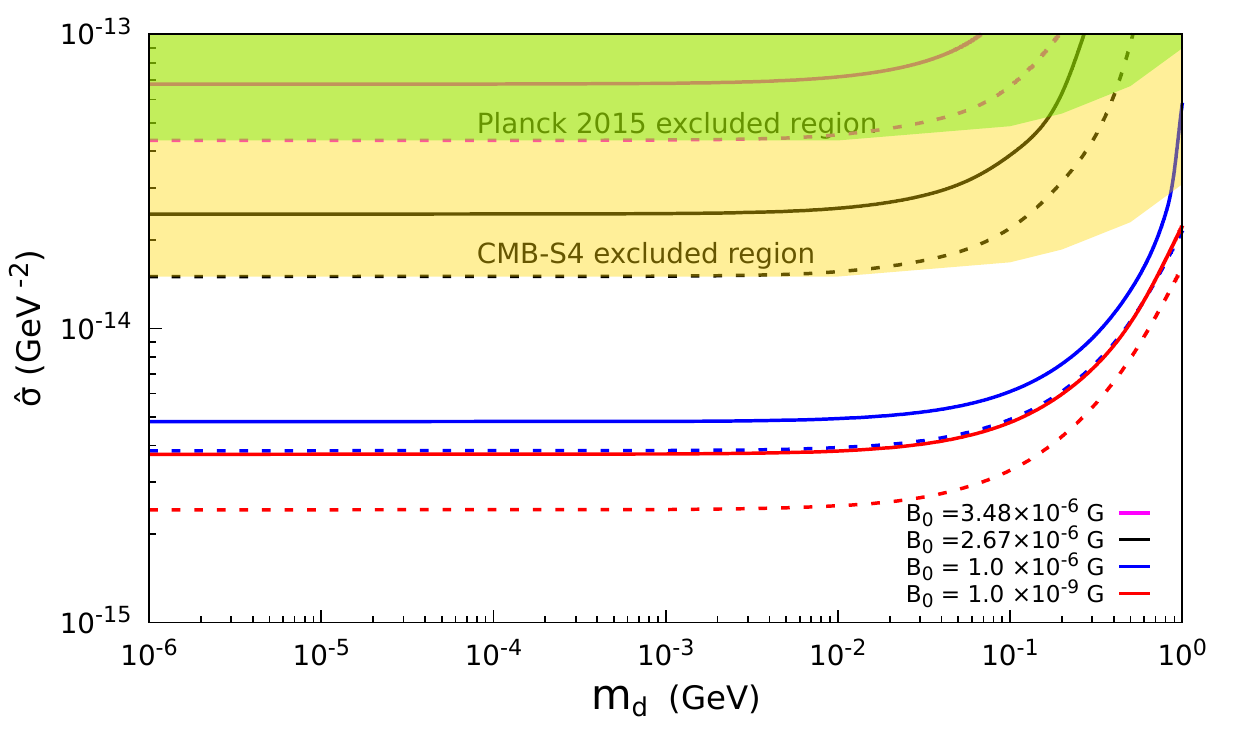}} 
    \end{center}
    \caption{The figure shows the minimal cross-section required to get $T_{21}\simeq-500$~mK (solid lines) and $T_{21}\simeq-300$~mK (dashed line) at $z=17$ as a function of mass for different strengths of PMFs. Here, we assume no x-ray heating of gas, and spin temperature is completely coupled to gas temperature, i.e. $T_{\rm gas}\simeq T_S$. The solid (dashed) magenta, black, blue and red line correspond to $B_0=3.48\times10^{-6}$~G, $2.67\times10^{-6}$~G, $10^{-6}$~G and $10^{-9}$~G respectively. The CMB-S4 (forecast) and Planck 2015 constraints on $\hat \sigma$ and $M_{\rm DM}$ with 95\% C.L. have been taken from the Refs. \cite{Kovetz2018, Boddy:2018G}. The green and gold regions are excluded by Planck 2015 and CMB-S4 forecast respectively.   ($1$ GeV$^{-2}=3.89 \times 10^{-28}$ cm$^2$)  }\label{fig:cross-mass}
\end{figure}

In this subsection, we analyse the effect of  $B_0$,  $M_{\rm DM}$ and  $\hat\sigma$ on gas and dark matter temperature. In  Fig. \eqref{fig:cross-mass}, we study constraints on  $M_{\rm DM}$ and $\hat\sigma$ for  $T_{21}\simeq-500$~mK ($T_{\rm gas}\simeq3.26$~K) and  $-300$~mK ($T_{\rm gas}\simeq5.2$~K). Here, we have taken $x_\alpha\gg1$ to plot $T_{21}$ profiles. Thus, from equation \eqref{Ts-1} one can get $T_S\approx T_{\rm gas}$ as $x_c$ is already $\ll1$ at required redshift due to the small number density of hydrogen, free electrons and protons. Subsequently, one can calculate $T_{21}$ from equation \eqref{t21f}. In  Fig. \eqref{fig:cross-mass}, we consider cases $B_0=3.48\times10^{-6}$~G, $2.67\times10^{-6}$~G, $10^{-6}$~G and $10^{-9}$~G and  solve equations (\ref{s4} with $\mathcal{E}=0$) and (\ref{11}, \ref{eq:baryon_temp}, \ref{eq:dm_temp} \& \ref{eq:vel}) for $T_{\rm gas}\simeq3.26$ and $5.2$~K at $z=17$ to get $\hat{\sigma}$ vs $M_{\rm DM}$ plots.  The solid and dashed lines represent the cases when $T_{21}\simeq-500$~mK and $-300$~mK, respectively. The  gold and green regions respectively show the CMB-S4 (forecast) and Planck 2015 upper constraint on $\hat\sigma-M_{\rm DM}$ with 95\% C.L.  \cite{Kovetz2018, Boddy:2018G}. The magenta, black, blue and red lines corresponds to $B_0=3.48\times10^{-6}$~G, $2.67\times10^{-6}$~G, $10^{-6}$~G and $10^{-9}$~G. As we increase the magnetic field strength from $10^{-9}$~G to $\sim10^{-6}$~G, larger value of $\hat\sigma$ is required for $M_{\rm DM} \in \{10^{-6},~ 1\}$~GeV to maintain $T_{21}\simeq-500$ or $-300$~mK at z=17. To get EDGES upper limit on $T_{21}$ (i.e. $-300$~mK), required $\hat{\sigma}$ is smaller compared to the case when $T_{21}=-500$~mK. This is  because we need to transfer less energy from gas to the dark matter to obtain EDGES upper limit on $T_{21}$. We get the upper limit on PMFs strength to $2.67\times 10^{-6}$~G by CMB-S4 (forecast) constraint on $\hat{\sigma}-M_{\rm DM}$ and maintaining $T_{21}\simeq-300$~mK at z=17. For $B_0=2.67\times 10^{-6}$~G, $M_{\rm DM} \gtrsim 10^{-2}$~GeV is excluded by CMB-S4 forecast for $T_{21}\simeq-300$~mK. By Planck 2015 constraint on $\hat{\sigma}-M_{\rm DM}$, the allowed maximum strength of PMFs is $3.48\times10^{-6}$~G by requiring EDGES upper constraint on $T_{21}$ at z=17. For $B_0=3.48\times10^{-6}$~G, mass of dark matter $\gtrsim 1\times10^{-2}$~GeV is excluded. Similarly, for the $B_0=10^{-6}$~G, $M_{\rm DM}\gtrsim 8\times10^{-1}$~GeV is excluded by CMB-S4 forecast.  When the dark matter mass approaches mass of hydrogen, the drag term in equation (\ref{eq:drag1}) also starts to contribute significantly in heating of the gas in addition to the magnetic heating. Therefore, higher mass of dark matter is excluded for higher magnetic field as shown by figure \eqref{fig:cross-mass}. As discussed in \cite{Munoz:2015bk}, when $M_{\rm DM}\sim 1$ GeV, the drag term heat up both the gas and dark matter in  such a way that we can not obtain $T_{\rm gas}= 3.26$ K at $z=17$ as required for the EDGES signal. There is a independent bound on the primordial magnetic field from CMB of the order of $\lesssim$~nG \cite{Trivedi:2012ssp, Trivedi:2013wqa}. This constraint, in our analysis, restricts value of $\hat\sigma\,$. Here, we note that in our analysis further decreasing value of $B_0$ below $10^{-9}$~G, does not change our result in significant way.


\subsection{Effect of primordial magnetic fields on the global 21 cm signal}{\label{B21}}
\begin{figure}
    \begin{center}
        {\includegraphics[width=4.5in,height=3in]{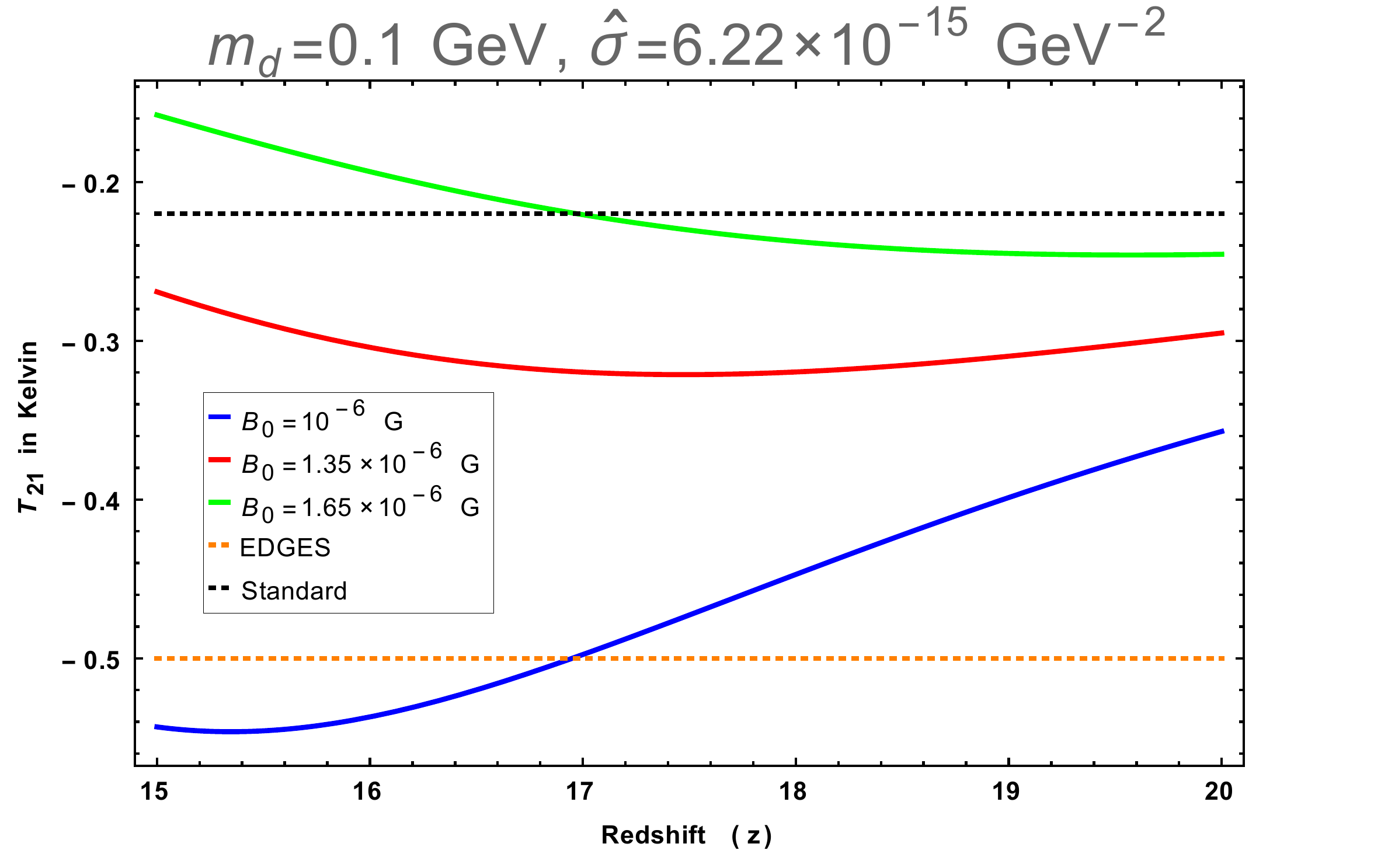}} 
    \end{center}
    \caption{21 cm differential brightness temperature (assuming infinite Ly$\alpha$ coupling) vs redshift when their is no x-ray heating.  The dotted black (orange) colour represents standard $\Lambda$CDM (EDGES) predictions for the global $T_{21}$ signal. Green, red and blue solid curves correspond to $B_0= 1\times 10^{-6}$, $1.35 \times 10^{-6}$ and $ 1.65\times 10^{-6}$~G respectively. Here, $M_{\rm DM}=10^{-1}$~GeV and $\hat{\sigma} = 6.22 \times 10^{-15} $~GeV$^{-2}$.}\label{plot:1a}
\end{figure}
\begin{figure}
    \begin{center}
        {\includegraphics[width=4.5in,height=3in]{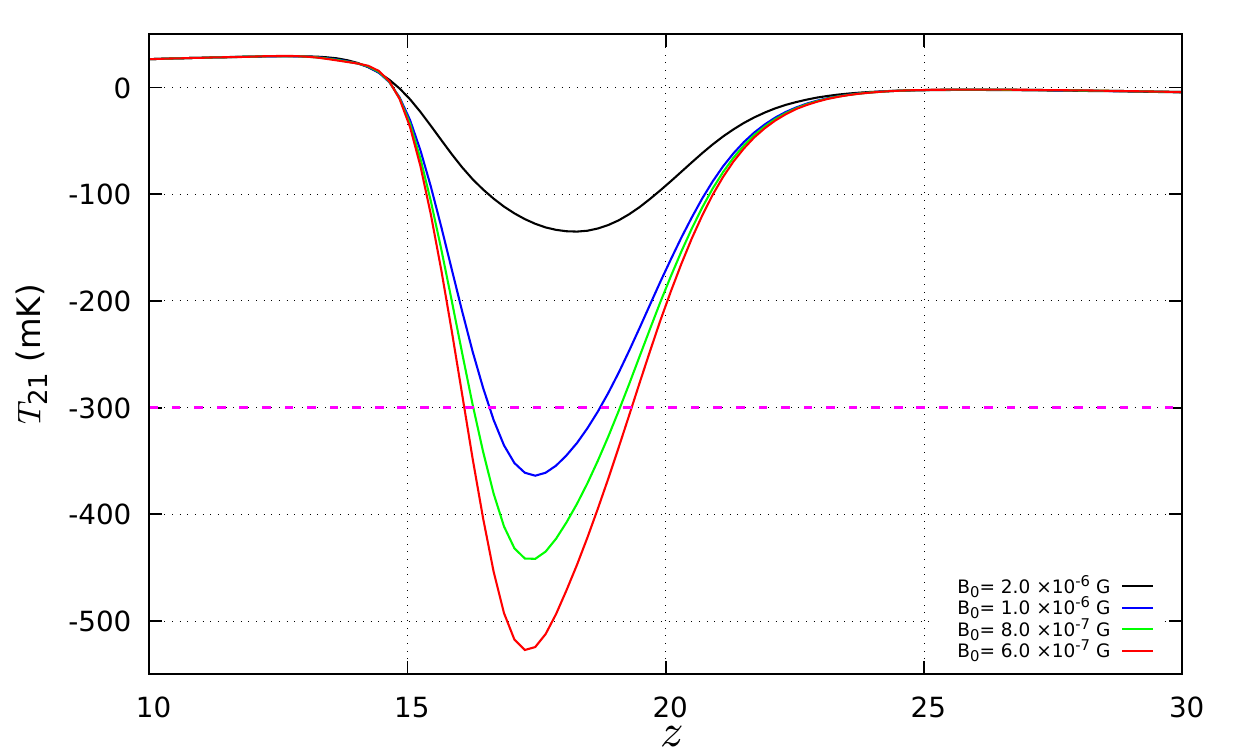}} 
    \end{center}
    \caption{$T_{21}$ plot with redshift when x-ray heating and finite Ly$\alpha$ coupling are considered \cite{Kovetz2018}. Black, blue, green and red solid curves correspond to $B_0= 2\times 10^{-6}$, $1 \times 10^{-6}$, $ 8\times 10^{-7}$ and $ 6\times 10^{-7}$~G respectively. The magenta dashed line is  corresponds to the EDGES upper bound on $T_{21}: -300$~mK. The values of $M_{\rm DM}$ and $\hat{\sigma}$ are same as considered  in figure \eqref{plot:1a}.}\label{plot:1b}
\end{figure}
%
%

We have discussed above that, with increase in the strength of the magnetic field the temperature of the gas increases for a fix $M_{\rm DM}$ and $\hat \sigma$. In figures \eqref{plot:1a} \& \eqref{plot:1b}, we plot 21 cm differential brightness temperature with redshift for different magnetic field strengths. These figures are obtained by keeping $M_{\rm DM}=10^{-1}$~GeV and $\hat{\sigma}=6.22\times 10^{-15}$~GeV$^{-2}$ constant. In figure \eqref{plot:1a}, to plot $T_{21}$ we assume infinite Ly$\alpha$ coupling ($x_\alpha\rightarrow\infty\Rightarrow T_S\simeq T_{\rm gas}$) and do not include the x-ray heating.  For $B_0=1\times10^{-6}$~G, the 21 cm line absorption signal reported by EDGES (i.e. $-500$~mK) can be explained. In figure  \eqref{plot:1b}, we include the x-ray heating and consider finite Ly$\alpha$ coupling ($x_\alpha$) \cite{Kovetz2018, Harker:2015M, Mirocha:2015G, Zygelman:2005}. As we decrease $B_0$ from $2\times10^{-6}$~G, the minimum value of  $T_{21}$ profile decreases. For the case when  $B_0=1\times10^{-6}$~G (blue solid line), minimum of $T_{21}$ profile is well below the EDGES upper limit on $T_{21}$ (i.e. $-300$~mK--- magenta dashed line). In figure \eqref{plot:1a}, when there is infinite Ly$\alpha$ coupling,  $T_{21}=-300$~mK  corresponds to  $B_0=1.35\times10^{-6}$~G.  Thus, we need to lower $B_0$ values  when the finite Ly$\alpha$ coupling is considered to get  desired value of $T_{21}$. As shown in figures \eqref{plot:1a} and \eqref{plot:1b}, brightness temperature is suppressed  by the increase of the strength of the magnetic field and it can even erase the standard 21 cm signal when the magnetic field strength increases above $\sim1 \times 10^{-6}$~G. This sets the upper limit on the strength of the magnetic field for $M_{\rm DM}=10^{-1}$~GeV and $\hat{\sigma}=6.22\times 10^{-15}~{\rm GeV}^{-2}$.


\section{Conclusions}\label{sec:conc}

Magnetic fields in \cite{Sethi:2004pe, Minoda:2018gxj} have shown to heat the gas during the cosmic dawn era by the ambipolar diffusion and the turbulence decay.  Since, it could erase the observed 21~cm absorption signal, one can calculate the upper bound on the magnetic field. One of the promising mechanisms to explain the absorption signal of the 21~cm line is to have interaction between the dark matter and baryons \cite{Barkana:2018lgd, Barkana:2018nd}. In this work, we have shown that in the presence of such an interaction the upper bound on the strength of magnetic fields can significantly be altered. The magnetic-energy converted to the thermal energy and it heats both the gas and dark matter when $\hat \sigma$ is non-zero. This is an extra heating effect of dark matter in addition to the drag heating. The drag term heats the dark matter and baryons; but in the lower range of dark matter mass ($\ll1$~GeV) it becomes small. To explain the observed anomaly in the 21 cm signal by the EDGES, a large baryon-dark matter scattering cross-section is required to balance the magnetic heating effect.  An earlier saturation occurs in baryon-dark matter cross-section in the presence of the strong magnetic fields. We have also explored the millicharged dark matter scenario. In this case, we are not able to reproduce the EDGES signal by considering the upper bound on $\hat{\sigma}-M_{\rm DM}$ by Planck 2015 and CMB-S4 (forecast) \cite{Kovetz2018, Boddy:2018G}. Recently, the similar results about ``millicharged" and ``Coulomb-like" dark matters also have been obtained in reference \cite{Driskell:2022}. They find that 100\% millicharged dark matter scenario can not reproduce the EDGES result for any parameter space. The inclusion of PMFs will further increase the gas temperature and reduce the amplitude of 21-cm absorptional signal. Therefore, it will further worsen the situation for millicharged dark matter scenario.

Considering upper bound on $\hat{\sigma}-M_{\rm DM}$ by Planck 2015 \cite{ Boddy:2018G} and EDGES upper constraint on $T_{21}$ ($-300$~mK) at $z=17$ \cite{Bowman:2018yin}, we found upper bound on the magnetic field strength: $B_0=3.48\times10^{-6}$~G, while considering CMB-S4 forecast constraint \cite{Kovetz2018} we get $B_0=2.67\times 10^{-6}$~G  for the dark matter mass $\lesssim 10^{-2}$~GeV. \\
\clearpage
\pagestyle{empty}
\cleardoublepage
\pagestyle{fancy}
\begin{savequote}[75mm]
    ``I seem to have been only like a boy playing on the seashore, and diverting myself in now and then finding a smoother pebble or a prettier shell than ordinary, whilst the great ocean of truth lay all undiscovered before me."
    \qauthor{Isaac Newton, \textit{``Memoirs of Newton" (1855), Vol II By David Brewster}}
\end{savequote}

\chapter{Summary and Future outlook}\label{chap6}
\vspace{-1.5cm}

\section{Summary}

The 21 cm signal is shown to be a prestigious probe in the cosmological laboratory to provide robust bounds on the physics of the early and late time Universe. The signal can give a good insight into the period when the galaxies and first stars were formed. In the thesis, I have analysed bounds on the present-day strength of primordial magnetic fields, sterile neutrino lifetime \& mixing angle with active neutrinos, and primordial black hole dark matter fraction using the global 21 cm signal during the cosmic dawn era. The 21 cm line corresponds to the wavelength of hyperfine transition between 1S singlet and triplet states of the neutral hydrogen atom. The corresponding frequency for the 21 cm line is 1420.4 MHz. For a transition at redshift z, the frequency can be mapped for a present-day observed frequency as $1420.4/(1+z)\,$. 

In the $\Lambda$CDM framework of cosmology, the evolution of the gas temperature and ionization fraction are well-established during the cosmic dawn era \cite{Seager1999}. The addition of any exotic source of energy can significantly impact the ionization and thermal history of the Universe. The change in the gas temperature can significantly modify the absorption feature in the global 21 cm signal during cosmic dawn \cite{Pritchard_2012}. This can provide constraints on the properties of such exotic sources of energy injection.

The EDGES collaboration has reported the 21 cm differential brightness temperature: $T_{21} =-500^{+200}_{-500}$~mK with 99 percent confidence limit centred at 78~MHz or redshift $z = 17.2$ \cite{Bowman:2018yin}. By considering $T_{\rm S}=T_{\rm gas}$, the observed brightness temperature translates to gas temperature as $T_{\rm gas}(z=17.2)=3.26^{+1.94}_{-1.58}$~K. In the $\Lambda$CDM framework, the  gas temperature at redshift  $z=17.2$ remains around 7~K. This corresponds to differential brightness temperature $T_{21}(z=17.2)\simeq -220$~mK--- equation \eqref{t21f} for $T_{\rm S}\simeq T_{\rm gas}$. To resolve the tension between the theoretical prediction based on $\Lambda$CDM model and EDGES observation, one requires to increase the ratio of $T_{\rm R}/T_{\rm S}$ in equation \eqref{t21f} over theoretical predictions in redshift range $15\leq z\leq 20$. This can be achieved either by increasing the background radiation or decrease the gas temperature. Both possibilities have been studied by several authors; for example, see the Refs. \cite{Ewall-Wice2018, Biermann:2014, Jana:2018, Feng2018, Lawson:2019,Natwariya:2021, Lawson:2013, Levkov:2020, Fixsen2011, Natwariya:2020, Brandenberger:2019, Chianese:2019,Bhatt2019pac, Tashiro:2014tsa, Barkana:2018lgd, SIKIVIE2019100289, Mirocha2019, Ghara2019}. However, such mechanisms to increase the background radio radiation or cooling the gas are debatable issues. One of such mechanisms to cool gas is baryon dark matter interaction \cite{Barkana:2018lgd}. This approach has been questioned by several authors   \cite{Munoz2018, FRASER2018159, Bransden1958, Barkana:2018nd, Berlin2018, Kovetz2018, Munoz2018a, Slatyer2018}. Here, it is to be noted that the authors do not consider heating of the gas by decaying or annihilating dark matter. Injection of electrons and photons by decaying or annihilating dark matter into IGM can heat the gas more than cooling of the gas \cite{Amico:2018, Mitridate:2018}.  Moreover, the EDGES measurement  has been also questioned in several articles \cite{Saurabh:2021, Saurabh:2019, Tauscher:2020, Hills:2018, Bradley:2019}.  Recently, SARAS 3 observation reported that the  EDGES observation is not of an astrophysical origin and it is rejected with the 95.3 percent confidence level \cite{Saurabh:2021}. In the Ref. \cite{Hills:2018}, the authors have questioned the fitting parameters for the foreground emission and data. There is a possibility that the absorption feature in the EDGES observation can be a ground screen artifact \cite{Bradley:2019}. The absorption amplitude may modify depending on modelling of  foreground \cite{Saurabh:2019, Tauscher:2020}. In Ref. \cite{Sims:2019}, the authors perform the Bayesian comparison of fitting models for EDGES data and argue that the highest evidence models favour an amplitude of $|T_{21}|<209$~mK. In the light of these controversies, it is require to verify the EDGES result by other observations. The future updated version of the hydrogen Epoch of Reionization Array (HERA)\footnote{\href{http://reionization.org/}{http://reionization.org/}}, Thirty Meter Telescope (TMT)\footnote{\href{http://tmt.org/}{http://tmt.org/}}, JWST, etc., will be able to probe the cosmic dawn era more precisely.  The following summarizes the results reported in the thesis:

\subsection{Bounds on dark matter candidates} \label{BDMC}

About 85 per cent of the total matter content in the Universe is dominated by dark matter. In the last decades, many dark matter models have been proposed to explain various astrophysical observations. However, the microscopic nature of dark matter is still unknown. During my doctoral research, I have considered sterile neutrinos and primordial black holes as dark matter candidates and constrain their properties using the absorption feature in 21 cm differential brightness temperature during the cosmic dawn era. As discussed earlier, here, we have taken 21~cm differential brightness temperature such that it does not change from its standard theoretical value ($\sim-220$~mK) by more than a factor of 1/4 (i.e. $-150$~mK) or 1/2 (i.e. $-100$~mK) at redshift 17.2\,.

\subsubsection{Sterile Neutrino Dark Matter}

In the warm dark matter models, one of the theoretically well-motivated candidates is KeV mass sterile neutrinos. We have constrained the sterile neutrino dark matter lifetime and mixing angle with active neutrino as a function of sterile neutrino mass \cite{Natwariya:2022}. Here, we have considered the two scenarios to get the bounds: First, IGM evolution without the heat transfer from the background radiation to gas mediated by Ly$\alpha$ photons (VDKZ18 effect). Next, we have considered additional VDKZ18 heating effects on the IGM gas. The following summarises our results for $T_{21} = -150$~mK\,: 

In the first scenario, the lower bound on the sterile neutrino lifetime varies from $8.3\times10^{27}$~sec to $9.4\times10^{25}$~sec by varying sterile neutrino mass from 2 KeV to 50 KeV.  While the upper bound on the mixing angle varies from  $6.8\times10^{-9}$ to  $6.1\times10^{-14}$ by varying sterile neutrino mass from 2 KeV to 50 KeV.

In the second scenario, the lower bound on the sterile neutrino lifetime varies from $1.5\times10^{28}$ sec to  $1.7\times 10^{26}$ sec by varying sterile neutrino mass from 2 KeV to 50 KeV. While the upper bound on the mixing angle varies from $3.8\times10^{-9}$ to $3.42\times10^{-14}$ by varying sterile neutrino mass from 2 KeV to 50 KeV.

\subsubsection{Primordial Black Hole Dark Matter}

Spinning primordial black holes can substantially affect the ionization and thermal history of the Universe.  Subsequently, it can modify the 21 cm absorption signal during cosmic dawn era by injecting energy due to Hawking evaporation. We study the upper  projected bounds on the fraction of dark matter in the form of PBHs as a function of mass and spin.  Our  projected constraints are stringent compared to DSNB, INTEGRAL observation of the 511~KeV line, IGRB, Planck, Leo T and COMPTEL. In the near future, AMEGO collaboration will be able to probe some parameter space in our considered mass range of PBHs. In the present work, we have considered the monochromatic mass distribution of PBHs. The allowed parameter space can also be explored for different PBHs mass distributions such as log-normal, power-law, critical collapse, etc. \cite{Arbey_2019}. We find the most robust lower projected constraint on the mass of PBHs, which is allowed to constitute the entire dark matter,  to $1.5\times10^{17}$~g, $1.9\times10^{17}$~g, $3.9\times10^{17}$~g and $6.7\times10^{17}$~g for PBH spins 0, 0.5, 0.9 and 0.9999, respectively. The lower bound on $M_{\rm PBH}$ for $\Omega_{\rm PBH}=\Omega_{\rm DM}$, for extremal spinning PBHs is nearly four times larger than non-spinning ones \cite{Natwariya:2021PBH}.

\subsection{Primordial Magnetic Fields}\label{sPMFs}

Observations suggest that the magnetic fields are ubiquitous in the Universe--- from the length scale of planets and stars to the cluster of galaxies \cite{Haverkorn:2008, Fletcher:2011, Carilli:2002, Brandenburg20051}. The origin and evolution of PMFs are one of the outstanding problems of modern cosmology (Ref. \cite{Subramanian:2016, Subramanian:2019} and references cited therein). Decaying PMFs can inject magnetic energy into the thermal energy of the IGM and heat the gas. As discussed earlier, the EDGES collaboration reported an absorption profile for the global 21 cm signal with an amplitude of $-500^{+200}_{-500}$~mK in the redshift range $15-20$. To explain the EDGES anomaly, one requires to enhance the background radio radiation above the CMB radiation or lower the gas temperature below $3.2$~K at redshift $\sim 17$. We have explored the upper bounds on the present-day strength of the PMFs in both scenarios by considering different models \cite{Natwariya:2021, Bhatt2019pac}.

\subsubsection{In the Presence of Excess Radio Radiation}

As discussed, one requires to enhance the background radiation above the CMBR to explain the EDGES anomaly. For excess radiation fraction to be LWA 1 limit, we have reported upper bounds on the present-day PMFs strength, $B_0$ on the scale of 1~Mpc. The following summarises our results for $T_{21}=-500$~mK (EDGES best fit result): 

We have reported $B_0\lesssim 3.7$~nG for spectral index $n_B=-2.99$ for excess radiation fraction to be LWA 1 limit. While for $n_B=-1$, the upper bound gets more stringent: $B_0\lesssim1.1\times10^{-3}$~nG. We also discuss the effects of first stars on IGM gas evolution and the allowed value of $B_0$. By decreasing excess radiation fraction below the LWA 1 limit, we get a more stringent bound on $B_0$ \cite{Natwariya:2021}.

\subsubsection{In the Presence of Baryon-Dark Matter Interaction}

One of the alternatives to explain the EDGES anomaly is by cooling the gas below 3.2 K. Since the dark matter is colder than the gas, adequate cooling of the gas can be obtained by introducing the baryon-dark matter interaction beyond the $\Lambda$CDM model. The introduction of baryon-dark matter interaction relaxes the upper bound on $B_0$ by transferring energy of the gas to the dark matter using drag between gas and dark matter.  Considering upper bound on $\hat{\sigma}-m_d$ by Planck 2015 and EDGES upper constraint on $T_{21}$ ($-300$~mK) at $z=17$, we found upper bound on the  present-day strength of PMFs: $B_0=3.48\times10^{-6}$~G, while considering CMB-S4 forecast constraint we get $B_0=2.67\times 10^{-6}$~G  for the dark matter mass $\lesssim 10^{-2}$~GeV. We have also discussed the bounds on $\hat{\sigma}-m_d$ by considering Planck 2018 upper bound on $B_0\sim10^{-9}$~G for EDGES best fit and upper bound on $T_{21}$ \cite{Bhatt2019pac}.

\clearpage
\pagestyle{empty}
\cleardoublepage
\pagestyle{fancy}
\appendix
\appendix
\chapter[Appendix]{}\label{appendA}
\section{Spin temperature of hydrogen}\label{appendA1}

In the presence of collisions, rate of change in the population of singlet state \cite{Field},
\begin{alignat}{2}
\frac{dn_{\rm S}}{dt}=-n_{\rm S}\,P_{\rm \scriptscriptstyle ST}^{\rm C}+n_{\rm T}\,P_{\rm \scriptscriptstyle TS}^{\rm C}\,.\label{coll}
\end{alignat}
In the steady state, the transition coefficients from equation \eqref{coll}: ${n_{\rm T}}/{n_{\rm S}}={P_{\rm \scriptscriptstyle ST}^{\rm C}}/{P_{\rm \scriptscriptstyle TS}^{\rm C}}\,$. In the presence of collisions, the spin temperature will be kinetic temperature of gas only. Therefore, from equation \eqref{r_pop},
\begin{alignat}{2}
{P_{\rm \scriptscriptstyle ST}^{\rm C}}=3\ \exp\left[-\frac{T_{\rm \scriptscriptstyle TS}}{T_{\rm gas}}\right]\times {P_{\rm \scriptscriptstyle TS}^{\rm C}}\simeq 3\ \left[1-\frac{T_{\rm \scriptscriptstyle TS}}{T_{\rm gas}}\right]\times {P_{\rm \scriptscriptstyle TS}^{\rm C}}\,.\label{a2}
\end{alignat}
As discussed in the section \eqref{sec212}, $T_{\rm gas},\ T_\alpha\gg T_{\rm \scriptscriptstyle TS}\,$: $\exp\left[-{T_{\rm \scriptscriptstyle TS}}/{T_{\rm gas}}\right]\simeq 1-{T_{\rm \scriptscriptstyle TS}}/{T_{\rm gas}}\,$. Similarly, for the Ly$\alpha$ radiation, $T_{\rm gas}$ and ${P_{\rm \scriptscriptstyle TS}^{\rm C}}$ will be replaced by $T_\alpha$ and $P_{\rm \scriptscriptstyle TS}^\alpha\,$, respectively, in equation \eqref{a2},
\begin{alignat}{2}
{P_{\rm \scriptscriptstyle ST}^\alpha}\simeq3\ \left[1-\frac{T_{\rm \scriptscriptstyle TS}}{T_\alpha}\right]\times {P_{\rm \scriptscriptstyle TS}^\alpha}\,.\label{a3}
\end{alignat}
In the hydrogen atom, there can be spontaneous and induced emissions by background radiation also,
\begin{alignat}{2}
P_{\rm \scriptscriptstyle TS}^{\rm R}=A_{10}+B_{10}\ I_\nu^{\rm R}\,,\label{a4}
\end{alignat}
here, $B_{10}\ I_\nu^{\rm R}$ is the induced emission due to background radiation and $I_\nu^{\rm R}$ is the specific intensity for 21 cm transition. Here, $A_{10}$ and $B_{10}$ are Einstein coefficients and their relation is given by $A_{10}=2\,\nu_{\rm \scriptscriptstyle TS}^2\,T_{\rm \scriptscriptstyle TS}\,B_{10}$\,. For the background radiation, in the Rayleigh-Jeans limit from equation \eqref{Inu}: $I_\nu^{\rm R}=2\,\nu_{\rm \scriptscriptstyle TS}^2\,T_{\rm R}\,$. Therefore, from equation \eqref{a4},
\begin{alignat}{2}
P_{\rm \scriptscriptstyle TS}^{\rm R}=\left(1+\frac{T_{\rm R}}{T_{\rm \scriptscriptstyle TS}}\right)\, A_{10}\,.\label{a5}
\end{alignat}
The induced transition from singlet to triplet due to background radiation \cite{Field},
\begin{alignat}{2}
P_{\rm \scriptscriptstyle ST}^{\rm R}=B_{01}\ I_\nu^{\rm R}=3\ B_{10}\ I_\nu^{\rm R}=3\ A_{10}\ \frac{T_{\rm R}}{T_{\rm \scriptscriptstyle TS}}\,. \label{a6}
\end{alignat}
Using equations \eqref{a5} and \eqref{a6},
\begin{alignat}{2}
\frac{P_{\rm \scriptscriptstyle ST}^{\rm R}}{P_{\rm \scriptscriptstyle TS}^R}\simeq3\ \left[1-\frac{T_{\rm \scriptscriptstyle TS}}{T_{\rm R}}\right] \,.\label{a7}
\end{alignat}
In the detailed balance between the population of $1$S singlet and triplet states (${dn_{\rm S}}/{dt}=0$), by solving the equation \eqref{pdn} with the use of equations \eqref{r_pop}, \eqref{a2}, \eqref{a3} and \eqref{a7}, we get,
\begin{equation}
\left[1-\frac{T_{\rm \scriptscriptstyle TS}}{T_{\rm S}}\right]=\frac{\left[1-\frac{T_{\rm \scriptscriptstyle TS}}{T_{\rm R}}\right]+x_\alpha\,\left[1-\frac{T_{\rm \scriptscriptstyle TS}}{T_{ \alpha}}\right] +x_c\,\left[1-\frac{T_{\rm \scriptscriptstyle TS}}{T_{\rm gas}}\right]}{1+x_\alpha+x_c}\,,\label{a8}
\end{equation}
here, $x_\alpha={P_{\rm \scriptscriptstyle TS}^{\rm \alpha}}/{P_{\rm \scriptscriptstyle TS}^{\rm R}}$ and $x_c={P_{\rm \scriptscriptstyle TS}^{\rm C}}/{P_{\rm \scriptscriptstyle TS}^{\rm R}}\,$. Solving the equation \eqref{a8}, we get \cite{Field, Pritchard_2012},
\begin{alignat}{2}
T_{\rm S}^{-1}=\frac{T_{\rm R}^{-1}+x_\alpha\,T_{\alpha}^{-1}+x_c\,T_{\rm gas}^{-1}}{1+x_\alpha+x_c}\,.
\end{alignat}

\section{Emergent brightness temperature}\label{appendA2}
Solving differential equation \eqref{RTe2} with initial conditions: when, $l=0\rightarrow$ $\tau_\nu=0$ and $I_\nu=I_{\nu_0}$ (figure \ref{Differential}),
\begin{equation}
I_\nu=S_\nu\,(1-e^{-\tau_\nu})+I_{\nu_0}\,e^{-\tau_\nu}\,.\label{a10}
\end{equation}
Here,  using equation \eqref{Inu},  $I_\nu=  2\,\nu^2\,T_{\rm R}'$ is the final/emergent specific intensity of light--- of frequency $\nu$. $S_\nu=  2\,\nu^2\,T_{\rm exc}$ is the specific intensity due to the medium having an excitation temperature, $T_{\rm exc}$, at a frequency of $\nu\,$. $I_{\nu_0}=  2\,\nu^2\,T_{\rm R}$ is the initial specific intensity of the light. As a result, we find the final/emergent brightness temperature as \cite{Field, Pritchard_2012},
\begin{equation}
T_{\rm R}'=T_{\rm exc}\,(1-e^{-\tau_\nu})+T_{\rm R}\,e^{-\tau_\nu}\,.\label{eq23}
\end{equation}

\section{Optical depth of hydrogen medium}\label{appendA3}

The radiative transfer equation in the presence of emission and absorption of a light with travelled distance $dl$ in the medium,
\begin{equation}
\frac{dI_\nu}{dl}=\frac{T_{\rm \scriptscriptstyle TS}}{4\,\pi}\ \phi(\nu)\,\left[\,n_{\rm \scriptscriptstyle T}\,A_{10}+ n_{\rm \scriptscriptstyle T}\,B_{10}\,I_\nu- n_{\rm \scriptscriptstyle S}\,B_{01}\,I_\nu\,\right]\,,\label{a12}
\end{equation}
here, $T_{\rm \scriptscriptstyle TS}=2\,\pi\,\nu_{\rm \scriptscriptstyle TS}$, and $\phi(\nu)$ represents line profile of the light beam. $T_{\rm \scriptscriptstyle TS}/(4\,\pi)$ represents the energy of light beam per unit solid angle. The first term in the bracket is due to the spontaneous emission from the triplet to the singlet state, and it is proportional to the population density of the triplet state. The second and third terms in the bracket are due to the stimulated/induced emission and absorption, respectively. 
Comparing equations \eqref{a12} and \eqref{RTe1}, we get,
\begin{equation}
\alpha_\nu=\frac{T_{\rm \scriptscriptstyle TS}}{4\,\pi}\ \phi(\nu)\, \left[\, n_{\rm \scriptscriptstyle S}\,B_{01}-n_{\rm \scriptscriptstyle T}\,B_{10}\,\right]\,.\label{a13}
\end{equation}
To get the optical depth of hydrogen medium, we can integrate equation \eqref{a13} over $dl$ (equation \ref{RTe3}), 
\begin{alignat}{2}
\tau_\nu=\frac{3\,A_{10}}{32\,\pi\,\nu_{\rm \scriptscriptstyle TS}^2}\times\frac{T_{\rm \scriptscriptstyle TS}}{T_{\rm S}}\times n_{\rm HI}\,\int\phi(\nu)dl\,,\label{a14}
\end{alignat}
here, we have used the relations: $A_{10}=2\,\nu_{\rm \scriptscriptstyle TS}^2\,T_{\rm \scriptscriptstyle TS}\,B_{10}$ and $B_{01}=3\ B_{10}\,$. As the neutral hydrogen number density: $n_{\rm HI}=n_{\rm \scriptscriptstyle S}+n_{\rm \scriptscriptstyle T}$, the singlet state population density can be approximated by $n_{\rm \scriptscriptstyle S}\simeq n_{\rm HI}/4\,$--- from equation \eqref{r_pop}. The ratio $n_{\rm \scriptscriptstyle T}/n_{\rm \scriptscriptstyle S}\,$, has given by equation \eqref{r_pop}. By solving the integral in equation \eqref{a14} for a line profile $\phi(\nu)=1/\Delta\nu$ with the Doppler shift in the frequency due to the moving medium with a proper velocity $v$ along the line of sight in the comoving coordinate ($\Delta r= (1+z)\,\Delta l\, $); we find the optical depth for hydrogen medium as \cite{Pritchard_2012},
\begin{alignat}{2}
\tau_\nu=\frac{3\,n_{\rm HI}}{32\,\pi\,\nu_{\rm \scriptscriptstyle TS}^3}\times\frac{T_{\rm \scriptscriptstyle TS}}{T_{\rm S}}\times \frac{A_{10}}{H}\times\left[\frac{H/(1+z)}{\partial v/\partial r}\right] \,.\label{a15}
\end{alignat}
Here, $\partial v/\partial r$ is the proper velocity gradient along the line of sight, and it can be taken as $H/(1+z)$ for high redshift or in the absence of peculiar velocity. Here, $n_{\rm HI}$ can be written as $x_{\rm HI}\,n_{\rm H}$, and $x_{\rm HI}$ is the neutral hydrogen fraction. The hydrogen number density can be expressed in the form of dimensionless baryon energy density: $n_{\rm H}\simeq8.5\times10^{-6}\ (1+\delta_{\rm b})\ \Omega_{\rm b}\,h^2\,(1+z)^3\,~{\rm cm^{-3}}\,$. Here, $\delta_{\rm b}=(\rho_{\rm b}-\bar{\rho}_{\rm b})/\bar{\rho}_{\rm b}$ is the baryon density contrast. $\rho_{\rm b}$ and $\bar{\rho}_{\rm b}$ are total and average baryon energy density, respectively. For the matter dominated era, we can take $H=H_0\,\sqrt{\Omega_{\rm m}}\ (1+z)^{3/2}\,$. Here, $H_0$ and $\Omega_{\rm m}$ are present-day values of Hubble parameter and dimensionless matter energy density parameter, respectively. After some manipulation, we get the final expression for optical depth of hydrogen medium for 21 cm line \cite{Zaldarriaga:2004, Mesinger:2007S, Mesinger:2011FS, Pritchard_2012, Mittal:2020},
\begin{alignat}{2}
\tau_\nu\simeq27\,x_{\rm HI}\,(1+\delta_{\rm b})\,(1+z)\,  \left(\frac{\rm mK}{T_{\rm S}}\right)\,\left(\frac{0.15}{\Omega_{\rm m }\,h^2}\,\frac{1+z}{10}\right)^{1/2}\left(\frac{\Omega_{\rm b}\,h^2}{0.023}\right)\,.\label{a16}
\end{alignat}
For a global 21 cm signal we can take $1+\delta_{\rm b}$  as $\sim1$.

\clearpage
\pagestyle{empty}
\cleardoublepage
\backmatter
\pagestyle{fancy}
\fancyfoot[RE,LO]{\textbf{\thepage}} 
\fancyfoot[CE,CO]{}
\fancyfoot[LE,RO]{\textbf{References}}
%
%
%
\setstretch{1.1}
\newcommand*{\doi}[1]{\href{http://dx.doi.org/#1}{DOI: #1}}
\renewcommand{\bibname}{References}
\bibliography{/home/pravin/Dropbox/Refrences/ref.bib}

\end{document}